\documentclass[12pt]{article}
\usepackage[utf8]{inputenc}
\usepackage[english]{babel}
\usepackage{amsmath}
\usepackage{mathrsfs}
\usepackage{amssymb}
\usepackage{amsthm}
\usepackage{amsfonts}
\usepackage{xspace}
\usepackage[normalem]{ulem}
\usepackage{graphicx}
\usepackage{multirow}
\usepackage{appendix}
\usepackage{enumerate}
\usepackage{url} % not crucial - just used below for the URL
\usepackage{authblk}
\usepackage{float}
\usepackage{xcolor}
\usepackage{subcaption}
\usepackage{enumitem}
\usepackage{array, makecell}

\usepackage{xcolor}
\usepackage{soul}
\usepackage{tikz}
\usetikzlibrary{calc}
\usetikzlibrary{arrows.meta,decorations.pathreplacing,calligraphy}
\usepackage{caption}
\usepackage{subcaption}

\usepackage[top=1in,bottom=1.5in,left=1in,right=1in]{geometry}

\usepackage{natbib}
\usepackage[unicode=true,bookmarks=true,bookmarksnumbered=false,bookmarksopen=false,breaklinks=false,pdfborder={0 0 1},backref=false,colorlinks=true]{hyperref}
\hypersetup{citecolor=blue}
\usepackage{url}

\newtheorem{theorem}{Theorem}

\newtheorem{proposition}{Proposition}

\newtheorem{assumption}{Assumption}

\newcommand{\bY}{\boldsymbol{Y}}
\newcommand{\bX}{\boldsymbol{X}}
\newcommand{\bW}{\boldsymbol{W}}
\newcommand{\bM}{\boldsymbol{M}}
\newcommand{\bc}{\boldsymbol{c}}
\newcommand{\bfX}{\mathbf{X}}
\newcommand{\bfQ}{\mathbf{Q}}

\newcommand{\bfI}{\mathbf{I}}

\newcommand{\bone}{\boldsymbol{1}}
\newcommand{\bzero}{\boldsymbol{0}}
\newcommand{\bbeta}{\boldsymbol{\beta}}
\newcommand{\btheta}{\boldsymbol{\theta}}
\newcommand{\balpha}{\boldsymbol{\alpha}}
\newcommand{\bpsi}{\boldsymbol{\psi}}
\newcommand{\bfSigma}{\mathbf{\Sigma}}

\title{\bf On the mixed-model analysis of covariance in cluster-randomized trials}
\author{\normalsize
Bingkai Wang$^{1}$, Michael O. Harhay$^{2,3}$, Jiaqi Tong$^{4,5}$, Dylan S. Small$^1$,  Tim P. Morris$^6$\\
\vspace{-0.1in}
and Fan Li$^{4,5}$
\vspace{10pt}

$^1$The Statistics and Data Science Department of the Wharton School, University of Pennsylvania, Philadelphia, PA, USA

$^2$Clinical Trials Methods and Outcomes Lab, Palliative and Advanced Illness Research (PAIR) Center, Perelman School of Medicine, University of Pennsylvania,
Philadelphia, PA, USA

$^3$Department of Biostatistics, Epidemiology and Informatics, Perelman School of Medicine, University of Pennsylvania,
Philadelphia, PA, USA

$^4$Department of Biostatistics, Yale School of Public Health, New Haven, CT, USA

$^5$Center for Methods in Implementation and Prevention Science, Yale School of Public Health, New Haven, CT, USA

$^6$MRC Clinical Trials Unit at UCL, London, UK}
\begin{document}
\def\spacingset#1{\renewcommand{\baselinestretch}%
{#1}\small\normalsize} \spacingset{1}

\date{\vspace{-5ex}}

\maketitle
\begin{abstract}
In the analyses of cluster-randomized trials, mixed-model analysis of covariance (ANCOVA) is a standard approach for covariate adjustment and handling within-cluster correlations. However, when the normality, linearity, or the random-intercept assumption is violated, the validity and efficiency of the mixed-model ANCOVA estimators for estimating the average treatment effect remain unclear. Under the potential outcomes framework, we prove that the mixed-model ANCOVA estimators for the average treatment effect are consistent and asymptotically normal under arbitrary misspecification of its working model. {If the probability of receiving treatment is $0.5$ for each cluster}, we further show that the model-based variance estimator under mixed-model ANCOVA1 (ANCOVA without treatment-covariate interactions) remains consistent, clarifying that the confidence interval given by standard software is asymptotically valid even under model misspecification.  {Beyond robustness, we discuss several insights on precision among classical methods for analyzing cluster-randomized trials, including the mixed-model ANCOVA, individual-level ANCOVA, and cluster-level ANCOVA estimators.} These insights may inform the choice of methods in practice. Our analytical results and insights are illustrated via simulation studies and analyses of three cluster-randomized trials. 
\end{abstract}
\noindent%
\vfill

\newpage
\spacingset{1.5}
\setcounter{page}{2}

\section{Introduction}

% (1) General description of CRTs
Cluster-randomized trials (CRTs) are increasingly used to study interventions and inform decision-making in real-world settings. Distinct from individually-randomized trials, CRTs randomize groups of individuals, such as a hospital, classroom, or village, to treatment conditions. Cluster-level treatment assignment is carried out when individual-level randomization is infeasible or when there are concerns about treatment contamination under individual randomization \citep{murray1998design,donner2000design}. A distinguishing feature of CRTs is that observations collected within the same cluster tend to be correlated, which is an important aspect that must be reflected during both the design and analysis stages. 

% \citep{cook2016statistical}
% and \citet{fu2006comparison} also 

% (2) Introduce contexts of mixed model ANCOVA
\subsection{{Statistical methods for analyzing CRTs: A concise review}}
{The mainstream methods for estimating the average treatment effect can be categorized into individual-data-based analysis and cluster-summary-based analysis. The former involves regression models fitted to the individual-level observations, whereas the latter directly works with the cluster-level summaries of individual-level observations.}

For individual-data-based analyses, the mixed-model analysis of covariance without treatment-covariate interactions \citep[mixed-model ANCOVA1, extending the terminology in Section 4 of][]{murray1998design} is one of the most popular methods for estimating the average treatment effect. It involves a regression model of individual-level outcomes on an intercept, treatment term, covariate terms, a cluster-level random intercept, and a random residual error. The average treatment effect is then estimated by the regression coefficient of the treatment term. In health sciences, mixed models are the most frequently used methods for CRTs since they can directly account for the intracluster correlation and, at the same time, adjust for cluster-level and individual-level covariates. A systematic review by \citet{fiero2016statistical} found that 52\% of the CRTs published between August 2013 and July 2014 used mixed models in their primary analyses. {In an updated review of 100 CRTs from the United Kingdom’s National Institute for Health Research online Journals Library (from January 1997 to July 2021), the vast majority (80\%) adopted a mixed model for analysis \citep{offorha2022statistical}. In social and political science, however, the ordinary-least-squares method \citep{green2008analysis} has been more common for individual-data-based analyses. This method can be cast as a special case of mixed-model ANCOVA1 by omitting the random intercept, and the robust sandwich variance estimator is used to account for the intracluster correlation. Extending the terminology in \citet{Tsiatis2008} from individually-randomized trials to CRTs, we refer to this method as individual-level ANCOVA1. Furthermore, treatment-covariate interaction terms can be incorporated to improve precision \citep{lin2013agnostic,su2021model}, leading to a method we refer to as individual-level ANCOVA2 (again following from \citet{Tsiatis2008}; see Table~\ref{tab:1} for a classification). Both mixed-model and individual-level ANCOVA1 fit into the general framework of generalized estimating equations \citep{liang1986longitudinal} with a constant working variance and an identity link function. With a categorical outcome, generalized linear mixed models are also used for analyzing CRTs, but the associated treatment effect estimator can usually only be interpreted without ambiguity under a correct mean model conditioning on the random intercept and covariates \citep{jiang2010large, mcneish2016modeling}.} More recently, advanced individual-data-based approaches have been developed to integrate nonparametric machine learning methods \citep{balzer2016targeted, wang2022model}, to address treatment non-compliance \citep{park2021assumption}, to account for missing outcome data \citep{prague2016accounting, balzer2021two,tong2023bayesian}, and to accommodate blocked randomization designs \citep{schochet2021design}. 

{Compared to individual-data-based analyses, cluster-summary-based analyses are relatively simple, since there is no need to model the intracluster correlation \citep{hayes2009cluster}, which is absorbed into the cluster-summary data. Therefore, standard methods for individually-randomized trials, such as ANCOVA1 based on cluster-level mean outcomes and covariates, can be directly used at the cluster level. Under this category, unbiased estimators were proposed to address a matched design \citep{imai2009essential}, a diverging number of individuals per cluster \citep{middleton2015unbiased}, and informative cluster sizes \citep{bugni2022inference}.
More recently, \cite{su2021model} contributed a comprehensive study of ordinary-least-squares estimators based on cluster summaries or individual data in estimating a class of weighted average treatment effect estimands.}

% However, how the data are summarized may impact the estimands and efficiency. For example, when the treatment effect varies by the cluster sizes, analyses of cluster averages and cluster totals can target different estimands since the weighting of each individual changes. In addition, \cite{su2021model} showed that regression analysis based on cluster totals can be more efficient than cluster averages. 

\subsection{{Robustness and precision of methods for analyzing CRTs}}
{
CRTs differ from individually-randomized trials because the observed data are collected at more than one level, therefore opening up the possibility for different specifications of regression models. A natural question is how these different specifications of regression models perform against one another regarding the estimation of the average treatment effect. With this question in mind, the present article aims to explore the properties of several popular methods in practice, which are listed in Table~\ref{tab:1}. These methods differ based on the working model specification and covariate adjustment methods. Among the three covariate adjustment methods, we refer to no adjustment as ``unadjusted'' (the fixed effects are only an intercept and a treatment term), linear adjustment with main effects of treatment and covariates as ``ANCOVA1'' (unadjusted plus covariate terms), and additional linear adjustment with treatment-covariate interaction terms as ``ANCOVA2'' (ANCOVA1 plus treatment-covariate interaction terms). 
This terminology is inherited from the influential work of \citet{Tsiatis2008} for individually-randomized trials. 
In Table~\ref{tab:1}, we aim to compare methods both horizontally and vertically. That is, for each method of covariate adjustment, we evaluate which working model may lead to the highest precision (a horizontal comparison); for each working model, we also investigate which covariate adjustment method may provide the most efficient average treatment estimator in CRTs (a vertical comparison).}

% The individual-level approaches cover three linear mixed models, where ``mixed-model unadjusted'' is a special case of the mixed-model ANCOVA1 with no covariate, and ``mixed-model ANCOVA2'' is an extension of the mixed-model ANCOVA1 with treatment-covariate interaction terms. Here, the ordinary least square estimators are also included as a special case of mixed models with no random effects.
% The cluster-level approaches involve the unadjusted estimator, ANCOVA1, and ANCOVA2 based on cluster averages. 
% For each method of covariate adjustment, we compare whether cluster-level approaches are better than individual-level approaches. For each approach category, we evaluate which form of covariate adjustment leads to the best performance. 

% To facilitate the discussion, we consider a superpopulation framework where the cluster size heterogeneity is uninformative such that all methods target the same estimand.

\begin{table}[htbp]
\renewcommand{\arraystretch}{1.4}
    \centering
    \caption{Categorization of several common methods for analyzing CRTs. }
    \label{tab:1}
    \begin{tabular}{lccc}
        \hline
        & \multicolumn{2}{c}{\makecell[c]{Individual-data-based\\  analysis}}    &  \makecell[c]{Cluster-summary-based\\  analysis}  \\
        \cline{2-4}
        & Mixed models & \makecell[c]{Ordinary least\\squares} & \makecell[c]{Ordinary least\\squares} \\
        \hline
    No covariate adjustment  & \makecell[c]{Mixed-model \\unadjusted}  & \makecell[c]{Individual-level \\ unadjusted} & \makecell[c]{Cluster-level \\ unadjusted}\\
    \vspace{5pt}
\makecell[l]{Linear adjustment with \\
main effects only}& \makecell[c]{Mixed-model\\ANCOVA1} & \makecell[c]{Individual-level \\ ANCOVA1} & \makecell[c]{Cluster-level \\ ANCOVA1}\\
\makecell[l]{Linear adjustment with \\ main effects and \\ treatment-covariate \\interactions} & \makecell[c]{Mixed-model \\ ANCOVA2} & \makecell[c]{Individual-level \\ ANCOVA2} & \makecell[c]{Cluster-level \\ ANCOVA2}\\
    \hline
    \end{tabular}
\end{table}

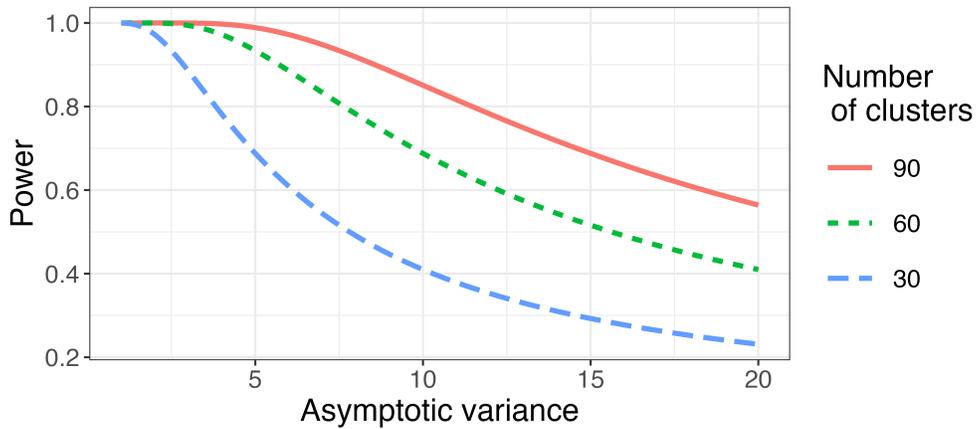
\begin{figure}
\centering
    \begin{subfigure}[b]{1\textwidth}
    \centering
        \begin{tikzpicture}
    \draw[draw=black](-7,-4.5) rectangle ++(16.2,5.5);
    \node[anchor=east] at (-0.5, 0) {Mixed-model ANCOVA:};
    \node[anchor=west] (working-model) at (0,0) {$Y_{ij} = \beta_0 + \beta_A A_i  + \beta_XX_i + \gamma_i + \epsilon_{ij}$};

    % \draw [-Latex](0.25,-1.4) -- (0.25,-0.4);
    % \node (Y) at (0.25, -1.6) {Outcome};
    % \draw [-Latex](2.5,-1.4) -- (2.5,-0.4);
    % \node (A) at (2.5, -1.6) {Treatment};
     
    \node[anchor=east] at (-0.5, -1) {True data generating distribution:};
    \node[anchor=west] (true-model) at (0,-1) {$Y_{ij} = A_i \exp\{\sin(X_{ij})\}\,\,+\,\, X_{i,j+1}\,\, +\,\,\, |\epsilon_{ij}|-\frac{\sqrt{2}}{\sqrt{\pi}}$};

    \draw [decorate,decoration = {calligraphic brace,mirror,amplitude=5pt, aspect=0.3},line width=2pt] (1.2,-1.5) --  (4,-1.5)
    node[pos=0.3,below=10pt,black,text width=3cm,text centered]{Non-linearity; heterogeneous treatment effect};

    \draw [decorate,decoration = {calligraphic brace,mirror,amplitude=5pt, aspect=0.45},line width=2pt] (5,-1.5) --  (6,-1.5)
    node[pos=0.45,below=10pt,black,text width=2.5cm,text centered]{Non-exchangeable correlation};

    \draw [decorate,decoration = {calligraphic brace,mirror,amplitude=5pt, aspect = 0.7},line width=2pt] (7,-1.5) --  (8.6,-1.5)
    node[pos=0.7,below=10pt,black,text width=2.5cm,text centered]{Non-normality};
\end{tikzpicture}
\caption{Possible Misspecifications of mixed-model ANCOVA. For individual $j$ in cluster $i$, $Y_{ij}$ is the outcome, $A_i$ is the cluster-level binary treatment assignment, $X_{ij} \sim N(0,1)$ is a covariate, $\gamma_i\sim N(0,\tau^2)$ is the random intercept, and $\epsilon_{ij} \sim N(0,\sigma^2)$ is the residual error.}\label{fig:illustration1}
    \end{subfigure}
    \hfill
    \vspace{10pt}
\begin{subfigure}[b]{\textwidth}
\centering
\includegraphics[width=0.8\textwidth]{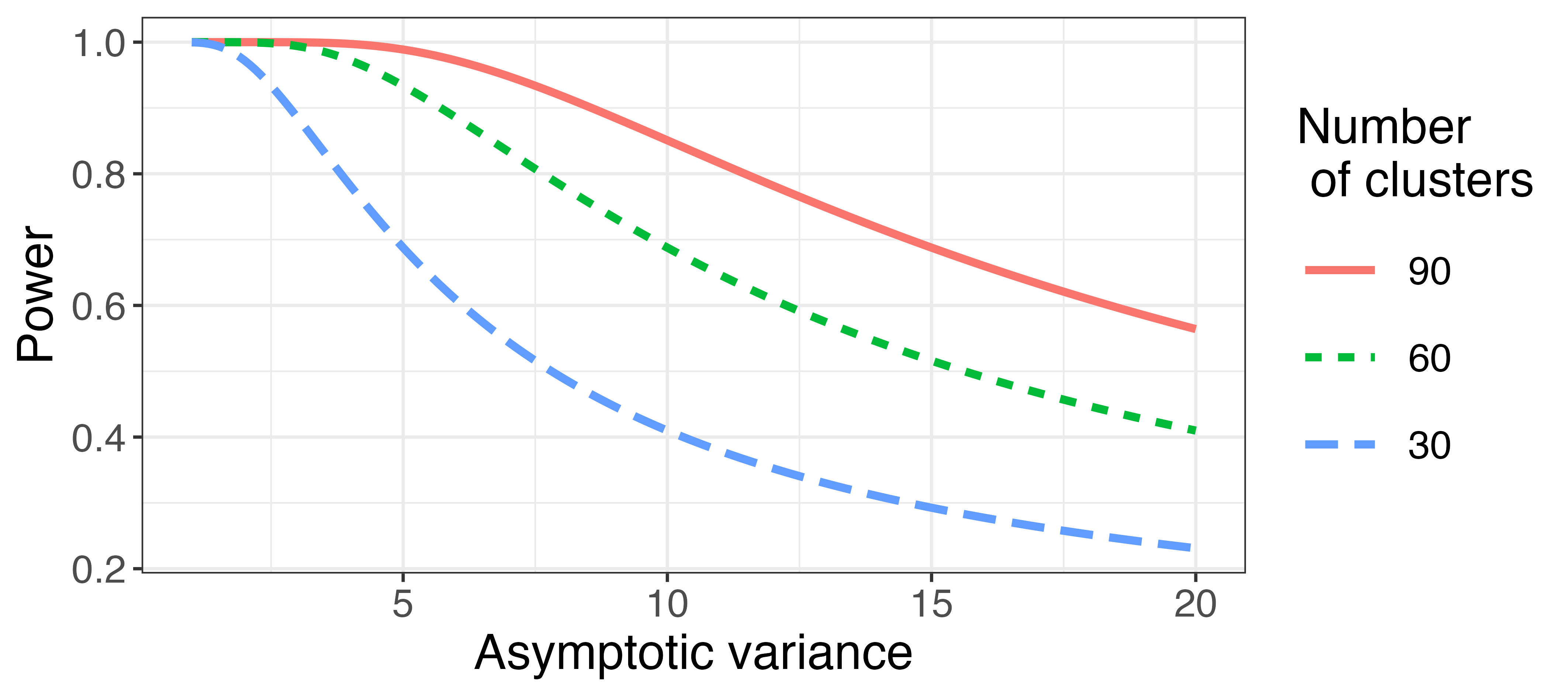}
\caption{Connection between the number of clusters, the asymptotic variance of an estimator, and power for detecting the effect size of 1 based on normal approximation and 0.05 type I error. }\label{fig:illustration2}
\end{subfigure}

    \caption{An Illustration of the important roles of robustness and precision in CRTs.}
    
\end{figure}

We focus on two performance metrics: robustness and precision. An approach is \textit{robust} if it leads to a consistent and asymptotically normal estimator for the average treatment effect under arbitrary misspecification of its working model. Since the true data-generating distribution is unknown and can be complex in a CRT, any parametric model can be potentially misspecified. 
{Figure~\ref{fig:illustration1} illustrates that a mixed-model ANCOVA1 can incorrectly assume linearity, homogeneous treatment effects, normality, and an exchangeable correlation structure even with a single covariate.} Given such complexity, model robustness is a desirable statistical property for any parametric model to achieve valid statistical inference in a CRT. In addition, we say an approach is more \textit{precise} or \textit{efficient} if its asymptotic variance is smaller. {A more precise approach can yield a narrower confidence interval and require fewer clusters to achieve sufficient power in a CRT. Figure~\ref{fig:illustration2} provides a simple visualization to demonstrate the connection between power, number of clusters, and the asymptotic variance in a CRT based on normal approximation. }

{Among the methods in Table~\ref{tab:1}, the robustness and precision of 
% cluster-summary-based and individual-data-based 
ordinary-least-squares estimators 
% for the average treatment effect 
have been investigated in the prior literature. For cluster-summary-based analyses of CRTs, the lessons learned from individually-randomized trials are directly applicable. For example, we can infer from \cite{Tsiatis2008} that all three cluster-summary-based approaches are robust, and that cluster-level ANCOVA1 does not reduce precision by covariate adjustment under equal randomization (i.e., each cluster has an equal chance of being assigned to treatment or control). Furthermore, the cluster-level ANCOVA2 estimator is always at least as precise as the cluster-level unadjusted estimator and the cluster-level ANCOVA1 estimator \citep{lin2013agnostic}.
For individual-data-based ordinary-least-squares estimators, \citet{su2021model} proved their robustness property, and further showed that the individual-level ANCOVA1 or ANCOVA2 estimator can sometimes be even less precise than the individual-level unadjusted estimator in cluster-randomized trials. Furthermore, they proved that the individual-data-based ordinary-least-squares estimator is less efficient than cluster-level ANCOVA2 based on scaled cluster totals. }

{Beyond these developments, we are unaware of any analytical result that formally clarifies the robustness and precision of the mixed-model estimators for the average treatment effect in CRTs, nor their comparisons with ordinary-least-squares estimators.}
For the robustness of mixed-model methods, when the random-effects distribution and/or the residual error distribution is misspecified, \citet{murray2006comparison} showed by simulations that mixed-model ANCOVA1 maintained a valid type I error under equal randomization, but did not provide theoretical justifications, meaning the generality of their observed results is unclear. More broadly, the robustness of generalized linear mixed models to misspecification of random-effect distributions has been extensively studied \citep{mcculloch2011misspecifying, neuhaus2013estimation, jiang2017asymptotic, drikvandi2017diagnosing}, under the strict assumption that the first moment is correctly specified conditioning on the random effects. We regard this assumption as `strict' because it rules out misspecification of the functional form of covariates, which is a strong restriction on the data-generating process. In general, the validity of mixed models for estimating the average treatment effect in CRTs remains unclear under arbitrary model misspecification (for example, the data-generating distribution in Figure~\ref{fig:illustration1}). In terms of precision, \cite{raudenbush1997statistical,li2016evaluation} showed by simulations that the mixed-model ANCOVA1 can improve precision compared to ignoring covariates when mixed-model ANCOVA1 is correctly specified; other simulation studies \citep{zhang2001linear, litiere2007type, litiere2008impact, mcculloch2011misspecifying} have found that non-normality could result in efficiency loss. 
{To date, no theoretical investigations have been carried out for the precision of the mixed-model ANCOVA1 or ANCOVA2 estimators, nor comparisons of precision between the mixed-model and ordinary-least-squares estimators (based on either individual data or cluster summaries). These apparent methodological gaps have motivated us to investigate new features of existing mixed-model estimators so that their routine use in CRTs can be better clarified, and additionally, a more clear recommendation on classical methods for analyzing CRTs can be offered.}

\subsection{Our contributions}

In this article, we first prove that the mixed-model unadjusted, ANCOVA1, and {ANCOVA2} estimators for the average treatment effect are all robust to arbitrary working model misspecification under certain regularity conditions. We further show that the model-based variance estimator for mixed-model ANCOVA1 remains consistent under arbitrary model misspecification under equal randomization. These insights significantly expand the results developed in \citet{wang2019analysis} from individually-randomized trials to CRTs, and justify the credibility of routinely-used mixed models in analyzing CRTs. In addition, we extend our discussions to accommodate stratified randomization.
%how to perform valid and anti-conservative inference.

{In addition to the robustness property, we also provide a comprehensive examination of the precision of mixed models. First, we contribute a surprising result that covariate adjustment by the mixed-model ANCOVA1 or ANCOVA2 may \textit{not} always increase the precision of the average treatment effect estimator, even under equal randomization. This result is in sharp contrast to findings for cluster-summary-based methods where covariate adjustment by ANCOVA2 leads to no loss in precision. 
Second, we show that accounting for the intracluster correlation by a simple random intercept in mixed-model ANCOVA1 or ANCOVA2 may \textit{not} always increase precision either. Assuming constant cluster sizes, individual-level ANCOVA1 or ANCOVA2 may even lead to greater precision gain than mixed-model ANCOVA1 or ANCOVA2 when adjusting for the same set of covariates. 
Third, we further demonstrate that, with equal cluster sizes, when covariate adjustment by the mixed-model ANCOVA1 or ANCOVA2 estimator leads to precision gain, the precision improvement does not exceed that from covariate adjustment by a cluster-level ANCOVA1 or ANCOVA2 estimator. 
In Figure~\ref{fig:summary-precision-comparison} of Section~\ref{sec: precision}, we provide a full summary of precision comparisons among different covariate-adjusted methods in Table~\ref{tab:1} under the constant cluster size assumption. 
When the cluster sizes vary, the situation becomes more complex. Compared to other methods, mixed models can be either more or less precise since they depend on the true outcome distribution, correlation structure, prognostic values of covariates, and the degree of cluster size variation.}

The remainder of the article is organized as follows. In Section~\ref{sec: def-assump}, we describe a super-population framework and present the structural assumptions for identifying the average treatment effect in CRTs. {This framework provides an asymptotic regime for us to investigate the properties of each average treatment effect estimator, and importantly, makes it feasible to compare across different methods in Table \ref{tab:1}.} Section~\ref{sec:main-result} gives our main results on robustness. Section~\ref{sec: precision} discusses the precision comparisons.
In Section~\ref{sec:simulation}, we demonstrate our theoretical results via simulation studies. 
%including that the mixed-model ANCOVA1 estimator can be less efficient than the cluster-level analysis or even the unadjusted analysis. 
In Section~\ref{sec:data-application}, we re-analyze three real-world CRTs to empirically illustrate the performance of all considered methods. Section \ref{sec:discussion} concludes.

\section{Notation, Assumptions and Mixed-Model ANCOVA}\label{sec: def-assump}
\subsection{Notation and general setup}\label{subsec:setup}
We consider a CRT with $m$ clusters. Each cluster $i$, $i=1,\dots, m,$ contains at least $n$ individuals, whereas only $N_i$ individuals are recruited and observed in the study, leading to potentially varying observed cluster sizes. We assume that $2 \leq N_i \le n$, namely, the number of recruited individuals in each cluster is upper bounded by $n$ but is at least $2$. For example, when clusters represent hospitals and each hospital can have more than $n=200$ patients that potentially satisfy the inclusion criteria, the study may only enroll $N_i \in[50, 100]$ patients from each hospital. In practice, $n$ can represent the number of individuals in the source population of interest and can be substantially larger than $N_i$. {We focus on estimating the average treatment effect based on the $n$ individuals per cluster, which corresponds to a constant source population size assumption and simplifies our estimands defined in Section 2.2. In Section~\ref{sec:discussion}, we discuss how our results may be extended to handle varying source population sizes.}

The defining feature of a CRT is that treatment is assigned at the cluster level instead of the individual level; individuals in the same cluster are therefore assigned to receive the same treatment. For each cluster $i=1,\ldots,m$, we define $A_i$ as the treatment indicator ($A_i=1$ if treated and 0 otherwise). For each individual $j =1,\ldots,n$ in cluster $i$, we define $Y_{ij}$ as the outcome (can either be continuous, binary or count) and $\bX_{ij} \in \mathbb{R}^p$ as a vector of baseline covariates. Here $\bX_{ij}$ can contain both individual-specific information ($X_{ij}$ varies across different individual $j$ in the same cluster $i$) and cluster-specific information ($X_{ij}$ is constant across individual $j$ in the same cluster $i$). We pursue the potential outcomes framework and assume consistency such that for each individual $j$ in cluster $i$,
\begin{align*}
    Y_{ij} = A_i Y_{ij}(1) + (1-A_i) Y_{ij}(0),
\end{align*}
where $Y_{ij}(a)$ is the potential outcome of individual $j$ in cluster $i$ if cluster $i$ were assigned treatment $a$ for $a \in \{0,1\}$. %We note that the counterfactual model is defined based on a hypothetical \textit{cluster-} rather than \textit{individual-}level intervention. 
We further define the complete (but not fully observed) data vector for individual $j$ in cluster $i$ as $\bW_{ij} = (Y_{ij}(1), Y_{ij}(0), \bX_{ij})$, and the complete data vector for cluster $i$ over $n$ individuals (in the source population) as $\bW_i = (\bW_{i1}, \dots, \bW_{in})$. To proceed, we make the following structural assumptions on $(\bW_1,\dots, \bW_m)$ and the assignment vector $(A_1,\dots, A_m)$.

\begin{assumption}(Super-Population Sampling and Cluster Randomization)\label{asp:1}\\
\emph{(a) $\{\bW_i, i = 1,\ldots, m\}$ are  independent and identically-distributed samples from the joint distribution $\mathcal{P}^{(\bW)}$ on the random vector $\bW = (\bW_{\bullet,1}, \ldots, \bW_{\bullet,n})$.\\
(b) Within $\mathcal{P}^{(\bW)}$, each $\bW_{\bullet,j}$ ($j = 1,\ldots, n$) follows a common distribution $\mathcal{P}$ on $(Y(1), Y(0), \bX)$. In other words, $\bW_{\bullet,1},\ldots,\bW_{\bullet,n}$ are marginally identically distributed.\\
(c) The cluster-level treatment assignment,  $\{A_i, i =1,\ldots, m\}$ are independent, identically distributed samples from a Bernoulli distribution $\mathcal{P}^{(A)}$ on $A$ with marginal probability ${P}(A=1)=\pi\in (0,1)$. Furthermore, $(A_1,\ldots, A_m)$ is independent of $(\bW_1, \ldots, \bW_m)$.}
\end{assumption}

% \vspace{10pt}
% \noindent\textbf{Assumption 1.}
% \begin{enumerate}[(a)]
%     \item $\bW_i, i = 1,\dots, m$ are  independent, identically-distributed samples from the joint distribution $P^{(\bW)}$ on $\bW = (\bW_{\cdot1}, \dots, \bW_{\cdot n})$.
    
%     \item 
%     Within $P^{(\bW)}$,
%     each $\bW_{\cdot j}, j = 1,\dots, n$ follows a common distribution $P$ on $ (Y(1), Y(0), \bX)$, i.e.,
%     $\bW_{\cdot1}, \dots, \bW_{\cdot n}$ are marginally identically distributed.
    
%     \item $A_i, i =1,\dots, m$ are independent, identically distributed samples from a Bernoulli distribution $P^{(A)}$ on $A$ with $P(A=1) = \pi, \pi \in (0,1)$.  Furthermore, $(A_1,\dots, A_m)$ is independent of $(\bW_1, \dots, \bW_m)$.
% \end{enumerate}

Assumption~\ref{asp:1}(a) implies that the data vector for each cluster is a random sample from a common distribution $\mathcal{P}^{(\bW)}$. Of note, this assumption is necessary to introduce the super-population framework, but does not assume homogeneous treatment effects across clusters, because the conditional distribution of $\bW_i$ given observed $(\bX_{i1},\ldots,\bX_{in})$ is still allowed to vary across clusters. Assumption~\ref{asp:1}(b) requires that, for each individual in the same cluster $i$, their complete data vectors follow the same marginal distribution, but they can be marginally correlated. Analogously, the conditional correlation structure among individuals of the same cluster can also vary across clusters. Finally, Assumption~\ref{asp:1}(c) is the treatment randomization assumption that holds under the cluster randomization design.

While Assumption~\ref{asp:1} elucidates conditions for the complete data vector, we require an additional assumption on the observed data vector. For $i=1,\dots,m$ and $j = 1,\dots, n$, we define a random variable $M_{ij}$ as an indicator of whether individual $j$ from cluster $i$ is enrolled in the study ($M_{ij}=1$) or not ($M_{ij}=0$). We further define $\mathcal{O}_i = \{j: M_{ij}=1\}$, the index set of enrolled individuals, which contains $N_i = \sum_{j=1}^n M_{ij}$ elements. {The observed data for cluster $i$ is therefore defined as $\mathbf{D}_i=\{(Y_{ij}, A_i, \bX_{ij}): j \in \mathcal{O}_i\}$.} Assumption~\ref{asp:2} underlies the connection between the complete data and the observed data.
%We make the following assumption on $N_i$:
%, whereas $M_{ij}$ is not observed

%Random
\begin{assumption}(Non-informative Enrollment)\label{asp:2}\\
\emph{Denote $\bM_i = (M_{i1},\dots, M_{in})$ as the collection of enrollment indicators. $\{\bM_i, i =1,\ldots, m\}$ are independent, identically distributed samples from a common distribution $\mathcal{P}^{(\bM)}$ on $\bM$. Furthermore, $(\bM_1, \ldots, \bM_m)$ is independent of $(\bW_1,\ldots, \bW_m)$ and $(A_1,\ldots, A_m)$.}
% \emph{Denote $\bM_i = (M_{i1},\dots, M_{in})$. The collection of observed cluster sizes, $\{N_i, i =1,\ldots, m\}$ are independent, identically distributed copies from a common distribution $\mathcal{P}^{(N)}$ with support being a subset of $\mathbb{N}=\{2,\dots, n\}$. Furthermore, $(N_1, \ldots, N_m)$ is independent of $(\bW_1,\ldots, \bW_m)$ and $(A_1,\ldots, A_m)$.}
\end{assumption}

% \vspace{10pt}
% \noindent \textbf{Assumption 2.} $N_i, i =1,\dots, m$ are independent, identically distributed samples from a distribution $P^{(N)}$ with support being a subset of $\{2,\dots, n\}$. Furthermore, $(N_1, \dots, N_m)$ are also independent of $(\bW_1,\dots, \bW_m)$ and $(A_1,\dots, A_m)$.

Assumption~\ref{asp:2} implies that, within each cluster, the enrollment of individuals, as well as the observed cluster size $N_i$, is random and independent of the remaining data information, including the potential outcomes, treatment, and baseline covariates. In addition, Assumption~\ref{asp:2} allows for unequal cluster sizes, whose randomness can be attributed to logistical and operational uncertainties across clusters but is otherwise unrelated to the potential outcomes. A similar non-informative cluster size assumption has been routinely invoked in the CRT literature, especially for purposes of sample size calculation \citep{eldridge2006sample}. Finally, Assumption~\ref{asp:2} will be violated when the cluster-specific average treatment effect depends on the cluster size \citep{seaman2014review}, or when informative enrollment of individuals by treatment conditions leads to selection bias \citep{li2021clarifying}. We will return to a discussion of these more challenging scenarios in Section \ref{sec:discussion}. 

\subsection{Causal estimands under the super-population framework}\label{subsec:estimand}
Our goal is to estimate the average treatment effect, defined as
\begin{align*}
    \Delta^* = E\{Y_{ij}(1)\} - E\{Y_{ij}(0)\},
\end{align*}
which compares the expected individual-level potential outcomes if a cluster were assigned treatment versus control. Given Assumptions \ref{asp:1} and \ref{asp:2}, the estimand can also be written as
\begin{align*}
    \Delta^* = E\{\overline{Y}_{i}(1)\} - E\{\overline{Y}_{i}(0)\},
\end{align*}
where $\overline{Y}_{i}(a) = n^{-1}\sum_{j=1}^n Y_{ij}(a)$ is the averaged potential outcomes for individuals in cluster $i$ if the cluster $i$ were assigned treatment $a$. Hence, the estimand $\Delta^*$ is also the average treatment effect of cluster averages across clusters.

Under the randomization-inference framework, \citet{su2021model} discussed two different estimands, which are the average treatment effect across enrolled individuals, $\Delta_I = {\sum_{i=1}^mN_i \delta_i}/{\sum_{i=1}^mN_i}$, 
and the average treatment effect across clusters, $\Delta_C= {\sum_{i=1}^m \delta_i}/{m}$, where $\delta_i = {N_i}^{-1}\sum_{j\in \mathcal{O}_j} \{Y_{ij}(1) - Y_{ij}(0)\}$ is the average of contrast in potential outcomes among the enrolled individuals in cluster $i$, and $Y_{ij}(a)$ and $N_i$ are treated as fixed quantities. These two estimands differ on how $\delta_i$ is weighted: $\Delta_I$ weights $\delta_i$ by the number of individuals enrolled in cluster $i$;  $\Delta_C$ assigns equal weight across clusters regardless of their sample sizes. {In general, these two estimands measure different quantities in finite samples, and their potential difference in magnitude critically depends on whether the cluster size is informative; see, for example, \citet{kahan2022estimands} for interpretation of both estimands, and \citet{kahan2023informative} for empirical evidence based on a case study}. Given our Assumptions \ref{asp:1} and \ref{asp:2} that underlie the super-population framework, we can show that $\Delta^* = E[\Delta_I] = E[\Delta_C]$, which unifies $\Delta_I$ and $\Delta_C$. {The underlying reasons for this simplification are the constant population size $n$, non-informative observed cluster sizes $N_i$, and the assumption that the potential outcomes are marginally identically distributed. To this extent, our estimand $\Delta^*$ can be interpreted as either the average treatment effect among the source population of interest (individual-average treatment effect) or the average treatment effect across clusters (cluster-average treatment effect).}

% If we only assume Assumption 1 instead of Assumption 2, then $\Delta^* = \Delta_2 \ne \Delta_1$. 

% For the same example given at the end of Section
% however, might not be scientifically meaningful under the super-population framework when $N_i/\sum_{i=1}^m N_i$ does not reflect the true proportion of cluster $i$ in the target population. 
% For example, if treatment works better for clusters with larger population, while fewer individuals are enrolled in these clusters because of high operational cost, then $N_i$ fail to reflect the true population size among clusters and, hence, $\Delta_1$ is smaller than the true average treatment effect across clusters .
% To avoid the possibility of  misinterpretation, we interpret $\Delta^*$ as the average treatment effect across clusters as for $\Delta_2$ above. 

\subsection{The mixed-model ANCOVA1 estimator}

Mixed-model ANCOVA1 is given by, for $i = 1,\ldots, m$ and $j =1,\dots,n$,
\begin{equation}\label{eq:ANCOVA1}
Y_{ij} = \beta_0 + \beta_AA_i + \bbeta_{\bX}^\top\bX_{ij} + \gamma_i +\epsilon_{ij},
\end{equation}
where 
%$Y_{ij}$ is the outcome, $A_i$ is the treatment status for cluster $i$, $\bX_{ij}$ is the collection of baseline covariates, 
$\gamma_i \sim N(0,\tau^2)$ is the random effect for cluster $i$, $\epsilon_{ij}\sim N(0,\sigma^2)$ is the residual error for individual $j$ in cluster $i$, and $(\beta_0, \beta_A, \bbeta_{\bX}, \sigma^2, \tau^2)$ represent unknown parameters. Typically in CRTs, mixed-model ANCOVA1 assumes that elements of $(\gamma_1,\ldots, \gamma_m, \epsilon_{11}, \ldots, \epsilon_{m,n})$ are mutually independent and are further independent of the treatment assignment $(A_1,\dots, A_m)$ and all covariates $(\bX_{11}, \dots, \bX_{m,n})$. Under this model, the proportion of total variance of $Y_{ij}$ that is attributable to the between-group variation, $\tau^2/(\tau^2+\sigma^2)$, is referred to as the intracluster correlation coefficient \citep{murray1998design}. When Equation~(\ref{eq:ANCOVA1}) involves no covariate, then the mixed-model ANCOVA1 estimator reduces to the mixed-model unadjusted estimator and hence treated as a special case.

We consider maximum likelihood estimators of $(\beta_0, \beta_A, \bbeta_{\bX}, \sigma^2, \tau^2)$ based on the observed data $\{(Y_{ij},A_i, \bX_{ij}):j\in \mathcal{O}_i, i =1,\dots,m\}$, and denote them by $(\widehat\beta_0, \widehat\beta_A, \widehat\bbeta_{\bX}, \widehat\sigma^2, \widehat\tau^2)$. We refer to \citet{jiang2017asymptotic} for full technical details on maximum likelihood estimation of linear mixed models, based on which we derive our key results. In CRT applications, the average treatment effect parameter, $\Delta^*$, is often estimated by $\widehat{\beta}_A$, which we denote as $\widehat\Delta_1$ and refer to as the mixed-model ANCOVA1 estimator hereafter. We also denote the model-based variance estimator for $\widehat\Delta_1$ as {$\widehat{Var}_{\textup{m}}(\widehat\Delta_1)$}, which is given by the second-row, second-column entry of
\begin{displaymath}
 \left\{\sum_{i=1}^m \bfQ_i^o{}^\top \widehat\bfSigma_i^{-1} \bfQ_i^o\right\}^{-1},
\end{displaymath}
where $\bfQ_i^o = (\bone_{N_i}, A_i \bone_{N_i}, \bfX_i^o)$ is the $N_i\times (p+2)$ design matrix for cluster $i$ and $\widehat\bfSigma_i = {m}/{(m-p-2)}(\widehat{\sigma}^2 \bfI_{N_i} + \widehat{\tau}^2 \bone_{N_i}\bone_{N_i}^\top) $ is the estimated covariance structure for cluster $i$ (with adjustment for the degrees of freedom), where $\bone_{N_i}$ is a $N_i$-dimensional column vector of ones, $ \bfI_{N_i}$ is the $N_i\times N_i$ identity matrix, and $\bfX_i^o = (\bX_{i,j_1},\dots, \bX_{i,j_{N_i}})^\top$ with $(j_1,\dots, j_{N_i})$ being the distinct elements of $\mathcal{O}_i$. %The adjustment for the degrees of freedom in $\widehat\bfSigma_i$ is known as the ``Between-Within method'' by \cite{li2015comparing}, who showed by simulation that this method can control the type I error of generalized linear mixed models in analyzing small sample CRTs.
%  following the method of \citealp{li2015comparing}
{An alternative variance estimator for $\Delta^*$ is the robust sandwich variance estimator (following Section 3.2 of \citealp{tsiatis2007}), which is the second-row, second-column entry of
\begin{align*}
    \left\{\sum_{i=1}^m \bfQ_i^o{}^\top \widehat\bfSigma_i^{-1} \bfQ_i^o\right\}^{-1}  \left\{\sum_{i=1}^m \bfQ_i^o{}^\top \widehat\bfSigma_i^{-1} (\bY_i^o - \bfQ_i^o\widehat{\bbeta} )(\bY_i^o - \bfQ_i^o\widehat{\bbeta} )^\top\widehat\bfSigma_i^{-1} \bfQ_i^o\right\} \left\{\sum_{i=1}^m \bfQ_i^o{}^\top \widehat\bfSigma_i^{-1} \bfQ_i^o\right\}^{-1},
\end{align*}
where $\bY_i^o = (Y_{i,j_1},\dots, Y_{i, j_{N_i}})$ is the vector of observed outcomes and $\widehat{\bbeta} = (\widehat\beta_0, \widehat\beta_A, \widehat\bbeta_{\bX}^\top)^\top$. We denote the robust variance estimator as $\widehat{Var}_{\textup{r}}(\widehat{\Delta}_1)$. Compared to the model-based variance estimator, the robust variance estimator is often used to achieve consistent variance estimation when the associated working model is misspecified.
}

{To derive our main theoretical results for the mixed-model ANCOVA1 estimator, we need additional regularity conditions on maximum likelihood estimation corresponding to mixed-model ANCOVA1 given by Equation \eqref{eq:ANCOVA1}. Specifically, the maximum likelihood estimation procedure is equivalent to solving the estimating equations $\sum_{i=1}^m\bpsi(\mathbf{D}_i;\btheta) = \bzero$, where $\bpsi(\mathbf{D}_i;\btheta)$ is the score function for cluster $i$ based on observed data $\mathbf{D}_i$ and unknown parameters $\btheta$. Assumption~\ref{asp:3} lists the regularity conditions on the function $\bpsi(\mathbf{D};\btheta)$.}

{\begin{assumption}\emph{(Regularity conditions)}\label{asp:3}\\
\emph{(a) The parameters $\btheta$ reside in a compact subset $\mathbf{\Theta}$ of the Euclidean space with $\sigma^2 > 0$.}\\
\emph{(b) There exists a unique $\underline{\btheta}$, an inner point of $\mathbf{\Theta}$, that satisfies $E[\bpsi(\mathbf{D}; \underline\btheta)] = \bzero$. }\\
\emph{(c) The estimating function has finite second moment, $E\left[\bpsi(\mathbf{D}; \underline\btheta)^\top\bpsi(\mathbf{D}; \underline\btheta)\right] <  \infty $.}\\
\emph{(d) $E\left[\frac{\partial }{\partial \btheta} \bpsi(\mathbf{D}; \btheta) \big |_{\btheta = \underline\btheta}\right]$ exists and is invertible.}\\
\emph{(e) Let $\psi_k$ be the $k$-th entry of $\bpsi$. Then $\frac{\partial}{\partial\btheta\partial\btheta^\top} \psi_k(\mathbf{d}, \btheta)$  is uniformly bounded by an integrable function $h(\mathbf{d})$ for every $\btheta$ in a small neighborhood of $\underline{\btheta}$ and $\mathbf{d}$ in the support of $\mathbf{D}$.}
\end{assumption}}
{In Assumption~\ref{asp:3}, (a) and (d) rule out degenerate cases for solving the likelihood equations, (b) requires that the maximum likelihood estimation has a single global maximizer asymptotically, whereas (c) and (e) are continuity and moment conditions for applying the central limit theorem. Collectively, these regularity conditions ensure that the maximum likelihood estimation has asymptotically stable performance and are similar to conditions invoked in Theorem~5.41 of \cite{vaart_1998} needed for proving asymptotic normality of an M-estimator. Of note, Assumption~\ref{asp:3} does not imply any component of the working model~(\ref{eq:ANCOVA1}) is correctly specified. In fact, when the random effect $\gamma_i$ is omitted from Equation~(\ref{eq:ANCOVA1}), Assumption 3 reduces to a simple condition that $(Y_{ij}(a), \bX_{ij})$ and $(\sum_{j\in \mathcal{O}_i}Y_{ij}(a), \sum_{j\in \mathcal{O}_i}\bX_{ij})$, $a\in\{0,1\}$, both have finite fourth moments and invertible covariance matrices. With $\gamma_i$, it is generally challenging to obtain an analytical solution of $\underline\btheta$ to $E[\bpsi(\mathbf{D}; \underline\btheta)] = \bzero$, and therefore we present these regularity conditions in terms of the estimating function $\bpsi$.}
% In the special case that $N_i$ is a constant across $i$ or equivalently under a balanced design, these conditions are (a) the distribution $\mathcal{P}^{\bW}$ has bounded second moments, (b) $Var(\bX)$ is non-singular, 
% %invertible, 
% (c) any pair of $\bW_{\bullet,1}, \ldots, \bW_{\bullet,N_i}$ are not identical with probability 1, and (d) $Var\{Y(a) - E[Y(a)|X]\} > 0$ for $a = 0,1$. For the general case that $N_i$ is a random variable but non-informative otherwise (as in Assumption~\ref{asp:2}), the regularity conditions are made instead on the derivative of the log-likelihood function corresponding to the mixed-model ANCOVA1 \eqref{eq:ANCOVA1}.

\subsection{{The mixed-model ANCOVA2 estimator}}
{
A variant of mixed-model ANCOVA1 further includes treatment-covariate interaction terms in the working model, and has the form
\begin{equation}\label{eq:ancova2}
    Y_{ij} = \beta_0 + \beta_AA_i + \bbeta_{\bX}^\top(\bX_{ij} - \overline{\bX}_{all}^o) + \bbeta_{A\bX}^\top A_i(\bX_{ij} - \overline{\bX}_{all}^o) + \gamma_i +\epsilon_{ij},
\end{equation}
where $\overline{\bX}_{all}^o = (\sum_{i=1}^m N_i)^{-1} \sum_{i=1}^m\sum_{j \in \mathcal{O}_i} \bX_{ij}$ is the mean covariates across all observed individuals, and all other variables are defined in Equation~(\ref{eq:ANCOVA1}). We refer to Equation~(\ref{eq:ancova2}) as mixed-model ANCOVA2. 
Compared to mixed-model ANCOVA1, mixed-model ANCOVA2 can capture the linear treatment effect heterogeneity that varies by treatment groups.
We compute the maximum likelihood estimator $(\widehat\beta_0, \widehat\beta_A, \widehat\bbeta_{\bX}, \widehat\bbeta_{A\bX}, \widehat\sigma^2, \widehat\tau^2)$ based on the mixed-model ANCOVA2 model, and estimate $\Delta^*$ by $\widehat{\beta}_A$, which we denote as $\widehat{\Delta}_2$. 
We compute its model-based variance variance estimator, $\widehat{Var}_{\textup{m}}(\widehat\Delta_2)$, and robust variance estimator, $\widehat{Var}_{\textup{r}}(\widehat\Delta_2)$, in the exact same way as for the mixed-model ANCOVA1 estimator $\widehat{\Delta}_1$ except that the design matrix includes additional columns with $\bfQ_i^o = (\bone_{N_i}, A_i\bone_{N_i}, \bfX_i^o - \overline{\bX}_{all}^o, A_i(\bfX_i^o - \overline{\bX}_{all}^o))$ and $\widehat{\bbeta} = (\widehat\beta_0, \widehat\beta_A, \widehat\bbeta_{\bX}^\top, \widehat\bbeta_{A\bX}^\top)^\top$.
In addition, we make the same regularity conditions (Assumption~\ref{asp:3}) for the estimating functions corresponding to Equation~(\ref{eq:ancova2}), but do not reproduce those conditions for brevity.
}

\section{Theoretical Results on Model Robustness}\label{sec:main-result}
% Our main results below hold under arbitrary model misspecification. The first result, in Section \ref{sec:robust}, is the robustness of the mixed-model ANCOVA1 point and variance estimator. Second, in Section \ref{subsec: precision-gain}, we clarify the precision gain by covariate adjustment via mixed-model ANCOVA1, and discuss possible efficiency improvement by instead analyzing cluster-specific means. In Section \ref{sec:stratification}, we extend the above results to stratified cluster randomization, which is standard practice for design-based control of covariates in CRTs with a relatively small number of clusters. 

\subsection{Robustness of the mixed-model ANCOVA1 estimator}\label{sec:robust}

% We first provide a result to establish the connection between the limit of regression coefficient in mixed-model ANCOVA1 and the average treatment effect estimand.

\begin{theorem}\label{thm1}
(a) Under Assumptions \ref{asp:1}, \ref{asp:2} and \ref{asp:3}, the mixed-model ANCOVA1 estimator $\widehat{\Delta}_1$ is consistent, i.e., $\widehat{\Delta}_1$ converges in probability to $\Delta^*$ as $m\rightarrow \infty$, and asymptotically normal, i.e., $\sqrt{n}(\widehat{\Delta}_1 - \Delta^*)$ converges weakly to a normal distribution $N(0, v_1)$, under arbitrary misspecification of its working model. The explicit form of $v_1$ is given in the Supplementary Material. {(b) In addition, $m\widehat{Var}_{\textup{r}}(\widehat{\Delta}_1)$ converges in probability to the true asymptotic variance $v_1$. (c) Finally, under equal randomization where $\pi = 0.5$, $m\widehat{Var}_{\textup{m}}(\widehat{\Delta}_1)$ also converges in probability to  $v_1$.} 
% , and therefore the model-based variance estimator $\widehat{Var}(\widehat{\Delta})$ remains valid.
\end{theorem}

% ,wang2019analysis,wang2021model
Theorem~\ref{thm1} provides a formal statement on the robustness of the mixed-model ANCOVA1 estimator for the average treatment effect in CRTs under arbitrary model misspecification. That is to say, even when the conditional mean structure, covariance structure of the random effects, and/or other aspects of residual error distribution are incorrect, the bias of the resulting estimator $\widehat{\Delta}_1$ vanishes with an increasing number of clusters. Theorem \ref{thm1}(a) extends the well-known results developed for the ANCOVA1 estimator under individually randomized trials \citep{YangTsiatis2001,wang2019analysis} to CRTs with correlated outcomes. It also provides a foundation to explain earlier simulation findings by \citet{murray2006comparison}, who observed that $\widehat{\Delta}$ had negligible bias when the data were simulated from an ANCOVA1 model with non-normal random effect and/or residual errors. 

% , and therefore the associated Wald test for the average treatment effect can adequately control for the type I error rate
{Theorem 1(b) implies that the robust variance estimator is valid for variance and can be used to control the type I error and construct valid confidence intervals.}
Under equal randomization of clusters (which is frequently the case in practice), Theorem~\ref{thm1}(c) implies that the model-based variance estimator is also robust to model misspecification. In other words, the standard error estimates returned by standard software for fitting linear mixed models yield (asymptotically) correct uncertainty statements. 
% The validity of the model-based variance based on ordinary least squares for simple ANCOVA1 estimator has been proved by \citet{wang2019analysis} in individually randomized trials, and Theorem \ref{thm1}(b) expands their finding to CRTs with a cluster-level assignment and correlated outcomes.
Taken together, the robustness of the point estimator and variance estimator implies that model-based inference via mixed-model ANCOVA1, e.g. the p-value and confidence interval output by standard statistical software, are asymptotically valid without requiring any parametric assumptions on the distribution of $(Y,A,\bX)$ {when the clusters are randomized by flipping a fair coin with $\pi = 0.5$}.

% , similar to observations made in individually randomized trials \citep{wang2019analysis}
%(Section 3.2 of \citealp{tsiatis2007})
% Under unequal randomization ($\pi \ne 0.5$), the model-based variance estimator may be biased. In this case, we define the sandwich variance estimator of $\widehat{\Delta}$ (following Section 3.2 of \citealp{tsiatis2007}) as the second-row, second-column entry of
% \begin{align*}
%     \left\{\sum_{i=1}^m \bfQ_i^o{}^\top \widehat\bfSigma_i^{-1} \bfQ_i^o\right\}^{-1}  \left\{\sum_{i=1}^m \bfQ_i^o{}^\top \widehat\bfSigma_i^{-1} (\bY_i^o - \bfQ_i^o\widehat{\bbeta} )(\bY_i^o - \bfQ_i^o\widehat{\bbeta} )^\top\widehat\bfSigma_i^{-1} \bfQ_i^o\right\} \left\{\sum_{i=1}^m \bfQ_i^o{}^\top \widehat\bfSigma_i^{-1} \bfQ_i^o\right\}^{-1},
% \end{align*}
% % \begin{equation*}
% %     \mathbf{B}^{-1}\left\{\sum_{i=1}^m \bfQ_i^o{}^\top \widehat\bfSigma_i^{-1} (\bY_i^o - \bfQ_i^o\widehat{\bbeta} )(\bY_i^o - \bfQ_i^o\widehat{\bbeta} )^\top\widehat\bfSigma_i^{-1} \bfQ_i^o\right\} \mathbf{B}^{-1},
% % \end{equation*}
% where $\bY_i^o = (Y_{i,j_1},\dots, Y_{i, j_{N_i}})$ is the vector of observed outcomes and $\widehat{\bbeta} = (\widehat\beta_0, \widehat\beta_A, \widehat\bbeta_{\bX}^\top)^\top$. Given our Assumptions \ref{asp:1} and \ref{asp:2}, the sandwich variance estimator is consistent to $v$ for all $\pi \in (0,1)$ and becomes asymptotically equivalent to the model-based variance estimator if the mixed-model ANCOVA1 model is correctly specified.

Intuitively, Theorem~\ref{thm1} is proved by ``translating'' a CRT to an unit-randomized trial, where each unit is a cluster and the observations collected within a unit are akin to repeatedly measured outcomes of the unit. Since all repeat measures are identically distributed according to Assumption~\ref{asp:1}(b), the average treatment effect across the repeat measures is identical to $\Delta^*$. When each cluster has a different number of enrolled individuals, we conceptualize it as a missing data problem where $M_{ij}$ is the non-missingness indicator, in which case the non-informative recruitment condition (Assumption~\ref{asp:2}) corresponds to missingness completely at random \citep{rubin1976inference}. Based on this conceptualization, Theorem \ref{thm1} is then proved by invoking the asymptotic results in \citet{vaart_1998}; a complete proof along with the explicit influence function of $\widehat{\Delta}_1$ is provided in the Supplementary Material.

\subsection{{Robustness of the mixed-model ANCOVA2 estimator}}

{
In parallel to the mixed-model ANCOVA1 estimator, the mixed-model ANCOVA2 estimator has a similar result on robustness stated below.
\begin{theorem}\label{thm:mixed-model-ANCOVA2}
(a) Under Assumptions \ref{asp:1}, \ref{asp:2} and \ref{asp:3}, the mixed-model ANCOVA2 estimator $\widehat{\Delta}_2$ is consistent for $\Delta^*$ and asymptotically normal, i.e., $\sqrt{n}(\widehat{\Delta}_2 - \Delta^*)$ converges weakly to a normal distribution $N(0, v_2)$, under arbitrary misspecification of its working model. (b) In addition, $m\widehat{Var}_{\textup{r}}(\widehat{\Delta}_2)$ converges in probability to the true asymptotic variance $v_2$. 
\end{theorem}
Like Theorem~\ref{thm1}, Theorem~\ref{thm:mixed-model-ANCOVA2} characterizes the same robustness property for the point estimator and validity of the robust variance estimator. 
However, the model-based variance estimator $\widehat{Var}_{\textup{m}}(\widehat{\Delta}_2)$ may be biased, even under $\pi = 0.5$. This is because the model-based variance does not account for the extra variability introduced by the treatment-covariate interaction terms. 
Therefore, we can carry out robust inference using $\widehat{\Delta}_2$ with its robust variance estimator instead of the model-based variance estimator. A second difference between $\widehat{\Delta}_1$ and $\widehat{\Delta}_2$ regards their asymptotic variances, $v_1$ and $v_2$ respectively, which we provide a comparison in Section 4.1.
}

{
For both aforementioned mixed-model ANCOVA estimators, an important special case is when the random effect $\gamma_i$ is omitted from the working models~(\ref{eq:ANCOVA1}) and (\ref{eq:ancova2}), i.e., assuming no intracluster correlation within the working model. Then the mixed-model unadjusted, ANCOVA1, and ANCOVA2 estimators reduce to the individual-level unadjusted, ANCOVA1, and ANCOVA2 estimators, respectively, obtained by ordinary-least-squares on individual-level data. \cite{su2021model} proved under a randomization-inference framework that these estimators are consistent and that their robust variance estimators remain valid in CRTs under arbitrary model misspecification. Under our super-population framework, we obtain the same result from Theorems~\ref{thm1} and~\ref{thm:mixed-model-ANCOVA2}.
}

\subsection{Extension to stratified randomization}\label{sec:stratification}

Stratified randomization refers to a restricted randomization procedure that achieves between-group balance on certain covariates within each pre-specified stratum, and has been frequently used in CRTs to minimize chance imbalance \citep{ivers2012allocation}. 
% Stratified randomization also applies to cluster-randomized trials where the randomization unit is the cluster.
For each cluster $i$, let $S_i$ be a categorical variable that encodes the randomization strata $\mathcal{S}$. For example, if cluster randomization is stratified by geographical location (urban versus rural), then the randomization strata are $\mathcal{S}$=\{urban, rural\} and $S_i\in\mathcal{S}$. We assume that the number of strata is fixed and the randomization proportion within strata remains $\pi \in (0,1)$. %Within each randomization stratum, stratified randomization allocates treatment by permuted blocks with fraction $\pi$ 1's (representing treatment) and fraction $1-\pi$ 0's (representing control) such that exact $\pi$ fraction of clusters receive treatment.
Under stratified randomization, Assumption~\ref{asp:1}(c) no longer holds since $(A_1,\dots,A_n)$ are correlated and are further correlated with $(S_1,\dots,S_n)$. 
For this design, Theorem \ref{thm3} below provides the asymptotic results for the mixed-model  ANCOVA1  and ANCOVA2 estimators.
% However, Theorem \ref{thm3} implies that the mixed-model ANCOVA1  and ANCOVA2 estimators retain their asymptotic validity and that stratified randomization can improve their precision. In addition, the model-based inference remains valid as long as $\pi = 0.5$ and provided that the strata variables are adjusted  as dummy variables. We summarize these results in the following Theorem \ref{thm3}.%(with one level dropped to avoid co-linearity with the intercept).

{
\begin{theorem}\label{thm3}
(a) Under stratified randomization, given Assumptions~\ref{asp:1}(a)--(b), \ref{asp:2}, \ref{asp:3}, $\widehat{\Delta}_1$ and $\widehat{\Delta}_2$ are both consistent for $\Delta^*$ and asymptotically normal under arbitrary misspecification of their working models. 
% , and regularity conditions in the Appendix
(b) Denoting $\widetilde{v}_1$ and $\widetilde{v}_2$ as asymptotic variances for $\widehat{\Delta}_1$ and $\widehat{\Delta}_2$ under stratified randomization, respectively, then stratified randomization does not lead asymptotic efficiency loss, that is, 
$\widetilde{v}_1 \le v_1$ and $\widetilde{v}_2 \le v_2$. (c) Finally, if the strata variable, $S_i$, is included in $\bX_{ij}$ and adjusted for in the working models as cluster-level dummy variables, then (i)~$\widetilde{v}_1 = v_1$ if $\pi = 0.5$; and (ii)~$\widetilde{v}_2 = v_2$.
\end{theorem}
}

{Theorem \ref{thm3}(a) implies that the mixed-model ANCOVA1 and ANCOVA2 estimators remain robust even with cluster-stratified randomization. In addition, compared to simple randomization, Theorem~\ref{thm3}(b) confirms that adjusting for $S_i$ in the randomization stage does not lead to asymptotic efficiency loss. This result provides an analytical explanation of the simulation results in \citet{li2016evaluation}, which showed that restricted randomization in CRTs often increases statistical power. Finally, Theorem \ref{thm3}(c) points out that adjusting for $S_i$ in mixed-model ANCOVA2 suffices to account for the potential precision gain from stratified randomization, which shows when the sandwich variance estimator is consistent under this design; for mixed-model ANCOVA1, an additional requirement of equal randomization is needed for this result. 
% under equal randomization, the asymptotic variance for mixed-model ANCOVA1 estimator under simple randomization and that under stratified randomization will coincide as long as $S_i$ is adjusted for in the working model. However, the asymptotic variance for mixed-model ANCOVA2 estimator under simple randomization and that under stratified randomization will always coincide as long as $S_i$ is adjusted for in the working model, without requirements on the randomization probability. 
Overall, Theorem~\ref{thm3} parallels to those developed in \citet{wang2021model} for individually-randomized trials.}

% Of note, a similar result can be stated for the cluster-level ANCOVA1 estimator, $\widehat{\Delta}^{\textup{(cl)}}$, as inferred by 
% %Corollary~\ref{corollary1} also holds for the cluster-level ANCOVA1 estimator as implied by  
% Corollary~1 of \cite{wang2021model}. The efficiency comparison between $\widehat{\Delta}$ and $\widehat{\Delta}^{\textup{(cl)}}$ under $\pi = 0.5$ and stratified randomization therefore follows Section~\ref{subsec: precision-gain}. We omit the formal statements for brevity.
% %As a result, Theorem~\ref{thm2} extends to stratified randomization if $S_i$ is included as dummy variables in the covariate vector $\bX_{ij}$.

\section{{Precision for Estimating the Average Treatment Effect}}\label{sec: precision}
\subsection{{Does covariate adjustment always improve precision in mixed-model ANCOVA estimators?}}
In CRTs, covariate adjustment by mixed-model ANCOVA1 may reduce precision compared to no covariate adjustment (that is, removing $\bbeta_{\bX}^\top\bX_{ij}$ from Equation \eqref{eq:ANCOVA1}), even under equal randomization. This finding is in sharp contrast to existing results for individually-randomized trials, where covariate adjustment by ANCOVA1 does not reduce the asymptotic efficiency under equal randomization. 
Heuristically, the efficiency loss of mixed-model ANCOVA1 can occur if we misspecify the true covariance structure, which can be different from the assumed exchangeable correlation structure in mixed-model ANCOVA1; such misspecification will compromise the ability of $\widehat{\bbeta}_{\bX}$ in capturing the true relationship between  $\sum_{j\in \mathcal{O}_i} Y_{ij}$ and  $\sum_{j\in \mathcal{O}_i} \bX_{ij}$, which can then inflate the variance of $\widehat{\Delta}_1$. We will empirically illustrate this result in the ensuing simulation study, where the mixed-model ANCOVA1 estimator can have 13\% larger variance compared to the unadjusted estimator under misspecification of its working model. {Following the same argument, covariate adjustment by mixed-model ANCOVA2 may also lead to precision loss compared to no covariate adjustment. In addition, the mixed-model ANCOVA2 estimator may be less precise than the mixed-model ANCOVA1 estimator. However, the following proposition outlines special cases where mixed-model ANCOVA1 and ANCOVA2 have the same asymptotic variance for estimating the average treatment effect.}

{\begin{proposition}\label{prop1}
    Given Assumptions~\ref{asp:1}, \ref{asp:2}, and \ref{asp:3}, and assuming equal randomization ($\pi  = 0.5$) and equal observed cluster sizes ($N_i = \widetilde{n}$ for all $i$), then 
    % the mixed-model ANCOVA1 estimator $\widehat{\Delta}_1$ and ANCOVA2 estimator $\widehat{\Delta}_2$ have the same asymptotic variance, i.e., 
    $v_1 = v_2$ if any of the following conditions hold: (a) mixed-model ANCOVA2 is correctly specified, (b) $\bX_i$ only contains cluster-level covariates, or (c) $\{(Y_{ij}, \bX_{ij}): j=1,\dots,n\}$ are independent of each other.
\end{proposition}}

{
In the special case of equal observed cluster sizes with equal randomization, 
Proposition 1 shows that adjusting for treatment-covariate interaction terms has no precision gain in mixed model ANCOVA2 under condition (a), (b), or (c). 
Here, any of these three conditions ensures that $\widehat{\bbeta}_{\bX}$ from mixed-model ANCOVA1 is asymptotically equal to $\widehat{\bbeta}_{\bX} + 0.5 \widehat{\bbeta}_{A\bX}$ from mixed-model ANCOVA2, a necessary condition for $v_1=v_2$. 
Unlike individually-randomized trials where the ANCOVA1 and ANCOVA2 estimators for the average treatment effect have the same asymptotic variance if $\pi = 0.5$, Proposition 1 additionally needs constant cluster size and condition (a), (b), or (c) because of the intracluster correlation and the random effect in the working model. Beyond these special cases, the comparison of $v_1$ and $v_2$ is indeterminate in general.
% $\widehat{\Delta}_1$ and $\widehat{\Delta}_2$ are asymptotic equivalent for estimating $\Delta^*$. This result is the direct analogue for that developed in \citet{Tsiatis2008} for ANCOVA1 and ANCOVA2 estimators under individually-randomized trials, and suggests that there is no gain in precision from adjusting for treatment-covariate interaction terms in CRTs. 
% In the general setting with variable observed cluster sizes, however, the ordering of precision among the unadjusted, ANCOVA1, and ANCOVA2  estimators for individually-randomized trials does not generalize to mixed model estimators for CRTs. 
% To give one example, consider the generating process 
% $$Y_{ij} = A_i\left(X_{ij} - \sum_{l=1}^n X_{il}\right) + \varepsilon_{ij}$$ 
% with $N_i = n$; then the mixed-model unadjusted estimator is more precise than the mixed-model ANCOVA1 and ANCOVA2 estimators. Furthermore, the mixed-model ANCOVA1 estimator is more precise than the ANCOVA2 estimator if $\pi < 0.5$, whereas the mixed-model ANCOVA1 is less efficient if $\pi > 0.5$. The mathematical details of this example are provided in the Supplementary Material. 
In Section~\ref{sec:simulation}, we construct examples to demonstrate that any of the mixed-model unadjusted, ANCOVA1, or ANCOVA2 estimators can have the highest precision depending on the data-generating distributions.}

{Although covariate adjustment in linear mixed models does not theoretically guarantee precision gain across all data-generating processes, in practice, we still recommend adjusting for pre-specified prognostic covariates whenever a mixed model is used. When the covariates are strongly prognostic, the precision gain from covariate adjustment is likely to overwhelm the potential precision loss from misspecification of certain aspects of the working model. Furthermore, compared to mixed-model ANCOVA1, although mixed-model ANCOVA2 can capture the treatment effect heterogeneity between treatment arms and has the potential to further improve precision for estimating $\Delta^*$, it might come at the cost of introducing additional model parameters. 
These additional model parameters will inevitably compromise the degrees of freedom and inflate the finite-sample variance of the average treatment effect estimator, especially when $m$ is small (a frequent limitation of many published CRTs).}
%Because of this tradeoff, mixed-model ANCOVA1 may be regarded as a simple and yet effective choice to leverage prognostic covariates for potential precision improvement in CRTs with a small to moderate number of clusters.}
% In Section 4.3, we show that cluster-level analyses will not lose precision by covariate adjustment and can serve as a better alternative for mixed models for improving precision.

{\subsection{Does the inclusion of a cluster-level random intercept always improve precision?}}
{In 2001, a Consolidated Standards of Reporting Trials (CONSORT) extension statement for CRTs \citep{campbell2004consort} was published to provide guidance on reporting for CRTs. In the statistical methods section, the guidance document recommended that ``\emph{statistical methods used to compare groups for primary outcome(s) should indicate how clustering was taken into account}.'' In light of this recommendation, mixed models have long been considered a natural candidate for analyzing CRTs because they include a cluster-level random intercept to explicitly account for clustering. Furthermore, the intracluster correlation coefficient that is implied from a mixed model is an interpretable design parameter that measures the level of clustering and is an essential input parameter for sample size calculation during trial planning \citep{eldridge2009intra}.}

{However, in theory, a random intercept may not always lead to precision gain for estimating the average treatment effect in CRTs. To elaborate on this point, let $\widehat{\Delta}_1^{(\textup{ols})}$ and $\widehat{\Delta}_2^{(\textup{ols})}$ be the individual-level ANCOVA1 and ANCOVA2 estimators for the average treatment effect, respectively. We denote their asymptotic variance as $v_1^{(\textup{ols})}$ and $v_2^{(\textup{ols})}$ respectively. By comparing $v_1^{(\textup{ols})}$ versus $v_1$ and $v_2^{(\textup{ols})}$ versus $v_2$, we would be able to evaluate the impact of including the random intercept on estimation precision. For analytical comparisons, we first focus on a special case of equal observed cluster sizes. The following proposition shows that, counter-intuitively, including a cluster-level random intercept does not improve the precision when cluster sizes are equal.
}

{\begin{proposition}\label{corollary2}
    Given Assumptions~\ref{asp:1}, \ref{asp:2}, and \ref{asp:3}, assuming equal observed cluster sizes with $N_i = \widetilde{n}$ for all $i$, then (a) $v_1^{(\textup{ols})} \le v_1$ if $\pi = 0.5$; (b) $v_2^{(\textup{ols})} \le v_2$.
\end{proposition}}

{Proposition \ref{corollary2} implies that, when a CRT is balanced in cluster sizes, one can obtain a more efficient average treatment effect estimator by dropping the random intercept and proceeding with the ordinary-least-squares estimator (allowing working models can be arbitrarily misspecified); the intracluster correlation can be accounted for by the robust variance estimators. The underlying intuition is that the random intercept may compromise the precision gain from covariate adjustment when the true correlation structure is not exchangeable. However, in the special case listed in Proposition~\ref{prop1}, adding a random intercept in the working models does not change the precision, i.e., $v_1^{(\textup{ols})} =v_1$ if $\pi = 0.5$, and $v_2^{(\textup{ols})} = v_2$.}
% In fact, when the mixed model is correctly specified and the cluster size varies, the individual-level ANCOVA1 (or ANCOVA2) estimator remains more precise than the mixed-model ANCOVA1 (or ANCOVA2) estimator. 

{For the general cases with variable cluster sizes, their variance comparison can often be indeterminate and will depend on the true outcome distribution, correlation structure, prognostic value of covariates, and variation in cluster size. Importantly, compared to the potential variance improvement due to covariate adjustment, the influence of a random intercept on precision is often much smaller since the intracluster correlation coefficient in CRTs rarely exceeds 0.2 in practice \citep{murray2003methods,campbell2005determinants}.
To showcase this point, we offer two simulation scenarios in Section~\ref{sec:simulation} and our data applications in Section~\ref{sec:data-application} where mixed model estimators can be either slightly more or slightly less efficient than individual-level ordinary-least-squares estimators in CRTs.}
% Finally, compared to the potential variance improvement due to covariate adjustment, the potential precision improvement from a random intercept may be smaller as the commonly reported intraclass correlation coefficient values in CRTs rarely exceed 0.2 \citep{murray2003methods,campbell2005determinants}, whereas the coefficient of determination ($R^2$) due to a prognostic covariate may often be higher. 
%Given these considerations, the mixed-model ANCOVA estimator remains a convenient and highly-accessible approach to adjust for covariates and account for clustering.}

{\subsection{Are individual-data-based analyses  always more precise than cluster-summary-based analyses?}}

Intuitively, analyses based on individual data use more information than analyses based on cluster summaries and, hence, should be more precise. Surprisingly, this is not always true for ANCOVA1 and ANCOVA2 methods in CRTs. To elaborate on this claim, we consider 
% cluster-level ANCOVA models based on cluster-specific means aggregated from individual-level information. Specifically, for $i = 1,\dots, m$, let $\overline{Y}_i^o = N_i^{-1}\sum_{j\in \mathcal{O}_i} Y_{ij}$ and $\overline{\bX}_i^o = N_i^{-1}\sum_{j\in \mathcal{O}_i}\bX_{ij}$ be the cluster-specific means, then 
the cluster-level ANCOVA1 working model, i.e., ANCOVA1 based on cluster-specific averages, defined as
\begin{equation}\label{eq:clANCOVA1}
    E\left[\overline{Y}_i^o|A_i, \overline{\bX}_i^o\right] = \alpha_0 + \alpha_AA_i + \balpha_{\overline{\bX}^o}^\top \overline{\bX}_i^o,
\end{equation}
where $\overline{Y}_i^o = N_i^{-1}\sum_{j\in \mathcal{O}_i} Y_{ij}$ and $\overline{\bX}_i^o = N_i^{-1}\sum_{j\in \mathcal{O}_i}\bX_{ij}$ are cluster-specific mean outcome and covariates, respectively. 
Under model \eqref{eq:clANCOVA1}, $\Delta^*$ is estimated by the ordinary-least-squares estimators of $\alpha_A$, which we refer to as $\widehat{\Delta}_1^{\textup{(cl)}}$. 
% This is essentially the ANCOVA1 model that has been extensively studied in individually-randomized trials and its statistical properties are relatively well-understood. For example, under Assumption~\ref{asp:1}, it is straightforward to infer from \cite{YangTsiatis2001} that $\widehat{\Delta}_1^{\textup{(cl)}}$ is consistent under arbitrary working model misspecification and further improves precision over a cluster-level unadjusted estimator through leveraging prognostic covariates when $\pi=0.5$ (equal randomization); under the same condition (including equal randomization with $\pi=0.5$), we can infer from \cite{wang2019analysis} that the model-based variance estimator under $\widehat{\Delta}_1^{\textup{(cl)}}$ is also valid without requiring the working model \eqref{eq:clANCOVA1} to be correctly specified.
{In parallel to the development of mixed-model estimators, we also consider a cluster-level ANCOVA2 working model:
\begin{equation}\label{eq:clANCOVA2}
    E\left[\overline{Y}_i^o|A_i, \overline{\bX}_i^o\right] = \alpha_0 + \alpha_AA_i + \balpha_{\overline{\bX}^o}^\top \left(\overline{\bX}_i^o - \frac{1}{m}\sum_{l=1}^m \overline{\bX}_l^o\right) + \balpha_{\overline{A\bX}^o}^\top A_i\left(\overline{\bX}_i^o - \frac{1}{m}\sum_{l=1}^m \overline{\bX}_l^o\right).
\end{equation}
Under model~(\ref{eq:clANCOVA2}), we also use the ordinary-least-squares estimator of $\alpha_A$ to estimate $\Delta^*$, which we denote as $\widehat{\Delta}_2^{(\textup{cl})}$.} Both $\widehat{\Delta}_1^{(\textup{cl})}$ and $\widehat{\Delta}_2^{(\textup{cl})}$ are defined the exact same way as in individually-randomized trials, and their statistical properties are hence relatively well-understood. Beyond the short review of their  properties in Section 1.2, complete results with technical details can be found in \cite{Tsiatis2008}, \citet{lin2013agnostic}, and \citet{wang2019analysis}.

% In CRTs, we can infer from \cite{Tsiatis2008} and \citet{lin2013agnostic} that $\widehat{\Delta}_2^{(\textup{cl})}$ is also consistent under arbitrary working model misspecification and improves precision over a cluster-level unadjusted estimator through covariate adjustment under any $\pi \in (0,1)$. Furthermore, the cluster-level ANCOVA2 estimator, $\widehat{\Delta}_2^{(\textup{cl})}$, is always asymptotically equivalent to the cluster-level ANCOVA1 estimator, $\widehat{\Delta}^{(\textup{cl})}$, under equal randomization ($\pi=0.5$), and $\widehat{\Delta}_2^{(\textup{cl})}$ is asymptotically more efficient for other values of $\pi$ \citep{Tsiatis2008}. 

In CRTs with equal observed cluster size, we show below that covariate adjustment by cluster-summary-based methods leads to equal or even more precision gain than covariate adjustment by individual-data-based methods.
%and each cluster has the same number of enrolled individuals.

{
\begin{proposition}\label{coro3}
Suppose $\widehat{\Delta}_1$, $\widehat{\Delta}_2$, $\widehat{\Delta}^{\textup{(ols)}}_1$, $\widehat{\Delta}^{\textup{(ols)}}_2$, $\widehat{\Delta}^{\textup{(cl)}}_1$, $\widehat{\Delta}^{\textup{(cl)}}_2$ are estimators for $\Delta^*$ based on the same set of covariates $\bX_{ij}$, and we denote their asymptotic variance as $v_1, v_2, v^{\textup{(ols)}}_1, v^{\textup{(ols)}}_2, v^{\textup{(cl)}}_1, v^{\textup{(cl)}}_2$, respectively.
Under Assumptions~\ref{asp:1}, \ref{asp:2}, \ref{asp:3}, and assuming equal observed cluster sizes with $N_i=\widetilde{n}$ for all $i$, then
(a) $v_1^{\textup{(cl)}} \le v_1$ and $v_1^{\textup{(cl)}} \le v_1^{\textup{(ols)}}$ if $\pi = 0.5$ and (b) $v_2^{\textup{(cl)}} \le v_2$ and $v_2^{\textup{(cl)}} \le v_2^{\textup{(ols)}}$.
\end{proposition}
To provide some intuition for Proposition \ref{coro3}, recall that the coefficient for the treatment term is defined at the cluster level across all six working models. Therefore, the associated asymptotic variance for each estimator turns out to be a function of $Var\{\overline{Y}_i^o(a) - \bc_a^\top \overline{\bX}_i^o\}$ for $a \in \{0,1\}$,  where $\overline{Y}_i^o(a) = N_i^{-1}\sum_{j\in \mathcal{O}_i} Y_{ij}(a)$, and $\bc_a$ being a constant vector; here both $\overline{Y}_i^o(a)$ and $\overline{\bX}_i^o$ are cluster-level variables. The optimal value of $\bc_a$ that minimizes the asymptotic variance is $Var(\overline{\bX}_i^o)^{-1}Cov(\overline{\bX}_i^o, \overline{Y}_i(a))$, which can actually be attained by the cluster-level ANCOVA methods. In contrast, the value of $\bc_a$ under the individual-data-based methods (mixed models or ordinary least squares) involves the individual-level components, $cov(\bX_{ij}, Y_{ij}(a))$ and  $Var(\bX_{ij})$, thereby deviating from the optimal $\bc_a$ and representing a source of efficiency loss. However, in the very special case when $Var(\overline{\bX}_i^o)^{-1}Cov(\overline{\bX}_i^o, \overline{Y}_i(a)) = Var(\bX_{ij})^{-1}cov(\bX_{ij}, Y_{ij}(a))$ (which holds under conditions listed in Proposition~\ref{prop1}), the \\individual-data-based and cluster-summary-based methods become asymptotically equivalent and have the same precision.  
}

{
The above precision comparisons are generally indeterminate when the observed cluster sizes vary. The main reason is that different methods use different ways to account for the variability of $N_i$. 
% For example, the relationship between $Var(N_i^{-1} \sum_{j \in \mathcal{O}_i} Y_{ij})$ and $E[N_i]^{-2}Var( \sum_{j \in \mathcal{O}_i} Y_{ij})$ is generally indeterminate, while the former quantity is involved in $v^{\textup{(cl)}}$ and the latter is related to $v^{\textup{(ols)}}$.
By increasing the variation of cluster sizes, the precision of all estimators tends to decrease, but by a different amount. In the special case where the mixed-model ANCOVA1 is correctly specified, the mixed-model ANCOVA1 estimator provides higher precision than the cluster-level ANCOVA1 estimator. Otherwise, their efficiency comparison depends on the degree of variability in cluster size, the magnitude of intracluster correlation, and the prognostic value of covariates. In the ensuing simulation study and data application, we give examples to demonstrate that individual-data-based methods can be either more precise or less precise than the cluster-summary-based approaches under different sets of design parameters.
% Finally, \citet{su2021model} recently proved that ordinary least regression based on individual data is asymptotically dominated by regression based on cluster totals regarding efficiency, under arbitrary model misspecification. Our results, however, suggest that the mixed-model ANCOVA1 and cluster-level analysis do not have dominating asymptotic efficiency over each other, under arbitrary model misspecification and unequal cluster sizes. 
}

{For cluster-summary-based methods, \cite{su2021model} proposed a new ANCOVA2 model based on scaled cluster totals, which estimates the average treatment effect by the ordinary-least-squares estimator for $\alpha_A$ in the regression model
$$\frac{mN_i\overline{Y}_i^o}{\sum_{i=1}^m N_i} \sim \alpha_0 + \alpha_A A_i + (\alpha_N  +\alpha_{AN}A_i) \frac{mN_i}{\sum_{i=1}^m N_i} + (\alpha_{\bX}+\alpha_{A\bX}A_i)^\top\left(\frac{mN_i\overline{\bX}_i^o}{\sum_{i=1}^m N_i} - \frac{mN_i\overline{\bX}_{all}^o}{\sum_{i=1}^m N_i}\right).$$
We denote this estimator as $\widehat\Delta_2^{\textup{(sct)}}$.
Under a finite-population, randomization inference framework, they showed that $\widehat\Delta_2^{\textup{(sct)}}$ is robust to model misspecification and more efficient than the individual-level ANCOVA2 estimator $\widehat\Delta_2^{\textup{(ols)}}$.  
Under our super population framework, we are able to reproduce this result and show the consistency of $\widehat\Delta_2^{\textup{(sct)}}$ to our estimand $\Delta^*$ (see Section D in the Supplementary Material for technical details). \cite{su2021model}  also discussed the cluster-level ANCOVA2 estimator $\widehat\Delta_2^{\textup{(cl)}}$, but did not compare its precision with $\widehat\Delta_2^{\textup{(sct)}}$ since they target different estimands in their framework (assuming informative cluster sizes). With our setup of non-informative cluster sizes, $\widehat\Delta_2^{\textup{(cl)}}=\widehat\Delta_2^{\textup{(sct)}}$ if the observed cluster size is constant, but otherwise they have different asymptotic variances that are indeterminate in general. In addition, $\widehat\Delta_2^{\textup{(sct)}}$ may be either more or less efficient than the mixed-model ANCOVA2 estimator $\widehat\Delta_2$, depending on the true data-generating distribution. 
We support these arguments with analytical examples provided in the Supplementary Material, but do not further investigate this new estimator hereafter.
}
% This result differs from \citet{su2021model}, who considered weighted least squares methods with weights $mN_i/\sum_{i=1}^m N_i$ to align the estimand for cluster-level ANCOVA2 based on cluster averages with that based on cluster totals. Under our causal framework, we do not need to appeal to this weight due to the unified estimands discussion in Section 2.2 and therefore our cluster-level ANCOVA2 based on cluster averages has a different asymptotic variance expression. Finally, compared to mixed models, cluster-level ANCOVA2 based on scaled cluster totals may also be less efficient when the cluster size varies, for which an example is provided in the Supplementary Material. 
% Furthermore, when treatment-covariate interaction terms are not included, the cluster-level ANCOVA1 estimator based on scaled cluster totals may also be less efficient than individual-level ANCOVA1 estimator. 

% Because the cluster-level ANCOVA models are frequently not optimally efficient under our framework, and they are used much less frequently in practice with CRTs, we do not further investigate this approach in the remainder of the article.

% the relationship across methods is indeterminate in our framework since each estimator accounts for the variability of $N_i$ differently. 

% For example, compared to the variance formula for cluster-level ANCOVA2 based on scaled cluster totals given by \cite{su2021model}, the variance under the super-population framework involves an extra term of $E[N]^{-2}\Delta^*Var(N)$.

{\subsection{Summary}}

\begin{figure}[htb]
\centering
\resizebox{0.85\textwidth}{!}{
\begin{tikzpicture}
\node at (-0.5,0) (mixed-ANCOVA1) [rectangle,draw,text width=2.5cm,text centered]  {Mixed-model ANCOVA1};

\node at (4.5,0) (ols-ANCOVA1) [rectangle,draw,text width=3cm,text centered]  {Individual-level ANCOVA1};

\node at (10,0) (cl-ANCOVA1) [rectangle,draw,text width=2.5cm,text centered]  {Cluster-level ANCOVA1};

\node at (-0.5,-4) (mixed-ancova2) [rectangle,draw,text width=2.5cm,text centered]  {Mixed-model ANCOVA2};

\node at (4.5,-4) (ols-ancova2) [rectangle,draw,text width=3cm,text centered]  {Individual-level ANCOVA2};

\node at (10,-4) (cl-ancova2) [rectangle,draw,text width=2.5cm,text centered]  {Cluster-level ANCOVA2};

\path [draw,-Latex] (-1.5,1) -- node [text width=10cm,above,text centered ] {Increasing or equivalent precision, when $\pi=0.5$} (11,1);

\path [draw,-Latex] (-1.5,-5) -- node [text width=10cm,below,text centered ] {Increasing or equivalent precision} (11,-5);

\path [draw,Latex-Latex] (mixed-ANCOVA1) -- node [text width=2.5cm,left,align=right] {Equivalent in special cases if $\pi=0.5$} (mixed-ancova2);

\path [draw,Latex-Latex] (ols-ANCOVA1) -- node [text width=2.5cm,left,align=right] {Equivalent if $\pi=0.5$}  (ols-ancova2);

\path [draw,-Latex] (cl-ANCOVA1) -- node [text width=5cm,left,align=right] {Equivalent if $\pi=0.5$, or increasing precision} (cl-ancova2);
% \path [draw,Latex-Latex] (cl-ANCOVA1) -- node [text width=2cm,right,align=left] {if $\pi=0.5$, less precise} (cl-ancova2);
%\node at (8,-2) (mixed-ancova2) [text width=2cm,right] {if $\pi=0.5$,\\ less precise} ;
\end{tikzpicture}
}
    \caption{Comparison of asymptotic precision among working models when each cluster has the same number of individuals ($N_i = \widetilde{n})$. When $N_i$ varies by clusters, most precision comparisons in this figure no longer hold except for the comparison between cluster-level ANCOVA1 and cluster-level ANCOVA2 and the comparison between individual-level ANCOVA1 and individual-level ANCOVA2. The special cases comparing mixed-model ANCOVA1 and mixed-model ANCOVA2 are listed in Proposition 1.}\label{fig:summary-precision-comparison}
\end{figure}
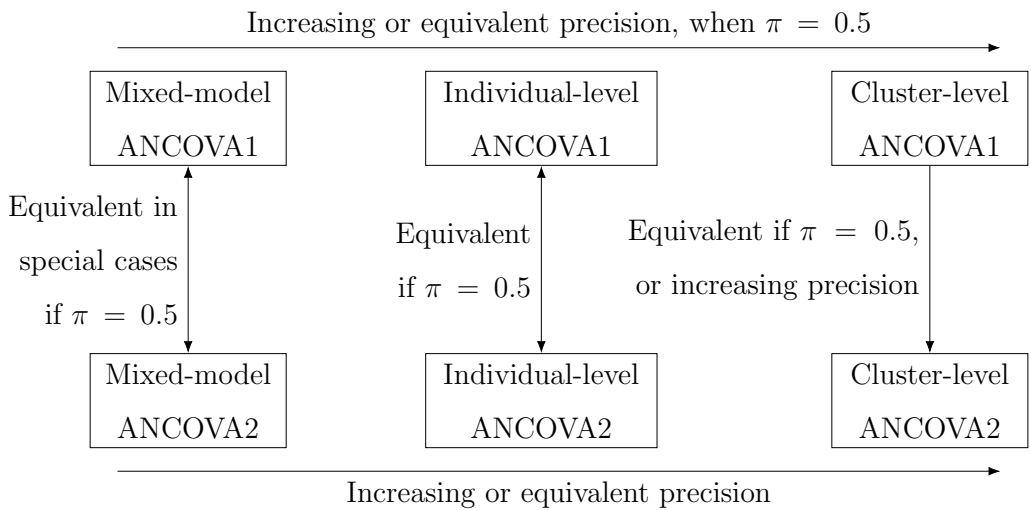

{
The results in Sections~\ref{sec:main-result} and~\ref{sec: precision} confirm that all six models can provide consistent estimators of the average treatment effect $\Delta^*$ in CRTs, and several analytical insights to the precision are offered in Section \ref{sec: precision}. For a quick reference, Figure~\ref{fig:summary-precision-comparison} summarizes the comparative relationships among different working models from the standpoint of precision, under the special case where the observed cluster sizes are equal. In this setting, the cluster-level ANCOVA2 estimator (and the cluster-level ANCOVA1 estimator if $\pi=0.5$) can be the most efficient one. 
% Each ANCOVA1 method is asymptotically equivalent to the corresponding ANCOVA2 method under equal randomization. 
Therefore, if the observed cluster sizes are equal or approximately equal, the cluster-level ANCOVA1 or ANCOVA2 estimator may be preferred due to efficiency. 
}

{
In the more general scenario where the observed cluster sizes vary, any of the considered methods may have the smallest asymptotic variance (allowing for ties) and can be a ``winner''; the exact comparative relationships will depend on the true data-generating distributions and the degree of cluster size variation. That is to say, no method is uniformly more precise than another. Therefore, mixed-model ANCOVA remains as useful as individual-level ANCOVA and cluster-level ANCOVA for analyzing CRTs with variable cluster sizes. Among them, mixed-model ANCOVA may be preferred in practice because (1) it provides a more efficient estimator when the correlation structure is approximately exchangeable, and (2) it offers an estimate of the intracluster correlation coefficient as a by-product of model fitting (an important quantity that has been recommended to report by the CONSORT extension statement for CRTs; see \citet{campbell2004consort}). If the number of clusters is limited or strong treatment effect heterogeneity is not anticipated, mixed-model ANCOVA1 is expected to exhibit better finite-sample properties because there is no additional cost of degrees of freedom to estimate interaction effect parameters. Under equal randomization of clusters, it is also worth emphasizing that the model-based variance estimator from mixed-model ANCOVA1 remains valid, and one can simply rely on the standard model output from standard statistical software for valid inference under arbitrary working model misspecification.
}

%In light of these results, if the number of clusters is small and strong treatment effect heterogeneity is not anticipated, ANCOVA1 methods may perform better than ANCOVA2 methods under equal randomization because there is no additional cost of degrees of freedom to estimate interaction effect parameters. If the correlation structure is approximately exchangeable, then the mixed models are naturally suitable and can often be an efficient choice for estimating the average treatment effect; otherwise, if the clustering structure is much more complex and challenging to capture analytically, the individual-level ANOVA methods via ordinary least squares may be better alternatives.}
%if each cluster has additional unobserved subclustering structures, e.g., during the spread of infectious diseases, ordinary least square estimators may be better alternatives..}

% However, the cluster-level ANCOVA2 estimator still has the advantage of no efficiency loss by covariate adjustment for any $\pi \in (0,1)$ \citep{Tsiatis2008, lin2013agnostic}. Therefore, we recommend using the cluster-level ANCOVA2 estimator to achieve robust inference and adjust for covariates.

\section{Simulation Study}\label{sec:simulation}

\subsection{Simulation design}
{
We report two simulation studies to demonstrate our main results.
In the first, we focus on the mixed-model unadjusted, ANCOVA1, and ANCOVA2 estimators and demonstrate their robustness and precision. For this purpose, we carefully design three scenarios where each of the three estimators has the highest precision in one scenario, thereby illustrating that (i) covariate adjustment in mixed models may or may not improve precision and (ii) treatment-covariate interaction terms in mixed models may or may not improve precision. Furthermore, their model robustness is demonstrated by all three scenarios. In the second simulation study, we compare the precision of ANCOVA1 and ANCOVA2 based on mixed models, individual-level ordinary least squares, or cluster-level ordinary least squares. We design two scenarios, with the first demonstrating the results in Figure~\ref{fig:summary-precision-comparison} and the second showing that the mixed models can be more precise than the others when the cluster sizes are variable.} In each scenario, we study both small- and large-sample behavior of the estimators by setting $m=20$ and $m=200$.

{
For the first simulation study, we generate
\begin{align*}
    &A_i\sim\text{Bernoulli}(\pi=0.5), \,Y_{ij} =4A_i\left( X_{ij} - {n}^{-1}\sum_{k=1}^n X_{ik}\right) + \gamma_i +  \epsilon_{ij}\quad  &\text{for Scenario 1,}\\
    &A_i\sim\text{Bernoulli}(\pi=0.4), \,Y_{ij} =2X_{ij}+4A_i\left( X_{ij} - {n}^{-1}\sum_{k=1}^n X_{ik}\right) + \gamma_i +  \epsilon_{ij}\quad  &\text{for Scenario 2,}\\
    &A_i\sim\text{Bernoulli}(\pi=0.6), \,Y_{ij} =4A_i\left(2 X_{ij} - {n}^{-1}\sum_{k=1}^n X_{ik}\right) + \gamma_i +  \epsilon_{ij}\quad  &\text{for Scenario 3,}
\end{align*}
where $X_{ij}\sim N(0,4)$, $\gamma_i \sim N(0,1)$ and  $\epsilon_{ij}\sim N(0,25)$. The random variables \\ $\{A_i, X_{i1},\ldots, X_{in},\gamma_i, \epsilon_{i1},\ldots, \epsilon_{in}\}$ are mutually independent and also independent across $i$.
For generating the observed data, we independently draw $N_i$ from a discrete uniform distribution on $\{4,\dots,12\}$ and then sample $N_i$ data vectors from $\{(Y_{ij}, A_i, X_{ij}): j = 1,\dots, 12\}$ without replacement. 
}

{
For the second simulation, we consider stratified randomization and independently generate a binary stratum indicator variable $S_i\sim\text{Bernoulli}(0.6)$ for each cluster $i$. We then assign $A_i$ under stratified randomization within each stratum defined by $S_i$ and let $\pi = 0.5$ for Scenario 1 and $\pi = 0.6$ for Scenario 2. Next, we generate $Y_{ij} = 5(S_i - 0.6+ X_{ij} ) A_i +2{n}^{-1}\sum_{k=1}^n X_{ik} + \epsilon_{ij}$ for Scenario 1 and $Y_{ij} = 5(S_i  + X_{ij} - 0.6) A_i + 5\gamma_i + \epsilon_{ij}$ for Scenario 2, where  $X_{ij}, \gamma_i, \epsilon_{ij}$ are as defined as in Scenario 1. We consider fixed cluster size $N_i = 8$ for Scenario 1 but random cluster size $N_i$ following a discrete uniform distribution on $\{2,\dots,18\}$ for Scenario 2. We then sample $N_i$ data vectors from $\{(Y_{ij}, A_i, X_{ij}): j = 1,\dots, 18\}$ without replacement. Compared to the first simulation study, we increase the variability of $N_i$ for Scenario 2 to illustrate its impact on precision comparisons.
}

For all scenarios in both simulation studies, we used $10,000$ repetitions. 
{
In the first simulation study, we estimate the average treatment effect (the true $\Delta^*$ is 0 for all scenarios) by the mixed-model unadjusted, ANCOVA1, and ANCOVA2 estimators. The mixed-model unadjusted estimator is obtained from the mixed-model ANCOVA1 estimator but with no adjustment for covariates. %which is asymptotically equivalent to the mixed-model ANCOVA1 estimator without covariate adjustment if $N_i $ is fixed. 
The mixed-model ANCOVA1 and ANCOVA2 estimators adjust for the covariate $X_{ij}$.} For each estimator, we consider the following performance metrics: bias, empirical standard error, averaged robust standard error, averaged model-based standard error, coverage probability of the 95\% confidence intervals (constructed using a normal approximation with the robust or model-based standard error), and relative efficiency versus the mixed-model unadjusted estimator. {In the second simulation study, we estimate the average treatment effect (the true $\Delta^*$ is 0) using six working models: mixed-model ANCOVA1 and ANCOVA2, individual-level ANCOVA1 and ANCOVA2, and cluster-level ANCOVA1 and ANCOVA2. 
The first four working models adjust for $\{X_{ij}, S_i\}$, while the last two working models adjust for $\{\overline{X}_i^o, S_i\}$.
% The individual-level ANCOVA1 and ANCOVA2 model formulations are the same as mixed-model ANCOVA1 and ANCOVA2 model formulations except that the random intercept is omitted. All six working models adjust for covariates $\{X_{ij}, S_i\}$. 
In the second simulation study, we consider the same performance metrics as in the first simulation study except that we drop the averaged model-based standard error and coverage probability, and use the relative efficiency versus the mixed-model ANCOVA1 (instead of the unadjusted) estimator. }

\subsection{Simulation results}
Table~\ref{table:simulation1} summarizes the results of the first simulation study. For all three mixed-model estimators, although their working models are misspecified, they have negligible bias and nominal coverage probability for the true average treatment effect across Scenarios 1 to 3, thereby illustrating their robustness.  {With a larger number of clusters ($m=200$), the robust standard errors for all estimators match their empirical standard errors. The model-based standard error for mixed-model ANCOVA1 is also valid in Scenario 1 where $\pi=0.5$, which confirms our theoretical results, while the model-based standard error for the mixed-model ANCOVA2 estimator may not always match the empirical standard error.} When $m=20$, all estimators have 0-4\% under-coverage due to the finite-sample bias of normal-based confidence intervals in CRTs; the under-coverage can be alleviated, for example, by considering a $t$-distribution with heavier tails \citep{li2016evaluation}.

{
Table~\ref{table:simulation1} also demonstrates the comparison of precision among mixed-model estimators. In Scenario 1, the mixed-model unadjusted estimator has the highest precision, showing that covariate adjustment by mixed models can even reduce the precision for estimating $\Delta^*$. 
Specifically, the data-generating distribution under Scenario~1 implies that 
% $Cov\left(\sum_{j=1}^n Y_{ij} \sum_{j=1}^n X_{ij}\right) = 0$ but $Cov(Y_{ij}, X_{ij}) \ne 0$, which means the aggregated covariate is not prognostic at the cluster level but is prognostic at the individual level. Since the treatment is assigned at the cluster level, only cluster-level correlation contributes to possible variance reduction in estimating the average treatment effect $\Delta^*$, which implies that 
covariate adjustment provides no variance reduction, and mixed-model ANCOVA1 and ANCOVA2 
% exploit both cluster and individual level correlations for estimating covariates coefficients and therefore lead to 
have 12-15\% efficiency loss. 
%by tapping into outcome-covariate correlations that are ancillary to the estimation of cluster-level treatment effect.
For Scenarios 2 and 3, we increase the prognostic value of $X_{ij}$ by adding $2X_{ij}$ and $4A_iX_{ij}$ in the outcome data-generating model, respectively; the addition of the first term favors mixed-model ANCOVA1 and the addition of the second term favors mixed-model ANCOVA2. Furthermore, we adjust the randomization probability $\pi$ to amplify the impact of this change. As a result, we observe that mixed-model ANCOVA1 leads to the highest precision in Scenario 2, whereas mixed-model ANCOVA2 leads to the highest precision in Scenario 3.
}
\begin{table*}
\centering
\caption{Simulation results for Scenarios 1--3 of the first simulation study with 20 or 200 clusters. The performance metrics are bias, empirical standard error (Emp SE), averaged robust standard error (Avg robust SE), averaged model-based standard error (Avg model-based SE), 
coverage probability of 95\% confidence intervals based on normal approximation and robust standard error (Robust CP) or model-based standard error (Model-based CP), and relative efficiency to the mixed-model unadjusted estimator (RE). Across all scenarios and estimators, the maximum Monte Carlo standard errors for bias, Emp SE, Avg robust SE, Avg model-based SE, robust CP, model-based CP, and RE are 0.031, 0.022, 0.003, 0.002, 0.003, 0.003, and 0.030, respectively. }\label{table:simulation1}
\resizebox{\textwidth}{!}{
\begin{tabular}{ccrrrrrrrr}
  \hline
Scenario & $m$ & \makecell[r]{Mixed model\\ estimator} & Bias & \makecell[r]{Emp \\ SE} & \makecell[r]{Avg robust\\ SE} & \makecell[r]{Avg model-\\based SE} & \makecell[r]{Robust\\CP} & \makecell[r]{Model-\\based\\CP} & RE \\ 
  \hline
  \multirow{6}{*}{1} & & unadjusted & 0.00 & 1.08 & 1.04 & 1.22 & 0.93 & 0.97 & 1.00 \\ 
& $20$ & ANCOVA1 & 0.00 & 1.17 & 1.14 & 1.12 & 0.93 &0.94 & 0.86 \\ 
 &  & ANCOVA2 &   0.00 & 1.15 & 1.15 & 1.11 & 0.94 &0.93 & 0.87 \\ 
 \cline{2-10}
& & unadjusted &  0.00 & 0.32 & 0.33 & 0.33 & 0.95 & 0.97 & 1.00 \\ 
& $200$ & ANCOVA1 &  0.00 & 0.35 & 0.35 & 0.35 & 0.95 &0.95 & 0.88 \\ 
& & ANCOVA2  & 0.00 & 0.35 & 0.35 & 0.37 & 0.95  & 0.96 & 0.87 \\ 
 \hline
\multirow{6}{*}{2} & & unadjusted &  0.01 & 1.43 & 1.36 & 1.57 & 0.92 & 0.97 & 1.00 \\  
& $20$ & ANCOVA1 & 0.00 & 1.15 & 1.11 & 1.11 & 0.94 & 0.94 & 1.55 \\  
 &  & ANCOVA2 & $-0.01$ & 1.23 & 1.21 & 1.09 & 0.93 & 0.91 & 1.35 \\ 
\cline{2-10}
 & & unadjusted & $-0.00$ & 0.43 & 0.43 & 0.48 & 0.95 & 0.96 & 1.00 \\ 
& $200$ & ANCOVA1 & $-0.00$ & 0.35 & 0.34 & 0.34 & 0.95 & 0.95 & 1.55 \\ 
 & & ANCOVA2 & $-0.00$ & 0.37 & 0.37 & 0.36 & 0.95 & 0.94 & 1.34 \\ 
 \hline
\multirow{6}{*}{3} & & unadjusted & 0.01 & 1.50 & 1.47 & 2.18 & 0.93 &0.99 & 1.00 \\ 
& $20$ & ANCOVA1 & 0.01 & 1.69 & 1.54 & 1.54 & 0.91 & 0.92 & 0.78 \\ 
 &  & ANCOVA2 & 0.02 & 1.37 & 1.42 & 1.19 & 0.94 &0.89 & 1.19 \\  
\cline{2-10}
 & & unadjusted & 0.01 & 0.46 & 0.46 & 0.67 & 0.95 & 0.99 & 1.00 \\ 
& $200$ & ANCOVA1 & 0.01 & 0.49 & 0.48 & 0.47 & 0.95 & 0.94 & 0.87 \\ 
 & & ANCOVA2 &0.01 & 0.43 & 0.43 & 0.39 & 0.95 & 0.93 & 1.15 \\ 
 \hline

\end{tabular}
}
\end{table*}

\begin{table*}
\centering
\caption{Simulation results for Scenarios 1--2 of the second simulation study with 20 or 200 clusters. The performance metrics are bias, empirical standard error (Emp SE), averaged robust standard error (Avg robust SE), coverage probability of 95\% confidence intervals based on normal approximation and robust standard error (Robust CP), and relative efficiency to the mixed-model ANCOVA1 estimator (RE). Across all scenarios and estimators, the maximum Monte Carlo standard errors for bias, Emp SE, Avg robust SE, CP, and RE are 0.017, 0.012, 0.001, 0.002, and 0.010, respectively. }\label{table:simulation2}
\resizebox{0.83\textwidth}{!}{

\begin{tabular}{ccrrrrrr}
  \hline
Scenario & $m$ & \makecell[r]{Estimator} & Bias & \makecell[r]{Emp \\ SE} & \makecell[r]{Avg robust \\SE} & \makecell[r]{Robust\\ CP\\} & RE \\ 
  \hline
  \multirow{12}{*}{1} & \multirow{6}{*}{$20$} & mixed-model ANCOVA1 & 0.01 & 0.71 & 0.73 &  0.94 & 1.00 \\
 &  & individual-level ANCOVA1 & 0.01 & 0.70 & 0.73 &   0.94 & 1.01 \\ 
 &  & cluster-level ANCOVA1 & 0.01 & 0.59 & 0.58 &   0.93 & 1.42 \\ 
 \cline{3-8}
 &  & mixed-model ANCOVA2 & 0.01 & 0.71 & 0.78   & 0.96 & 1.00 \\ 
 &  & individual-level ANCOVA2 & 0.01 & 0.70 & 0.78 &  0.96 & 1.01 \\ 
 &  & cluster-level ANCOVA2 & 0.00 & 0.59 & 0.63 &   0.95 & 1.42 \\ 
 \cline{2-8}
 &\multirow{6}{*}{$200$} & mixed-model ANCOVA1 & 0.00 & 0.22 & 0.22 & 0.95 & 1.00 \\ 
 &  &  individual-level ANCOVA1 & 0.00 & 0.22 & 0.22  & 0.95 & 1.02 \\ 
 &  &  cluster-level ANCOVA1  & 0.00 & 0.18 & 0.18  & 0.95 & 1.52 \\ 
\cline{3-8}
 &  &  mixed-model ANCOVA2 & 0.00 & 0.22 & 0.22 & 0.95 & 1.00 \\ 
 &  &  individual-level ANCOVA2 & 0.00 & 0.22 & 0.22  & 0.95 & 1.02 \\ 
 &  &  cluster-level ANCOVA2  & 0.00 & 0.18 & 0.18  & 0.95 & 1.52 \\ 
  \hline
  \multirow{12}{*}{2} & \multirow{6}{*}{$20$} & mixed-model ANCOVA1 & 0.02 & 1.62 & 1.71  & 0.94 & 1.00 \\ 
 &  & individual-level ANCOVA1 & 0.02 & 1.66 & 1.71 &   0.94 & 0.96 \\ 
 &  & cluster-level ANCOVA1 & 0.02 & 1.71 & 1.73 &  0.94 & 0.90 \\ 
 \cline{3-8}
 &  & mixed-model ANCOVA2 & $-0.01$ & 1.56 & 1.83 &  0.96 & 1.08 \\
 &  & individual-level ANCOVA2 & $-0.01$ & 1.60 & 1.84   & 0.96 & 1.03 \\ 
 &  & cluster-level ANCOVA2 &    $-0.00$ & 1.70 & 1.90  & 0.96 & 0.91 \\ 
 \cline{2-8}
 &\multirow{6}{*}{$200$} &  mixed-model ANCOVA1 & $-0.01$ & 0.77 & 0.77 & 0.97 &1.00\\ 
 &  &  individual-level ANCOVA1 & 0.00 & 0.51 & 0.54 &  0.96 & 0.95 \\ 
 &  &  cluster-level ANCOVA1  & 0.00 & 0.52 & 0.55 &   0.96 & 0.91 \\
\cline{3-8}
 &  &  mixed-model ANCOVA2 & $-0.00$ & 0.49 & 0.52 &   0.96 & 1.03 \\ 
 &  &  individual-level ANCOVA2 & $-0.01$ & 0.50 & 0.53 &  0.96 & 0.98 \\ 
 &  &  cluster-level ANCOVA2  & 0.00 & 0.52 & 0.54 &  0.96 & 0.91 \\
 \hline
\end{tabular}

}
\end{table*}

% Results under Scenario~1 demonstrate that the mixed-model ANCOVA1 estimator can be less efficient than the mixed-model unadjusted estimator. Specifically, the data generating distribution under Scenario~1 implies that $Cov(\overline{Y}^o, \overline{X}^o) = 0$ but $Cov(Y, X) = 0.875$, which means the aggregated covariate is not prognostic at the cluster level but is prognostic at the individual level. Since the treatment is assigned at the cluster level, only $Cov(\overline{Y}^o, \overline{X}^o)$ is related to possible variance reduction in estimating the average treatment effect. For Scenario 1, $Cov(\overline{Y}^o, \overline{X}^o) = 0$ implies that covariate adjustment provides no variance reduction. In contrast, mixed-model ANCOVA1 exploits both $Cov(\overline{Y}^o, \overline{X}^o)$ and $Cov(Y, X)$ for estimating $\beta_X$ and therefore leads to 4\% efficiency loss by tapping into correlations that are ancillary to the estimation of cluster-level treatment effect. %, hence, falsely identifies correlations, which . 
% The magnitude of efficiency loss depends on the variance of $X$ and can be as high as 37\% if the variance of $X$ is increased to 100 (for fixed $\beta_X$). In this scenario, the cluster-level ANCOVA1 is also less efficient than the mixed-model unadjusted estimator, since the unadjusted mixed-model can be considered correctly specified (by marginalizing over $X$) and the variation in cluster sizes results in further efficiency loss of the cluster-level ANCOVA1 estimator as discussed in Section~\ref{subsec: precision-gain}.

{Table~\ref{table:simulation2} summarizes the results of the second simulation study. All six methods retain their robustness under stratified randomization. In Scenario 1 where $\pi = 0.5$, the robust variance is consistent as implied by Theorem~\ref{thm3}(c), while the estimators are slightly conservative under Scenario 2 with $\pi = 0.6$, as expected. In terms of precision, Scenario 1 demonstrates Figure~\ref{fig:summary-precision-comparison} under equal observed cluster sizes. Since $\pi = 0.5$, the ANCOVA1 and ANCOVA2 estimators are equivalent. The individual-level ANCOVA1 estimators are slightly more precise than the mixed-model ANCOVA1 estimators but less precise than cluster-level ANCOVA1 estimators. In contrast, the true model of Scenario 2 is correctly specified by mixed-model ANCOVA2, which indeed has the highest precision among all estimators. Due to the variability of cluster sizes, individual-level and cluster-level ANCOVA1 and ANCOVA2 lead to 2--9\% variance inflation compared to mixed-model ANCOVA1.}

\section{Applications to Three Cluster-Randomized Trials}\label{sec:data-application}
\subsection{Background and contexts}\label{subsec: two-trials}
\emph{Task Shifting and Blood Pressure Control in Ghana} (TASSH) is a CRT evaluating the effectiveness of a nurse-led task shifting strategy for hypertension control through systolic blood pressure (SBP) reduction \citep{ogedegbe2018health}. Thirty-two community health centers were randomly assigned to receive treatment (provision of health insurance coverage plus TASSH, 389 patients within 16 clusters) or usual care (provision of health insurance coverage only, 368 patients within 16 clusters). Each cluster recruited a different number of individuals, ranging from 17 to 31. 
%with mean 23.66 and standard error 2.89. 
We focus on the primary outcome of the study, the change in SBP from baseline to 12 months; the included baseline covariates are age, SBP, Diastolic Blood Pressure, and the location of the health center (rural or urban).

\emph{Improving Early Childhood Development in Zambia} (IECDZ) is a CRT assessing the effect of a community-based early childhood development (ECD) program on physical and cognitive development \citep{rockers2018two}. Thirty clusters of villages were equally randomized to receive treatment (ECD, 195 caregiver-child dyads within 15 clusters) or control (no intervention, 182 caregiver-child dyads within 15 clusters) with each cluster including 2--26 caregiver-child dyads. %(mean 12.57 and standard error 5.15).
We focus on the continuous outcome, height-for-age z-score (HAZ), at the year-2 follow-up, which was used to determine children's stunting status (HAZ $< -2$) in the primary analysis of the study. We adjust for the baseline covariates age, baseline HAZ, as well as child motor score.

The \emph{Work, Family, and Health Study} (WFHS) is a CRT designed to enhance the understanding of the impact of workplace practices and policies on employees' work, family, and health outcomes \citep{WFHS}. We use data from one study site, a Fortune 500 company, where 56 study groups were randomly assigned to receive a workplace intervention (423 employees in 29 clusters) or usual practice (400 employees in 27 clusters) with each cluster including 3--50 employees. 
%(mean 14.70, standard error 8.62). 
We focus on the control over work hours (CWH) at the 6-month follow-up, which is a continuous measure ranging from 1 to 5 and demonstrated the largest treatment effect \citep{kelly2014changing}. Baseline CWH and job function (core or supporting) are adjusted for as baseline covariates.

% To achieve treatment balance, TASSH and IECDZ used matched-pairs for randomization, while WFHS adopted a modified biased-coin design. For the purpose of demonstration, we perform no adjustment for these randomization schemes and deem that simple randomization were used; such a simplification might result in conservative inference \citep{imai2009essential, wang2021model}. Missing outcomes present in all three trials, where the missing proportions are 15\%, 1\%, and 20\% for TASSH, IECDZ, and WFHS, respectively.

% Note that both of these trials have a reasonably small number of clusters close to the `small' number considered in our simulation studies.

\subsection{Data analysis results}
{
In each CRT, we estimate the average treatment effect using the mixed-model unadjusted, ANCOVA1, and ANCOVA2 estimators, individual-level ANCOVA1 and ANCOVA2 estimators, and cluster-level ANCOVA1 and ANCOVA2 estimators.} For illustration purposes, we assume that simple randomization is used in all CRTs. Furthermore, individuals with missing outcomes (15\%, 1\%, and 20\% for TASSH, IECDZ, and WFHS) are removed from the analysis and missing baseline variables are imputed once by the mean of non-missing observations \citep{sullivan2018should}.

\begin{table*}
\centering
\caption{Summary of data analyses results: point estimate of the average treatment effect (Est), the robust estimator for standard error (SE), 95\% confidence interval (CI), and proportion variance reduction compared to the unadjusted estimator (PVR). Positive (negative) PVR indicates that covariate adjustment leads to variance reduction (inflation).}\label{table:data-analysis}
\begin{tabular}{crrrrr}
  \hline
Study name & Estimators & Est & SE & 95\% CI & PVR \\ 
  \hline
& mixed-model unadjusted & $-1.29$ & 2.11 & ($-5.43$, 2.85) & - \\ 
 &  mixed-model ANCOVA1 & $-2.22$ & 2.06 & ($-6.27$, 1.82) & 4\% \\ 
&  individual-level ANCOVA1 & $-2.23$ & 2.07 & ($-6.28$, 1.83) & 4\% \\ 
TASSH & cluster-level ANCOVA1 & $-1.54$ & 1.81 & ($-5.08$, 2.01) & 26\% \\ 
&  mixed-model ANCOVA2 & $-2.17$ & 2.26 & ($-6.61$, 2.26) & $-15\%$ \\ 
&  individual-level ANCOVA2 & $-2.21$ & 2.26 & ($-6.64$, 2.22) & $-15\%$ \\ 
 & cluster-level ANCOVA2 & $-1.07$ & 1.98 & ($-4.95$, 2.82) & 12\% \\ 
   \hline
& mixed-model unadjusted & 0.08 & 0.14 & ($-0.19$, 0.35) & - \\ 
 &  mixed-model ANCOVA1 & 0.08 & 0.15 & ($-0.22$, 0.38) & $-20\%$ \\ 
&  individual-level ANCOVA1 & 0.07 & 0.16 & ($-0.25$, 0.38) & $-30\%$ \\
IECDZ & cluster-level ANCOVA1 & 0.08 & 0.14 & ($-0.19$, 0.36) & $-2\%$ \\ 
&  mixed-model ANCOVA2 & 0.08 & 0.16 & ($-0.24$, 0.40) & $-36\%$ \\ 
&  individual-level ANCOVA2 & 0.07 & 0.17 & ($-0.26$, 0.40) & $-47\%$ \\ 
 & cluster-level ANCOVA2 & 0.10 & 0.15 & ($-0.20$, 0.39) & $-18\%$ \\ 
   \hline
& mixed-model unadjusted & 0.16 & 0.07 & (0.01, 0.30) & - \\ 
 &  mixed-model ANCOVA1 & 0.21 & 0.05 & (0.12, 0.30) & 59\% \\ 
&  individual-level ANCOVA1 & 0.21 & 0.05 & (0.12, 0.30) & 62\% \\ 
WFHS & cluster-level ANCOVA1 & 0.25 & 0.05 & (0.14, 0.35) & 48\% \\ 
&  mixed-model ANCOVA2 & 0.21 & 0.05 & (0.12, 0.31) & 57\% \\ 
&  individual-level ANCOVA2 & 0.21 & 0.05 & (0.12, 0.30) & 60\% \\ 
 & cluster-level ANCOVA2 & 0.25 & 0.05 & (0.14, 0.35) & 45\% \\ 
   \hline
\end{tabular}
\end{table*}

Table~\ref{table:data-analysis} presents the data analysis results.
For the TASSH study, covariate adjustment by the cluster-level ANCOVA1 results in a 26\% variance reduction and a 14\% narrower confidence interval compared to the mixed-model unadjusted estimator. 
% {On the other hand, covariate adjustment by the cluster-level ANCOVA2 results in 12\% variance reduction. 
% This implies that, if researchers plan to use cluster-level ANCOVA1 for their primary analyses,  approximately 10\% fewer clusters may be needed compared to using an unadjusted analysis for achieving the same statistical power. 
{Mixed-model ANCOVA1 and individual-level ANCOVA1 perform similarly, resulting in a small variance reduction of 4\% compared to the mixed-model unadjusted estimator. 
This variance reduction is smaller than the variance reduction from cluster-level ANCOVA1, likely due to the small variation in cluster sizes (the coefficient of variation of cluster sizes is 0.16), thereby close to the setting of Proposition~\ref{coro3} with constant observed cluster sizes.
Compared to ANCOVA1 estimators, ANCOVA2 estimators have 19\% larger variance, likely due to the loss of degrees of freedom when estimating the interaction terms that may be absent from the true underlying data-generating distribution. 
%. This observation may reflect a lack of strong treatment effect heterogeneity between arms, which leads to inflated variance in estimating the average treatment effect via ANCOVA2 methods especially when the number of clusters is not large ($m=32$).
}

% In contrast, the mixed-model ANCOVA1 estimator only has a slightly larger variance estimate than the mixed-model unadjusted estimator, which may suggest a small efficiency loss due to covariate adjustment in the presence of a misspecified intracluster correlation structure.%which can come from small sample randomness or reflect a minor efficiency loss.

For the IECDZ study, the mixed-model unadjusted estimator has the smallest variance, which resembles findings from Scenario~1 of our first simulation study. Possible reasons for this efficiency loss due to covariate adjustment are that the covariates are not prognostic for the cluster-level mean outcome (no regression coefficient estimate for covariates is statistically significant under cluster-level ANCOVA1), and the cluster sizes are only moderately variable (coefficient of variation of cluster sizes is 0.41). {Like the TASSH study, methods based on cluster summaries are more precise than methods based on individual data, and ANCOVA2 methods are less precise than ANCOVA1 methods.}

% Compared to cluster-level ANCOVA1, mixed-model ANCOVA1 is 13\% less efficient for estimating the average treatment effect, which again might be attributed to its misspecification of the intracluster correlation structure. 
% the cluster-level ANCOVA1 estimator has 4\% smaller variance than the unadjusted estimator. The mixed-model ANCOVA1 estimator, however, has 6.8\% larger variance than the unadjusted estimator; such a precision loss may occur if mixed-model ANCOVA1 cannot reduce variance as discussed in Section~\ref{subsec: precision-gain}, or it may result from the uncertainty due to the limited number of clusters. 

In the analysis of WFHS, all {six} covariate-adjusted estimators have substantial precision gain compared to the mixed-model unadjusted estimator, likely because the baseline CWH is highly prognostic of the follow-up outcome. {In this example, we observe that methods based on individual data perform similarly and return a more precise average treatment effect estimator than methods based on cluster summaries. This might be because the mixed-model specifications are close to the true data-generating distribution, and the cluster sizes are highly variable (the coefficient of variation of cluster sizes is $0.59$). }

\section{Discussion}\label{sec:discussion}
{In this paper, we have revisited mixed-model ANCOVA methods to provide new results about their properties in the context of CRTs. Although the mixed-model ANCOVA has become a standard and commonly-used approach to analyze CRTs (especially in the health sciences studies), to date, there has been little formal investigation on its asymptotic properties for estimating the average treatment effect, defined under the potential outcomes framework, when model assumptions fail to hold. Our primary contribution here has been to establish the consistency and asymptotic normality of the mixed-model unadjusted, ANCOVA1, and ANCOVA2 estimators under arbitrary misspecification of their working models. Under equal randomization ($\pi=0.5$), we further showed that the model-based variance estimator from mixed-model ANCOVA1 remains consistent and henceforth the standard error estimate returned by current software routines yields asymptotically correct uncertainty statements, even if the working model is incorrect. However, the robust variance estimator is required for making inference on the average treatment effect when using mixed-model ANCOVA2. As the ANCOVA models are only working models, our results are applicable when the outcome $Y_{ij}$ is either continuous, binary, or count. These robustness properties are reassuring and serve to provide new justifications for conducting mixed-model ANCOVA analyses of CRTs by alleviating concerns about model misspecification.}

{Beyond the robustness properties, we additionally provide a comparison of estimation precision among the several classical methods in Table~\ref{tab:1} for analyzing CRTs. Under the assumption of equal observed cluster sizes, Figure~\ref{fig:summary-precision-comparison} gives the efficiency ordering among the considered covariate-adjusted estimators. 
% First, for each model, ANCOVA1 and ANCOVA2 specifications lead to asymptotically equivalent average treatment effect estimators when the randomization probability is $\pi=0.5$. Second, mixed models are asymptotically no more efficient than the individual-level ordinary least squares (linear regression with individual-level data), which are asymptotically no more efficient than cluster-level ordinary least squares (linear regression with cluster-level mean data). 
One takeaway message from Figure~\ref{fig:summary-precision-comparison} is that the cluster-level ANCOVA1 or ANCOVA2 estimators, whenever feasible, may be the preferred approach from an efficiency perspective if the cluster sizes are equal or approximately so. 
% However, as a caveat, when there is a limited number of clusters, there might be insufficient degrees of freedom at the cluster-level to adjust for a handful of baseline covariates via cluster-level ANCOVA methods. 
In the more general settings where cluster sizes are at least moderately variable, the precision comparisons among these methods are generally indeterminate, and mixed-model estimators can be more efficient than the other approaches, for example, when there is only mild model misspecification. Because unequal cluster size is the norm rather than the exception in CRTs \citep{eldridge2004lessons}, we believe that mixed-model ANCOVA estimators remain as useful for analyzing CRTs as individual-level ANCOVA or cluster-level ANCOVA estimators. 
For all considered methods, we list their potential advantages and disadvantages in Table~\ref{tab:discussion-features}.
% A brief summary of features of each method is provided in Table 5.
% Another potential advantage that mixed-model ANCOVA brings over other methods in Table \ref{tab:1} is that it directly offers a model-based estimate the intracluster correlation coefficient, which is an important quantity that may help inform the design of future studies. In addition, it has been well-established that the robust variance estimators (that are required under individual-level ANCOVA methods) can be biased toward zero if the number of clusters is small \citep{li2015small}. Since the model-based variance estimator is already consistent under mixed-model ANCOVA1, this approach bypasses the need to bias-correct the robust variance estimator for analyzing small CRTs and is expected to provide better control of the type I error rate. 
In practice, we recommend pre-specifying and adjusting for baseline covariates that are anticipated to be prognostic of the outcomes, whichever ANCOVA estimator is used.% we find an interesting caveat that covariate adjustment via mixed-model ANCOVA1 or ANCOVA2 does not always guarantee a more precise average treatment effect estimator, which are in sharp contrast to lessons learned from individually-randomized trials. This suggests that, for practical consideration, one should continue to pre-specify and adjust for baseline covariates that are anticipated to be prognostic of the outcomes rather than any covariates that happen to be measured. A brief summary of features of each method is provided in Table 5.
}

\begin{table}[htbp]
\centering
\caption{Potential advantages and disadvantages of several commonly used methods for estimating the average treatment effect in CRTs. All approaches lead to consistent point estimators for $\Delta^*$ under arbitrary model misspecification; this common advantage is not listed for brevity.}\label{tab:discussion-features}
\footnotesize
\begin{tabular}{>{\centering\arraybackslash}m{2.2cm}m{6.5cm}m{6.5cm}}
\hline
\raggedright Methods &  Potential advantages & Potential disadvantages \\
 \hline
Mixed-model ANCOVA & 
\begin{itemize}[leftmargin=10pt]
\itemsep0em 
\item Model-based inference for mixed-model ANCOVA1 is asymptotically valid under arbitrary model misspecification with 1:1 randomization
\item An estimate of the intracluster correlation coefficient available as a by-product of model fitting
\item Flexible to adjust both individual-level and cluster-level covariates
\vspace{-0.2in}
\end{itemize} & 
\begin{itemize}[leftmargin=10pt]
\itemsep0em
\item Covariate adjustment or the random intercept may cause precision loss
\end{itemize}\\
\hline
Individual-level ANCOVA & 
\begin{itemize}[leftmargin=10pt]
\itemsep0em 
\item Flexible to adjust both individual-level and cluster-level covariates
\end{itemize}
& 
\begin{itemize}[leftmargin=10pt]
\itemsep0em 
\item Covariate adjustment may cause precision loss
\item The robust sandwich variance estimator required for valid inference may exhibit negative bias with a small number of clusters
\item No direct estimate of the intracluster correlation coefficient
\vspace{-0.2in}
\end{itemize}\\
\hline
Cluster-level ANCOVA & 
\begin{itemize}[leftmargin=10pt]
\itemsep0em 
\item Covariate adjustment guarantees no precision loss asymptotically
\item Most asymptotically efficient under equal observed cluster sizes
\item Model-based inference for cluster-level ANCOVA1 is asymptotically valid under arbitrary model misspecification with 1:1 randomization
\vspace{-0.2in}
\end{itemize}
& 
\begin{itemize}[leftmargin=10pt]
\itemsep0em 
\item Can only adjust for cluster-level covariates and cluster-level summary of individual-level covariates
\item Limited cluster-level degrees of freedom for covariate adjustment when the number of clusters is small
\item No direct estimate of the intracluster correlation coefficient
\vspace{-0.2in}
\end{itemize}\\
\hline
\end{tabular}
\end{table}

%\citealp{das1979asymptotic}; 
For mixed-model ANCOVA, an alternative approach for point estimation is through maximizing the restricted maximum likelihood (REML), which is known to reduce the bias of the variance component estimators. If the mixed model is correctly specified, the maximum likelihood estimator and REML estimator are asymptotically equivalent (\citealp[c.f. p.17 in][]{jiang2017asymptotic}), but differences may arise when the working model is misspecified. To provide a preliminary assessment of REML estimators, we have replicated the first simulation study and data applications using the REML estimator in the Supplementary Material. In our simulations, the REML estimator demonstrates a slighter better performance than the maximum likelihood estimator in terms of coverage when the number of clusters is small ($m=20$), whereas these two approaches present negligible differences with a larger number of clusters. In our data applications, the REML variance estimator tends to be smaller than the maximum likelihood variance estimator, especially when the covariates do not appear prognostic. The limited empirical evidence sheds light on the robustness of the REML estimator, although a formal proof of its robustness under arbitrary model misspecification is subject to additional research.

{Our super-population framework assumes a constant source population size, violations of which necessitate alternative definitions of estimands. As a possible extension, we can let $n_i$ denote the source population size of cluster~$i$, and modify Assumptions~1 and~2 to accommodate variable source population sizes (see, for example, \citealp{wang2022model} for such an extension). Under this setup, we can either apply equal weight to each cluster to obtain the cluster-average estimand, $E[\sum_{i=1}^{n_i}\{Y_{ij}(1)-Y_{ij}(0)\}/n_i]$, or apply equal weight to each individual to get the individual-average estimand, $E[\sum_{i=1}^{n_i}\{Y_{ij}(1)-Y_{ij}(0)\}]/E[n_i]$. Interpretation of these two estimands in general settings can be found in \cite{kahan2022estimands}. In the special case that the source population size $n_i$ is independent of the treatment effect contrasts (and assuming that the potential outcomes are marginally identically distributed), these two estimands can still be unified to $E[Y_{ij}(1)-Y_{ij}(0)]$, and our theoretical results remain valid. In cases where the source population size $n_i$ is correlated with the treatment effect contrasts, the cluster-average and individual-average estimands are generally different, and the mixed-model ANCOVA estimators may no longer be robust without further modifications; an example modification of the mixed-model estimators via the g-computation formula to achieve robust inference can be found in \citet{wang2022model}.
Furthermore, \citet{su2021model} provides a unified theory for individual-level ANCOVA and cluster-level ANCOVA estimators under informative source population size with $N_i = n_i$. A comprehensive comparison of all possible ANCOVA estimators in this more general setting will be pursued in future studies.}

{Our results rest on a key identification assumption of non-informative enrollment. When the enrollment procedure is informative, namely, $\bM_i$ is correlated with the outcomes, treatment, or covariates, the observed individuals may present selection bias, thereby causing different treatment effects between the enrolled population and the underlying source population. Therefore, the mixed-ANCOVA estimators may be biased for the average treatment effect of the source population \citep{li2021clarifying}. Handling informative enrollment is challenging and represents an important direction for future research. %but there are several ad hoc solutions, including principal stratification , focusing on inference for the enrolled population using the methods from \cite{su2021model}, or taking a step back by assuming non-informative sampling given an informative cluster size \citep{bugni2022inference}.}
Finally, our asymptotic scheme assumes the number of clusters goes to infinity while the population of interest for each cluster is upper-bounded. In some scenarios such as clinical settings, while the number of individuals per cluster may dominate the number of clusters, our asymptotic results can still be applicable by specifying a large enough upper bound for the cluster size. For establishing the asymptotic results assuming that both $n$ and $m$ approach infinity, additional regularity assumptions are needed on $\mathcal{P}^{(\bW)}$ to accommodate the diverging dimension of data for each cluster. For this purpose, \cite{su2021model} provided subtle conditions for the ordinary-least-squares estimators under a randomization-based inference framework; however, for mixed-model estimators, the analogous assumptions may be more complicated and we leave related development under a super-population framework for future research.}

% When the cluster sizes are informative, namely, $N_i$ is no longer independent of the outcomes, treatment or covariates \citep{seaman2014review}, the average treatment effect among the enrolled individuals is no longer representative of the whole source population of interest ($n$ individuals of each cluster). For example, consider two clusters with an equal population size but opposite cluster-specific treatment effects. If $N_i$ is larger in the cluster where the treatment is beneficial, then the average treatment effect across all enrolled individuals will be positive, even though the average treatment effect among the whole population of the two clusters remains null. Similarly, when selective recruitment of individuals leads to selection bias, \citet{li2021clarifying} has shown that covariate adjustment via mixed-model ANCOVA1 is often insufficient for unbiased estimation of the average treatment effect, and implies that valid estimation requires access to baseline information among the non-enrolled population. These practical challenges, in fact, often speak to the inherent limitation of the cluster randomization design, rather than the mixed-model ANCOVA1 estimator itself. Addressing informative cluster sizes and selection bias in CRTs is beyond the scope of this article, and will be pursued in a separate study. 

\section*{Acknowledgements}
Research in this article was partially supported by the Patient-Centered Outcomes Research Institute\textsuperscript{\textregistered} (PCORI\textsuperscript{\textregistered} Awards ME-2020C1-19220 to M.O.H. and ME-2020C3-21072, ME-2022C2-27676 to F.L). M.O.H. and F.L. are also funded by the United States National Institutes of Health (NIH), National Heart, Lung, and Blood Institute (grant number R01-HL16820). All statements in this report, including its findings and conclusions, are solely those of the authors and do not necessarily represent the views of the NIH or PCORI\textsuperscript{\textregistered} or its Board of Governors or Methodology Committee.
T.P.M. was funded by MRC grants MC\_UU\_00004/07 and MC\_UU\_00004/09. B.W. was funded by National Institute of Allergy and Infectious Diseases (NIAID) grants R01-AI148127, K99-AI173395.

{\small
\bibliographystyle{apalike}
\bibliography{references}

\begin{thebibliography}{}

\bibitem[Balzer et~al., 2016]{balzer2016targeted}
Balzer, L.~B., Petersen, M.~L., van~der Laan, M.~J., and Collaboration, S.
  (2016).
\newblock Targeted estimation and inference for the sample average treatment
  effect in trials with and without pair-matching.
\newblock {\em Statistics in Medicine}, 35(21):3717--3732.

\bibitem[Balzer et~al., 2023]{balzer2021two}
Balzer, L.~B., van~der Laan, M., Ayieko, J., Kamya, M., Chamie, G., Schwab, J.,
  Havlir, D.~V., and Petersen, M.~L. (2023).
\newblock Two-stage {TMLE} to reduce bias and improve efficiency in cluster
  randomized trials.
\newblock {\em Biostatistics}, 24(2):502--517.

\bibitem[Bugni et~al., 2023]{bugni2022inference}
Bugni, F., Canay, I., Shaikh, A., and Tabord-Meehan, M. (2023).
\newblock Inference for cluster randomized experiments with non-ignorable
  cluster sizes.
\newblock {\em arXiv preprint arXiv:2204.08356}.

\bibitem[Campbell et~al., 2004]{campbell2004consort}
Campbell, M.~K., Elbourne, D.~R., and Altman, D.~G. (2004).
\newblock Consort statement: extension to cluster randomised trials.
\newblock {\em BMJ}, 328(7441):702--708.

\bibitem[Campbell et~al., 2005]{campbell2005determinants}
Campbell, M.~K., Fayers, P.~M., and Grimshaw, J.~M. (2005).
\newblock Determinants of the intracluster correlation coefficient in cluster
  randomized trials: the case of implementation research.
\newblock {\em Clinical Trials}, 2(2):99--107.

\bibitem[Donner and Klar, 2000]{donner2000design}
Donner, A. and Klar, N. (2000).
\newblock {\em Design and Analysis of Cluster Randomization Trials in Health
  Research}.
\newblock London: Arnold.

\bibitem[Drikvandi et~al., 2017]{drikvandi2017diagnosing}
Drikvandi, R., Verbeke, G., and Molenberghs, G. (2017).
\newblock Diagnosing misspecification of the random-effects distribution in
  mixed models.
\newblock {\em Biometrics}, 73(1):63--71.

\bibitem[Eldridge et~al., 2004]{eldridge2004lessons}
Eldridge, S.~M., Ashby, D., Feder, G.~S., Rudnicka, A.~R., and Ukoumunne, O.~C.
  (2004).
\newblock Lessons for cluster randomized trials in the twenty-first century: a
  systematic review of trials in primary care.
\newblock {\em Clinical Trials}, 1(1):80--90.

\bibitem[Eldridge et~al., 2006]{eldridge2006sample}
Eldridge, S.~M., Ashby, D., and Kerry, S. (2006).
\newblock Sample size for cluster randomized trials: effect of coefficient of
  variation of cluster size and analysis method.
\newblock {\em International Journal of Epidemiology}, 35(5):1292--1300.

\bibitem[Eldridge et~al., 2009]{eldridge2009intra}
Eldridge, S.~M., Ukoumunne, O.~C., and Carlin, J.~B. (2009).
\newblock The intra-cluster correlation coefficient in cluster randomized
  trials: a review of definitions.
\newblock {\em International Statistical Review}, 77(3):378--394.

\bibitem[Fiero et~al., 2016]{fiero2016statistical}
Fiero, M.~H., Huang, S., Oren, E., and Bell, M.~L. (2016).
\newblock Statistical analysis and handling of missing data in cluster
  randomized trials: a systematic review.
\newblock {\em Trials}, 17(1):1--10.

\bibitem[Green and Vavreck, 2008]{green2008analysis}
Green, D.~P. and Vavreck, L. (2008).
\newblock Analysis of cluster-randomized experiments: A comparison of
  alternative estimation approaches.
\newblock {\em Political Analysis}, 16(2):138--152.

\bibitem[Hayes and Moulton, 2009]{hayes2009cluster}
Hayes, R.~J. and Moulton, L.~H. (2009).
\newblock {\em {Cluster Randomised Trials}}.
\newblock Taylor \& Francis Group, LLC, Boca Raton, FL.

\bibitem[Imai et~al., 2009]{imai2009essential}
Imai, K., King, G., and Nall, C. (2009).
\newblock The essential role of pair matching in cluster-randomized
  experiments, with application to the {M}exican universal health insurance
  evaluation.
\newblock {\em Statistical Science}, 24(1):29--53.

\bibitem[Ivers et~al., 2012]{ivers2012allocation}
Ivers, N.~M., Halperin, I.~J., Barnsley, J., Grimshaw, J.~M., Shah, B.~R., Tu,
  K., Upshur, R., and Zwarenstein, M. (2012).
\newblock Allocation techniques for balance at baseline in cluster randomized
  trials: a methodological review.
\newblock {\em Trials}, 13(1):1--9.

\bibitem[Jiang, 2010]{jiang2010large}
Jiang, J. (2010).
\newblock {\em Large sample techniques for statistics}.
\newblock Springer.

\bibitem[Jiang, 2017]{jiang2017asymptotic}
Jiang, J. (2017).
\newblock {\em Asymptotic Analysis of Mixed Effects Models: Theory,
  Applications, and Open Problems}.
\newblock CRC Press.

\bibitem[Kahan et~al., 2023a]{kahan2023informative}
Kahan, B.~C., Li, F., Blette, B., Jairath, V., Copas, A., and Harhay, M.
  (2023a).
\newblock Informative cluster size in cluster-randomised trials: A case study
  from the trigger trial.
\newblock {\em Clinical Trials}.

\bibitem[Kahan et~al., 2023b]{kahan2022estimands}
Kahan, B.~C., Li, F., Copas, A.~J., and Harhay, M.~O. (2023b).
\newblock Estimands in cluster-randomized trials: choosing analyses that answer
  the right question.
\newblock {\em International Journal of Epidemiology}, 52(1):107--118.

\bibitem[Kelly et~al., 2014]{kelly2014changing}
Kelly, E.~L., Moen, P., Oakes, J.~M., Fan, W., Okechukwu, C., Davis, K.~D.,
  Hammer, L.~B., Kossek, E.~E., King, R.~B., Hanson, G.~C., et~al. (2014).
\newblock Changing work and work-family conflict: Evidence from the work,
  family, and health network.
\newblock {\em American Sociological Review}, 79(3):485--516.

\bibitem[Li et~al., 2016]{li2016evaluation}
Li, F., Lokhnygina, Y., Murray, D.~M., Heagerty, P.~J., and DeLong, E.~R.
  (2016).
\newblock An evaluation of constrained randomization for the design and
  analysis of group-randomized trials.
\newblock {\em Statistics in Medicine}, 35(10):1565--1579.

\bibitem[Li et~al., 2022]{li2021clarifying}
Li, F., Tian, Z., Bobb, J., Papadogeorgou, G., and Li, F. (2022).
\newblock Clarifying selection bias in cluster randomized trials.
\newblock {\em Clinical Trials}, 19(1):33--41.

\bibitem[Liang and Zeger, 1986]{liang1986longitudinal}
Liang, K.-Y. and Zeger, S.~L. (1986).
\newblock Longitudinal data analysis using generalized linear models.
\newblock {\em Biometrika}, 73(1):13--22.

\bibitem[Lin, 2013]{lin2013agnostic}
Lin, W. (2013).
\newblock Agnostic notes on regression adjustments to experimental data:
  Reexamining freedman’s critique.
\newblock {\em The Annals of Applied Statistics}, 7(1):295--318.

\bibitem[Liti{\`e}re et~al., 2007]{litiere2007type}
Liti{\`e}re, S., Alonso, A., and Molenberghs, G. (2007).
\newblock Type {I} and type {II} error under random-effects misspecification in
  generalized linear mixed models.
\newblock {\em Biometrics}, 63(4):1038--1044.

\bibitem[Liti{\`e}re et~al., 2008]{litiere2008impact}
Liti{\`e}re, S., Alonso, A., and Molenberghs, G. (2008).
\newblock The impact of a misspecified random-effects distribution on the
  estimation and the performance of inferential procedures in generalized
  linear mixed models.
\newblock {\em Statistics in medicine}, 27(16):3125--3144.

\bibitem[McCulloch and Neuhaus, 2011]{mcculloch2011misspecifying}
McCulloch, C.~E. and Neuhaus, J.~M. (2011).
\newblock Misspecifying the shape of a random effects distribution: why getting
  it wrong may not matter.
\newblock {\em Statistical Science}, 26(3):388--402.

\bibitem[McNeish and Stapleton, 2016]{mcneish2016modeling}
McNeish, D. and Stapleton, L.~M. (2016).
\newblock Modeling clustered data with very few clusters.
\newblock {\em Multivariate behavioral research}, 51(4):495--518.

\bibitem[Middleton and Aronow, 2015]{middleton2015unbiased}
Middleton, J.~A. and Aronow, P.~M. (2015).
\newblock Unbiased estimation of the average treatment effect in
  cluster-randomized experiments.
\newblock {\em Statistics, Politics and Policy}, 6(1-2):39--75.

\bibitem[Murray and Blitstein, 2003]{murray2003methods}
Murray, D.~M. and Blitstein, J.~L. (2003).
\newblock Methods to reduce the impact of intraclass correlation in
  group-randomized trials.
\newblock {\em Evaluation Review}, 27(1):79--103.

\bibitem[Murray et~al., 1998]{murray1998design}
Murray, D.~M. et~al. (1998).
\newblock {\em Design and Analysis of Group-Randomized Trials}, volume~29.
\newblock Oxford University Press, USA.

\bibitem[Murray et~al., 2006]{murray2006comparison}
Murray, D.~M., Hannan, P.~J., Pals, S.~P., McCowen, R.~G., Baker, W.~L., and
  Blitstein, J.~L. (2006).
\newblock A comparison of permutation and mixed-model regression methods for
  the analysis of simulated data in the context of a group-randomized trial.
\newblock {\em Statistics in Medicine}, 25(3):375--388.

\bibitem[Neuhaus et~al., 2013]{neuhaus2013estimation}
Neuhaus, J.~M., McCulloch, C.~E., and Boylan, R. (2013).
\newblock Estimation of covariate effects in generalized linear mixed models
  with a misspecified distribution of random intercepts and slopes.
\newblock {\em Statistics in medicine}, 32(14):2419--2429.

\bibitem[Offorha et~al., 2022]{offorha2022statistical}
Offorha, B.~C., Walters, S.~J., and Jacques, R.~M. (2022).
\newblock Statistical analysis of publicly funded cluster randomised controlled
  trials: a review of the national institute for health research journals
  library.
\newblock {\em Trials}, 23(1):115.

\bibitem[Ogedegbe et~al., 2018]{ogedegbe2018health}
Ogedegbe, G., Plange-Rhule, J., Gyamfi, J., Chaplin, W., Ntim, M., Apusiga, K.,
  Iwelunmor, J., Awudzi, K.~Y., Quakyi, K.~N., Mogaverro, J., et~al. (2018).
\newblock Health insurance coverage with or without a nurse-led task shifting
  strategy for hypertension control: A pragmatic cluster randomized trial in
  {G}hana.
\newblock {\em PLoS Medicine}, 15(5):e1002561.

\bibitem[Park and Kang, 2023]{park2021assumption}
Park, C. and Kang, H. (2023).
\newblock Assumption-lean analysis of cluster randomized trials in infectious
  diseases for intent-to-treat effects and network effects.
\newblock {\em Journal of the American Statistical Association},
  118(542):1195--1206.

\bibitem[Prague et~al., 2016]{prague2016accounting}
Prague, M., Wang, R., Stephens, A., Tchetgen~Tchetgen, E., and DeGruttola, V.
  (2016).
\newblock Accounting for interactions and complex inter-subject dependency in
  estimating treatment effect in cluster-randomized trials with missing
  outcomes.
\newblock {\em Biometrics}, 72(4):1066--1077.

\bibitem[Raudenbush, 1997]{raudenbush1997statistical}
Raudenbush, S.~W. (1997).
\newblock Statistical analysis and optimal design for cluster randomized
  trials.
\newblock {\em Psychological Methods}, 2(2):173.

\bibitem[Rockers et~al., 2018]{rockers2018two}
Rockers, P.~C., Zanolini, A., Banda, B., Chipili, M.~M., Hughes, R.~C., Hamer,
  D.~H., and Fink, G. (2018).
\newblock Two-year impact of community-based health screening and parenting
  groups on child development in {Z}ambia: Follow-up to a cluster-randomized
  controlled trial.
\newblock {\em PLoS Medicine}, 15(4):e1002555.

\bibitem[Rubin, 1976]{rubin1976inference}
Rubin, D.~B. (1976).
\newblock Inference and missing data.
\newblock {\em Biometrika}, 63(3):581--592.

\bibitem[Schochet et~al., 2022]{schochet2021design}
Schochet, P.~Z., Pashley, N.~E., Miratrix, L.~W., and Kautz, T. (2022).
\newblock Design-based ratio estimators and central limit theorems for
  clustered, blocked {RCTs}.
\newblock {\em Journal of the American Statistical Association},
  117(540):2135--2146.

\bibitem[Seaman et~al., 2014]{seaman2014review}
Seaman, S., Pavlou, M., and Copas, A. (2014).
\newblock Review of methods for handling confounding by cluster and informative
  cluster size in clustered data.
\newblock {\em Statistics in Medicine}, 33(30):5371--5387.

\bibitem[Su and Ding, 2021]{su2021model}
Su, F. and Ding, P. (2021).
\newblock Model-assisted analyses of cluster-randomized experiments.
\newblock {\em Journal of the Royal Statistical Society, Series B},
  83(5):994--1015.

\bibitem[Sullivan et~al., 2018]{sullivan2018should}
Sullivan, T.~R., White, I.~R., Salter, A.~B., Ryan, P., and Lee, K.~J. (2018).
\newblock Should multiple imputation be the method of choice for handling
  missing data in randomized trials?
\newblock {\em Statistical Methods in Medical Research}, 27(9):2610--2626.

\bibitem[Tong et~al., 2023]{tong2023bayesian}
Tong, G., Li, F., Chen, X., Hirani, S.~P., Newman, S.~P., Wang, W., and Harhay,
  M.~O. (2023).
\newblock A bayesian approach for estimating the survivor average causal effect
  when outcomes are truncated by death in cluster-randomized trials.
\newblock {\em American Journal of Epidemiology}, 192(6):1006--1015.

\bibitem[Tsiatis, 2007]{tsiatis2007}
Tsiatis, A. (2007).
\newblock {\em Semiparametric Theory and Missing Data}.
\newblock Springer Science \& Business Media.

\bibitem[Tsiatis et~al., 2008]{Tsiatis2008}
Tsiatis, A., Davidian, M., Zhang, M., and Lu, X. (2008).
\newblock Covariate adjustment for two-sample treatment comparisons in
  randomized clinical trials: A principled yet flexible approach.
\newblock {\em Statistics in Medicine}, 27(23):4658--4677.

\bibitem[van~der Vaart, 1998]{vaart_1998}
van~der Vaart, A. (1998).
\newblock {\em Asymptotic Statistics}.
\newblock Cambridge Series in Statistical and Probabilistic Mathematics.
  Cambridge University Press.

\bibitem[Wang et~al., 2019]{wang2019analysis}
Wang, B., Ogburn, E.~L., and Rosenblum, M. (2019).
\newblock Analysis of covariance in randomized trials: More precision and valid
  confidence intervals, without model assumptions.
\newblock {\em Biometrics}, 75(4):1391--1400.

\bibitem[Wang et~al., 2023a]{wang2022model}
Wang, B., Park, C., Small, D.~S., and Li, F. (2023a).
\newblock Model-robust and efficient inference for cluster-randomized
  experiments.
\newblock {\em arXiv preprint arXiv:2210.07324}.

\bibitem[Wang et~al., 2023b]{wang2021model}
Wang, B., Susukida, R., Mojtabai, R., Amin-Esmaeili, M., and Rosenblum, M.
  (2023b).
\newblock Model-robust inference for clinical trials that improve precision by
  stratified randomization and covariate adjustment.
\newblock {\em Journal of the American Statistical Association},
  118(542):1152--1163.

\bibitem[{{Work, Family and Health Study (WFHS)}}, 2018]{WFHS}
{{Work, Family and Health Study (WFHS)}} (2018).
\newblock Work, family and health network.
\newblock {\em Inter-university Consortium for Political and Social Research
  [distributor]}.

\bibitem[Yang and Tsiatis, 2001]{YangTsiatis2001}
Yang, L. and Tsiatis, A. (2001).
\newblock Efficiency study of estimators for a treatment effect in a
  pretest-posttest trial.
\newblock {\em The American Statistician}, 55(4):314--321.

\bibitem[Zhang and Davidian, 2001]{zhang2001linear}
Zhang, D. and Davidian, M. (2001).
\newblock Linear mixed models with flexible distributions of random effects for
  longitudinal data.
\newblock {\em Biometrics}, 57(3):795--802.

\end{thebibliography}


\begin{thebibliography}{}

\bibitem[Su and Ding, 2021]{su2021model}
Su, F. and Ding, P. (2021).
\newblock Model-assisted analyses of cluster-randomized experiments.
\newblock {\em Journal of the Royal Statistical Society, Series B},
  83(5):994--1015.

\bibitem[Tsiatis, 2007]{tsiatis2007}
Tsiatis, A. (2007).
\newblock {\em Semiparametric Theory and Missing Data}.
\newblock Springer Science \& Business Media.

\bibitem[van~der Vaart, 1998]{vaart_1998}
van~der Vaart, A. (1998).
\newblock {\em Asymptotic Statistics}.
\newblock Cambridge Series in Statistical and Probabilistic Mathematics.
  Cambridge University Press.

\bibitem[Wang et~al., 2023]{wang2021model}
Wang, B., Susukida, R., Mojtabai, R., Amin-Esmaeili, M., and Rosenblum, M.
  (2023).
\newblock Model-robust inference for clinical trials that improve precision by
  stratified randomization and covariate adjustment.
\newblock {\em Journal of the American Statistical Association},
  118(542):1152--1163.

\end{thebibliography}
}

\end{document}

\def\spacingset#1{\renewcommand{\baselinestretch}%
{#1}\small\normalsize} \spacingset{1}
%\title{The mixed-model ANCOVA model for cluster-randomized clinical trials: robustness and efficiency}
\title{\bf Supplementary Material for\\
``On the  mixed-model analysis of covariance in cluster-randomized trials''
}
% \author{\small
% Bingkai Wang$^{1}$,
% Michael O. Harhay$^{2,3}$, Dylan S. Small$^1$, Tim P. Morris$^4$, and Fan Li$^5$
% \vspace{10pt}

% $^1$The Statistics and Data Science Department of the Wharton School, University of Pennsylvania, Philadelphia, PA, USA

% $^2$Clinical Trials Methods and Outcomes Lab, Palliative and Advanced Illness Research (PAIR) Center, Perelman School of Medicine, University of Pennsylvania,
% Philadelphia, PA, USA

% $^3$Department of Biostatistics, Epidemiology and Informatics, Perelman School of Medicine, University of Pennsylvania,
% Philadelphia, PA, USA

% $^4$MRC Clinical Trials Unit at UCL, London, UK

% $^5$Department of Biostatistics, Yale University School of Public Health, New Haven, CT, USA}
\author{\normalsize
Bingkai Wang$^{1}$, Michael O. Harhay$^{2,3}$, Jiaqi Tong$^{4,5}$, Dylan S. Small$^1$,  Tim P. Morris$^6$\\
\vspace{-0.1in}
and Fan Li$^{4,5}$
\vspace{10pt}

$^1$The Statistics and Data Science Department of the Wharton School, University of Pennsylvania, Philadelphia, PA, USA

$^2$Clinical Trials Methods and Outcomes Lab, Palliative and Advanced Illness Research (PAIR) Center, Perelman School of Medicine, University of Pennsylvania,
Philadelphia, PA, USA

$^3$Department of Biostatistics, Epidemiology and Informatics, Perelman School of Medicine, University of Pennsylvania,
Philadelphia, PA, USA

$^4$Department of Biostatistics, Yale School of Public Health, New Haven, CT, USA

$^5$Center for Methods in Implementation and Prevention Science, Yale School of Public Health, New Haven, CT, USA

$^6$MRC Clinical Trials Unit at UCL, London, UK}
% \def\spacingset#1{\renewcommand{\baselinestretch}%
% {#1}\small\normalsize} \spacingset{1}

% \date{\vspace{-5ex}}
\date{\vspace{-5ex}}

\maketitle

\renewcommand\thesection{\Alph{section}}
\setcounter{section}{0}

% \noindent%
% {\it Keywords: covariate adjustment, mixed model, stratified randomization}  
% \vfill
% t saw 
% \newpage
% \spacingset{1.5}
% \setcounter{page}{1}
\spacingset{1.5}

    Section \ref{sec:proof} proves Theorems 1-3 presented in the main paper. Section~\ref{sec:variance} provides the asymptotic variances for all considered estimators, and Section~\ref{sec: comparison} compares these estimators. Section~\ref{ssec: cluster-totals}
    further discuss the cluster-level ANCOVA2 estimators based on scaled cluster totals.
    Section \ref{sec:REML} compares ML and REML estimation by replicating the simulation study and data application.

\section{Proofs}\label{sec:proof}
\subsection{Proof of Theorem 1}
\begin{proof}
For each cluster $i$, let $j_{i,1}< \dots < j_{i,N_i}$ be the ordered list of indices such that the observed outcomes are $\bY_i^o = (Y_{i,j_{i,1}}, \dots, Y_{i, j_{i,N_i}})$.  We define $\mathbf{D}_{\bM_i} = [\boldsymbol{e}_{j_{i,1}} \,\,\, \boldsymbol{e}_{j_{i,2}} \,\,\, \dots \,\,\, \boldsymbol{e}_{j_{i,N_i}}] \in \mathbb{R}^{n\times N_i}$, where $\boldsymbol{e}_j \in \mathbb{R}^n$ has the $j$-th entry 1 and the rest 0. For $\mathbf{D}_{\bM_i}$, we use the subscript $\bM_i$ to indicate that it is a deterministic function of $\bM_i$.  Then we have $\bY_i^o = \mathbf{D}_{\bM_i}^\top \bY_i$, $\bfX_i^o = \mathbf{D}_{\bM_i}^\top \bfX_i$, $\bone_{N_i} = \mathbf{D}_{\bM_i}^\top \bone_n$ and $\mathbf{D}_{\bM_i}\bone_{N_i} = \bM_i$, where $\bfX_i = (\bX_{i1}, \dots, \bX_{in})^\top$.

The mixed-model ANCOVA1 for the population of interest can be re-written in matrix notation as:
\begin{align*}
    \bY_i = \beta_0 \bone_{n} + A_i \beta_A  \bone_{n} + \bfX_i\bbeta_{\bX} + \gamma_i\bone_{n} + \bepsilon_i, 
\end{align*}
where $\bY_i = (Y_{i1}, \dots, Y_{in})^\top$, $\bfX_i = (\bX_{i1}, \dots, \bX_{in})^\top 
\in \mathbb{R}^{n \times p}$, $\bepsilon_i = (\epsilon_{i1}, \dots, \epsilon_{in})^\top$. Marginalizing over the distribution of random effects, we have, given the mixed-model ANCOVA1
\begin{align*}
    \bY_i|(A_i,\bfX_i) \sim N(\bfQ_i \bbeta, \bfSigma),
\end{align*}
where $\mathbf{Q}_i = (\bone_{n}, A_i \bone_{n}, \bfX_i) \in \mathbb{R}^{n \times (p+2)}$, $\bbeta = (\beta_0, \beta_A, \bbeta_{\bX}^\top)^\top \in \mathbb{R}^{p+2}$ and $\bfSigma = \sigma^2 \bfI_{n} + \tau^2 \bone_{n} \bone_{n}^\top$. We denote the parameters in the above model as $\btheta = (\bbeta^\top, \sigma^2, \tau^2)^\top \in \mathbb{R}^{p+4}$. Then, for the observed outcome, we have $\bY_i^o|(A_i,\bfX_i^o,N_i) \sim N(\mathbf{D}_{\bM_i}^\top \bfQ_i \bbeta, \mathbf{D}_{\bM_i}^\top \bfSigma \mathbf{D}_{\bM_i})$ under the mixed-model ANCOVA1 working model and Assumption 2. We note that, although $\bM_i$ is not observed, $\mathbf{D}_{\bM_i}^\top \bfQ_i = (\bone_{N_i}, A_i \bone_{N_i}, \bfX_i^o)$ and $\mathbf{D}_{\bM_i}^\top \bfSigma \mathbf{D}_{\bM_i} = \sigma^2 \bfI_{N_i} + \tau^2 \bone_{N_i} \bone_{N_i}^\top$ are only functions of observed data $(A_i,\bfX_i^o,N_i)$, which allows the maximum likelihood estimator (MLE) to be well-defined.

% In dependently for each $i$, we define $(Y_{i,N_i+1}(a), \dots, Y_{i, N}(a))$ and $(\bX_{i,N_i+1}, \dots, \bX_{i, N})$ such that $(Y_{ij}(1), Y_{ij}(0), \bX_{ij}), j = N_i+1, \dots, N$ are independent, identically distributed samples from $P$, i.e., the distribution that $(Y_{ij}(1), Y_{ij}(0), \bX_{ij}), j = 1, \dots, N_i$ follow. We a

% Based on the complete data vector $(\bW_{i1},\dots, \bW_{in})$ and $(A_1,\dots,A_n)$,
% we define $\bY_i^*  = (Y_{i1}, \dots, Y_{in})^\top$, $\bfX_i^* = (\bX_{i1}, \dots, \bX_{in})^\top$, $\bfQ_i^* = (\bone_{n}, A_i \bone_{n}, \bfX_i^*)$ and $\bfSigma^* = \sigma^2 \bfI_{n} + \tau^2 \bone_{n} \bone_{n}^\top$. Letting $\mathbf{D}_{\bM_i} =   (\bfI_{N_i},\bzero_{N_i, n-N_i} )^\top \in \mathbb{R}^{N\times N_i}$, we have $\bY_i = \mathbf{D}_{\bM_i}^\top \bY_i^*$, $\bfX_i = \mathbf{D}_{\bM_i}^\top \bfX_i^*$, $\bfQ_i = \mathbf{D}_{\bM_i}^\top\bfQ_i^*$ and $\bfSigma_i = \mathbf{D}_{\bM_i}^\top\bfSigma^* \mathbf{D}_{\bM_i}$.
Based on the observed data, the log-likelihood function conditioning on $\{A_i, \bfX_i, N_i\}$ is defined as 
\begin{align*}
    &l(\btheta; \{\bY_i^o\}_{i=1}^m|\{A_i, \bfX_i, N_i\}_{i=1}^m)\\
    &=  C -\frac{1}{2}\sum_{i=1}^m\left\{\log(|\mathbf{D}_{\bM_i}^\top\bfSigma \mathbf{D}_{\bM_i}|) + (\bY_i^o - \mathbf{D}_{\bM_i}^\top\bfQ_i\bbeta)^\top  (\mathbf{D}_{\bM_i}^\top\bfSigma \mathbf{D}_{\bM_i})^{-1} (\bY_i^o - \mathbf{D}_{\bM_i}^\top\bfQ_i\bbeta) \right\} \\
    &=  C -\frac{1}{2}\sum_{i=1}^m\left\{\log(|\mathbf{D}_{\bM_i}^\top\bfSigma \mathbf{D}_{\bM_i}|) + (\bY_i - \bfQ_i\bbeta)^\top \mathbf{D}_{\bM_i} (\mathbf{D}_{\bM_i}^\top\bfSigma \mathbf{D}_{\bM_i})^{-1}\mathbf{D}_{\bM_i}^\top (\bY_i - \bfQ_i\bbeta) \right\} 
\end{align*}
where $C$ is a constant independent of the parameters $\btheta$. The derivative of the log-likelihood function is then
\begin{align*}
   & \frac{\partial l(\btheta; \{\bY_i\}_{i=1}^m|\{A_i, \bfX_i, N_i\}_{i=1}^m)}{\partial \btheta}&= -\sum_{i=1}^m \left(\begin{array}{c}
    \bfQ_i^{\top}\bfV_i(\bY_i - \bfQ_i\bbeta)  \\
    -tr(\bfV_i) + (\bY_i - \bfQ_i\bbeta)^\top \bfV_i^2(\bY_i - \bfQ_i\bbeta)\\
  -\bone_n^\top\bfV_i\bone_n + (\bY_i - \bfQ_i\bbeta)^\top \bfV_i\bone_n\bone_n^\top \bfV_i(\bY_i - \bfQ_i\bbeta)    
   \end{array}\right),
\end{align*}
where $\bfV_i = \mathbf{D}_{\bM_i} (\mathbf{D}_{\bM_i}^\top\bfSigma \mathbf{D}_{\bM_i})^{-1} \mathbf{D}_{\bM_i}^\top \in \mathbb{R}^{n\times n}$ and $tr(\bfV_i)$ is the trace of $\bfV_i$. 
We hence define the estimating function as
\begin{equation}\label{eq:psi}
  \bpsi(\bY, A, \bfX, \bM; \btheta) =   \left(\begin{array}{c}
    \bfQ^{\top}\bfV(\bY - \bfQ\bbeta)  \\
    -tr(\bfV) + (\bY - \bfQ\bbeta)^\top \bfV^2(\bY - \bfQ\bbeta)\\
  -\bone_n^\top\bfV\bone_n + (\bY - \bfQ\bbeta)^\top \bfV\bone_n\bone_n^\top \bfV(\bY - \bfQ\bbeta)    
   \end{array}\right).
\end{equation}
The MLE for $\btheta$ is define as a solution to the estimating equation
$$\sum_{i=1}^n \bpsi(\bY_i, A_i, \bfX_i, \bM_i;\btheta)=\bzero.$$ 

We next prove the consistency of $\widehat{\beta}_A$ to $\Delta^*$, under arbitrary misspecification of its working model. By Assumption 1, $\Delta^*=E\{Y_{ij}(1)\}-E\{Y_{ij}(1)\} = E[Y(1)]-E[Y(0)]$. Given the regularity conditions, similar to Example 19.8 of \cite{vaart_1998}, the estimating function $\bpsi(\bY, A, \bfX, \bM;\btheta)$ is Glivenko-Cantelli, and, hence, by Theorem 5.9 of \cite{vaart_1998}, $\widehat{\btheta} \xrightarrow{P} \underline\btheta$, where $\underline\btheta = (\underline\beta_0, \underline\beta_A, \underline\bbeta_{\bX}^{\top}, \underline\sigma^2, \underline\tau^2)^\top$ solves $E[\bpsi(\bY, A, \bfX, \bM;\btheta)] = \bzero$. To compute $\underline\beta_A$, by Assumption 2,  $E[\bpsi(\bY, A, \bfX, \bM;\underline\btheta)] = \bzero$ implies
\begin{align*}
    & \bone_n^\top E[\underline\bfV] E[\bY - \underline\beta_0 \bone_n - A \underline\beta_A \bone_n - \bfX \underline\bbeta_{\bX} ] = 0, \\
    & \bone_n^\top E[\underline\bfV] E[A(\bY - \underline\beta_0 \bone_n - A \underline\beta_A \bone_n - \bfX \underline\bbeta_{\bX})] = 0, 
\end{align*}
where $\underline\bfV = \mathbf{D}_{\bM} (\mathbf{D}_{\bM}^\top\underline\bfSigma \mathbf{D}_{\bM})^{-1} \mathbf{D}_{\bM}^\top$ with  $\underline\bfSigma = \underline\sigma^2 \bfI_n + \underline\tau^2 \bone_n\bone_n^\top$.
The above two equations imply that
\begin{displaymath}
\bone_n^\top E[\underline\bfV]E[\bY(1) - \bY(0) - \underline\beta_A \bone_n] = 0.
\end{displaymath}
By Assumption 1, we have $E[\bY(a)] = \bone_n E[Y(a)]$ for $a = 0,1$. Then
\begin{displaymath}
\bone_n^\top E[\underline\bfV] \bone_n E[Y(1) - Y(0) - \underline\beta_A ] = 0.
\end{displaymath}
To show $\underline\beta_A = E[Y(1)] - E[Y(0)]$, it suffices to prove $E[\bone_n^\top \underline\bfV \bone_n^\top]> 0$.
Direct algebra gives that
\begin{align*}
    \bfV = \bfD_{\bM}\left( \frac{1}{\sigma^2} \bfI_N - \frac{\tau^2}{\sigma^2(\sigma^2 + N \tau^2)} \bone_N\bone_N^\top \right)\bfD_{\bM}^\top,
\end{align*}
which implies that $\bone_n^\top \underline\bfV \bone_n^\top = \frac{N}{\underline{\sigma}^2 + N \underline{\tau}^2}$. Since $N \ge 2$, 
$\bone_n^\top \underline\bfV \bone_n^\top> 0$ as long as $\underline\sigma^2 >0$. This is implied by the regularity conditions (1) and (2), which completes the proof of consistency.

We next prove the asymptotic normality. By the regularity conditions, Theorem 5.41 of \cite{vaart_1998} implies that
\begin{align*}
    \sqrt{m}(\widehat{\btheta}-\underline\btheta) = \frac{1}{\sqrt{m}}\sum_{i=1}^m\bfB^{-1} \bpsi(\bY_i, A_i, \bfX_i, \bM_i;\underline\btheta) + o_p(\bone),
\end{align*}
where $\bfB = E\left[\frac{\partial }{\partial \btheta} \bpsi(\bY, A, \bfX, \bM; \btheta) \big |_{\btheta = \underline\btheta}\right]$.  
Then, by computing $\bfB^{-1}$, we get
\begin{align*}
    \sqrt{m}(\widehat{\Delta}_1-\underline\Delta) = \frac{1}{\sqrt{m}}\sum_{i=1}^mIF(\bY_i, A_i, \bfX_i, \bM_i;\underline\btheta) + o_p(\bone),
\end{align*}
where
\begin{equation}\label{eq:if}
    IF(\bY, A, \bfX, \bM;\underline\btheta) = \frac{A-\pi}{\pi(1-\pi)\bone_n^\top E[\underline\bfV] \bone_n}\bone_n^\top \underline\bfV (\bY - \bfQ \underline\bbeta)
\end{equation}
is the influence function for $\widehat\Delta$, which is also the second component of $\bfB^{-1} \bpsi(\bY, A, \bfX, \bM;\btheta)$. Assumptions 1-2 in the main manuscript imply that $IF(\bY_i, A_i, \bfX_i, \bM_i;\underline\btheta), i =1,\dots, m$ are independent and identically distributed. The regularity conditions (3) and (4) imply that the influence function has bounded second moments. Hence, by the Central Limit Theorem, we have $\sqrt{m}(\widehat{\Delta}_1-\underline\Delta) \xrightarrow{d } N(0, v_1)$ with $v_1 = E[IF^2(\bY, A, \bfX, \bM;\underline\btheta)]$.

The model-robust variance estimator $m \widehat{Var}_{\textrm{r}}(\widehat{\Delta}_1)$ is the second-row, second-column entry of $$\widehat{\bfB}^{-1} m^{-1}\sum_{i=1}^m \bpsi(\bY_i, A_i, \bfX_i, \bM_i; \widehat\btheta)\bpsi(\bY_i, A_i, \bfX_i, \bM_i; \widehat\btheta)^\top \widehat{\bfB}^{-1},$$
where $\widehat{\bfB}^{-1} = m^{-1}\sum_{i=1}^m \frac{\partial }{\partial \btheta} \bpsi(\bY_i, A_i, \bfX_i, \bM_i; \btheta) \big |_{\btheta = \widehat\btheta}$. To prove the consistency of the model-robust variance estimator, we first show $\widehat{\bfB}^{-1} = \bfB + o_p(1)$. Denoting $\dot{\bpsi}_{ij}(\widehat\btheta)$ as the transpose of the $j$th row of $\frac{\partial \bpsi(\bY_i, A_i, \bfX_i, \bM_i;\btheta)}{\partial \btheta}\big|_{\btheta = \widehat\btheta}$, we apply the multivariate Taylor expansion to get
\begin{align*}
    m^{-1} \sum_{i=1}^m \dot{\bpsi}_{ij}(\widehat\btheta) -     m^{-1} \sum_{i=1}^m \dot{\bpsi}_{ij}(\underline\btheta) =  m^{-1} \sum_{i=1}^m \ddot{\bpsi}_{ij}(\widetilde\btheta) (\widehat\btheta-\underline\btheta)
\end{align*}
for some $\widetilde{\btheta}$ on the line segment between $\widehat{\btheta}$ and $\underline\btheta$ and $ \ddot{\bpsi}_{ij}$ being the derivative of $\dot{\bpsi}_{ij}$. By the regularity condition 5 and $\widehat{\btheta} \xrightarrow{P} \btheta$, we have $m^{-1} \sum_{i=1}^m \ddot{\psi}_{ij}(\widetilde\btheta)(\widehat\btheta-\underline\btheta) = O_p(1)o_p(1) = o_p(1)$. Therefore, $\widehat{\bfB}^{-1} = \bfB + o_p(1)$. Following a similar proof, we can obtain that 
\begin{align*}
   & m^{-1}\sum_{i=1}^m \bpsi(\bY_i, A_i, \bfX_i, \bM_i; \widehat\btheta)\bpsi(\bY_i, A_i, \bfX_i, \bM_i; \widehat\btheta)^\top\\
   &= E[\bpsi(\bY, A, \bfX, \bM; \underline\btheta)\bpsi(\bY, A, \bfX, \bM; \underline\btheta)^\top] + o_p(1).
\end{align*}
By continuous mapping theorem, we have
\begin{align*}
    & \widehat{\bfB}^{-1} m^{-1}\sum_{i=1}^m \bpsi(\bY_i, A_i, \bfX_i, \bM_i; \widehat\btheta)\bpsi(\bY_i, A_i, \bfX_i, \bM_i; \widehat\btheta)^\top \widehat{\bfB}^{-1} \\
    &= \bfB^{-1} E[\bpsi(\bY, A, \bfX, \bM; \underline\btheta)\bpsi(\bY, A, \bfX, \bM; \underline\btheta)^\top] \bfB^{-1} + o_p(1),
\end{align*}
which implies the consistency of the model-robust variance estimator.

We next prove the validity of the model-based variance estimator under $\pi = 0.5$.
To simplify the asymptotic variance $v$, we observe that, by the last equation of  \\ $E[\bpsi(\bY, A, \bfX, \bM;
\underline\btheta)] = \bzero$, we have
\begin{equation}\label{eq: last-est-eq}
 E[\bone_n^\top \underline\bfV (\bY - \bfQ \underline\bbeta)(\bY - \bfQ \underline\bbeta)^\top\underline\bfV\bone_n] = \bone_n^\top E[\underline\bfV] \bone_n.
\end{equation}
Hence
\begin{align*}
    v_1 &= \frac{E[(A-\pi)^2\{\bone_n^\top \underline\bfV (\bY - \bfQ \underline\bbeta)\}^2]}{\pi^2(1-\pi)^2(\bone_n^\top E[\underline\bfV] \bone_n)^2} \\
    &= \frac{1}{(1-\pi)^2\bone_n^\top E[\underline\bfV] \bone_n} + \frac{(1-2\pi)E[A\{\bone_n^\top \underline\bfV (\bY - \bfQ \underline\bbeta)\}^2]}{\pi^2(1-\pi)^2(\bone_n^\top E[\underline\bfV] \bone_n)^2}.
\end{align*}
If $\pi = 0.5$, then the asymptotic variance simply reduces to $4(\bone_n^\top E[\underline\bfV] \bone_n)^{-1}$. 

Recall the model-based variance estimator for $\widehat{\bbeta}$ is 
\begin{displaymath}
 \widehat{Var}(\widehat{\bbeta}) = \left(\sum_{i=1}^m \bfQ_i^{o\top} \widehat{\bfSigma}_i^{-1} \bfQ_i^o\right)^{-1},
\end{displaymath}
where $\mathbf{Q}_i^o = \mathbf{D}_{\bM_i}^\top \bfQ_i = (\bone_{N_i}, A_i \bone_{N_i}, \bfX_i^o)$ and $\widehat{\bfSigma}_i = \frac{m}{m-p-2} (\widehat{\sigma}^2 \bfI_{N_i} + \widehat{\tau}^2 \bone_{N_i}\bone_{N_i}^\top)$ with $\widehat{\sigma}^2, \widehat{\tau}^2$ being the MLE for variance components paramaters, $\sigma^2, \tau^2$, in the mixed-model ANCOVA, respectively. We next show that $m\widehat{Var}(\widehat{\bbeta}) \xrightarrow{P} (E[\bfQ^{\top}  \underline\bfV \bfQ])^{-1}$. By the Woodbury matrix identity, we have
\begin{displaymath}
 \widehat{\bfSigma}_i^{-1} = \frac{m-p-2}{m} \widehat{\sigma}^{-2}\left(\bfI_{N_i} - \frac{\widehat{\tau}^{2} }{\widehat{\sigma}^{2}+ N_i\widehat{\tau}^{2} }\bone_{N_i}\bone_{N_i}^\top\right).
\end{displaymath}
Using the formula of $\underline\bfV$ and the result that $\widehat{\sigma}^2 = \underline{\sigma}^2 + o_p(1)$ and $\widehat{\tau}^2 = \underline{\tau}^2 + o_p(1)$ (as implied by $\widehat{\btheta} \xrightarrow{P} \underline\btheta$), we have
\begin{align*}
 \frac{1}{m}\sum_{i=1}^m \bfQ_i^{o\top} \widehat{\bfSigma}_i^{-1} \bfQ_i^o &=   \widehat{\sigma}^{-2}\frac{m-p-2}{m^2}\sum_{i=1}^m\left(\bfQ_i^{o\top}\bfQ_i^o - \frac{\widehat{\tau}^{2} }{\widehat{\sigma}^{2}+ N_i\widehat{\tau}^{2} }\bfQ_i^{o\top}\bone_{N_i}\bone_{N_i}^\top\bfQ_i^o\right) \\
  &= \widehat{\sigma}^{-2}\frac{m-p-2}{m^2}\sum_{i=1}^m\left(\bfQ_i^{o\top}\bfQ_i^o - \frac{\underline\tau^{2}}{\underline\sigma^{2} + N_i\underline\tau^{2}}\bfQ_i^{o\top}\bone_{N_i}\bone_{N_i}^\top\bfQ_i^o\right) + \mathbf{r} \\
  &= \frac{\underline{\sigma}^{2}}{\widehat\sigma^2} \frac{m-p-2}{m^2}\sum_{i=1}^m\bfQ_i^\top\bfD_{\bM_i}\left( \frac{1}{\underline\sigma^2} \bfI_{N_i} - \frac{\underline\tau^{2}}{\underline\sigma^2(\underline\sigma^{2} + N_i\underline\tau^{2})} \bone_{N_i}\bone_{N_i}^\top \right)\bfD_{\bM_i}^\top\bfQ_i + \mathbf{r} \\
  &= \frac{\underline{\sigma}^{2}}{\widehat\sigma^2}  \frac{m-p-2}{m^2}\sum_{i=1}^m \bfQ_i^\top \underline{\bfV}_i^{-1} \bfQ_i +  \mathbf{r} \\
  &= (1+o_p(1))(E[\bfQ^{\top}  \underline\bfV \bfQ] + o_p(1)) + \mathbf{r}
\end{align*}
where 
\begin{align*}
    \mathbf{r} =  \widehat{\sigma}^{-2}\frac{m-p-2}{m^2}\sum_{i=1}^m \frac{\widehat{\sigma}^{2}\underline{\tau}^{2} - \underline\sigma^{2}\widehat\tau^{2}}{(\widehat{\sigma}^{2} + N_i\widehat{\tau}^{2})(\underline\sigma^{2} + N_i\underline\tau^{2})}\bfQ_i^{o\top}\bone_{N_i}\bone_{N_i}^{\top}\bfQ_i^o.
\end{align*}
If we can show that $\mathbf{r} = o_p(1)$, then by the Continuous Mapping Theorem, we get $m\widehat{Var}(\widehat{\bbeta}) \xrightarrow{P} (E[\bfQ^{\top}  \underline\bfV \bfQ])^{-1}$. To show $\mathbf{r} = o_p(1)$, by Assumption 1, regularity condition (2) and the fact that $\underline\sigma^2 >0$, we get
\begin{align*}
    ||\mathbf{r}|| &\le \widehat{\sigma}^{-2}\frac{m-p-2}{m^2}\sum_{i=1}^m \frac{|\widehat{\sigma}^{2}\underline{\tau}^{2} - \underline\sigma^{2}\widehat\tau^{2}|}{(\widehat{\sigma}^{2} + N_i\widehat{\tau}^{2})(\underline\sigma^{2} + N_i\underline\tau^{2})}||\bfQ_i^{o\top}\bone_{N_i}\bone_{N_i}^\top\bfQ_i^o|| \\
    &\le \frac{|\widehat{\sigma}^{2}\underline{\tau}^{2} - \underline\sigma^{2}\widehat\tau^{2}|}{\widehat\sigma^{4}\underline\sigma^{2}} \frac{m-p-2}{m}\frac{1}{m}\sum_{i=1}^m ||\bone_{N_i}^\top\bfQ_i^o||^2 \\
    &=\frac{o_p(1)}{(\underline{\sigma}^2 + o_p(1))^2\underline{\sigma}^2}O_p(1) \\
    &=o_p(1).
\end{align*}
Hence $m\widehat{Var}_{\textrm{m}}(\widehat{\Delta}_1) \xrightarrow{P} 1/\{\pi(1-\pi)\bone_n^\top E[\underline\bfV] \bone_n\}$, which is the second-row second-column entry of $(E[\bfQ^{\top } \underline\bfV \bfQ])^{-1}$. If $\pi = 0.5$, then $m\widehat{Var}_{\textrm{m}}(\widehat{\Delta}_1)  \xrightarrow{P} 4(\bone_n^\top E[\underline\bfV] \bone_n)^{-1}$, which happens to be the true asymptotic variance of $\widehat{\Delta}$ under arbitrary misspecification of the working ANCOVA model that generates $\widehat{\Delta}_1$.
\end{proof}

\subsection{Proof of Theorem 2}
\begin{proof} 
We inherit all notations from the proof of Theorem 1. The mixed-model ANCOVA2 for the population of interest can be re-written in matrix notation as:
\begin{align*}
    \bY_i = \beta_0 \bone_{n} + A_i \beta_A  \bone_{n} + (\bfX_i - \bone_n\overline{\bX}_{all}^o{}^\top)\bbeta_{\bX} + A_i(\bfX_i - \bone_n\overline{\bX}_{all}^o{}^\top)\bbeta_{A\bX}\gamma_i\bone_{n} + \bepsilon_i. 
\end{align*}
Then, the derivative of the log-likelihood function is 
\begin{align*}
    -\sum_{i=1}^m \left(\begin{array}{c}
    \widehat{\bfQ}_i^{\top}\bfV_i(\bY_i - \widehat{\bfQ}_i\bbeta)  \\
    -tr(\bfV_i) + (\bY_i - \widehat{\bfQ}_i\bbeta)^\top \bfV_i^2(\bY_i - \widehat{\bfQ}_i\bbeta)\\
  -\bone_n^\top\bfV_i\bone_n + (\bY_i - \widehat{\bfQ}_i\bbeta)^\top \bfV_i\bone_n\bone_n^\top \bfV_i(\bY_i - \widehat{\bfQ}_i\bbeta)    
   \end{array}\right),
\end{align*}
where $\widehat{\bfQ}_i = (\bone_n, A_i\bone_n,  \bfX_i - \bone_n\overline{\bX}_{all}^o{}^\top,  A_i(\bfX_i - \bone_n\overline{\bX}_{all}^o{}^\top))$. To get rid of $\overline{\bX}_{all}^o$ in the estimating equations, we define a parameter $\bmu_{\bX}$, and the above estimating equations are equivalent to
\begin{align*}
    -\sum_{i=1}^m \left(\begin{array}{c}
    \bfQ_i^{\top}\bfV_i(\bY_i - \bfQ_i\bbeta)  \\
    -tr(\bfV_i) + (\bY_i - \bfQ_i\bbeta)^\top \bfV_i^2(\bY_i - \bfQ_i\bbeta)\\
  -\bone_n^\top\bfV_i\bone_n + (\bY_i - \bfQ_i\bbeta)^\top \bfV_i\bone_n\bone_n^\top \bfV_i(\bY_i - \bfQ_i\bbeta)    \\
  N_i \bmu_{\bX} - \bfX_i^\top \bM_i
   \end{array}\right),
\end{align*}
where $\bfQ_i = (\bone_n, A_i\bone_n,  \bfX_i - \bone_n\bmu_{\bX}^\top,  A_i(\bfX_i - \bone_n\bmu_{\bX}^\top))$, since the last estimating equation $ \sum_{i=1}^m N_i \bmu_{\bX} - \bfX_i^\top \bM_i$ gives a solution $\widehat{\bmu}_{\bX} = \overline{\bX}_{all}^o$. Therefore, we can define the estimating function as
\begin{equation}\label{eq:psi2}
  \bpsi(\bY, A, \bfX, \bM; \btheta) =   \left(\begin{array}{c}
    \bfQ^{\top}\bfV(\bY - \bfQ\bbeta)  \\
    -tr(\bfV) + (\bY - \bfQ\bbeta)^\top \bfV^2(\bY - \bfQ\bbeta)\\
  -\bone_n^\top\bfV\bone_n + (\bY - \bfQ\bbeta)^\top \bfV\bone_n\bone_n^\top \bfV(\bY - \bfQ\bbeta)    \\
   N \bmu_{\bX} - \bfX^\top \bM
   \end{array}\right),
\end{equation}
where $ \btheta = (\beta_0,\beta_A, \beta_{\bX}^\top, \beta_{A\bX}^\top, \sigma^2, \tau^2, \bmu_{\bX}^\top)$. Following the same proof as for Theorem 1, we obtain $\btheta \xrightarrow{P} \underline{\btheta}$ and $\sqrt{m}(\widehat{\btheta}-\underline\btheta) = \frac{1}{\sqrt{m}}\sum_{i=1}^m\bfB^{-1} \bpsi(\bY_i, A_i, \bfX_i, \bM_i;\underline\btheta) + o_p(\bone),$ with $\underline{\btheta} = (\underline\beta_0,\underline\beta_A, \underline\beta_{\bX}^\top, \underline\beta_{A\bX}^\top, \underline\sigma^2, \underline\tau^2, \underline\bmu_{\bX}^\top)$ satisfying $E[\bpsi(\bY, A, \bfX, \bM; \underline\btheta)]=\bzero$, which implies the asymptotic normality. 

We next prove the consistency, i.e., $\underline\beta_A = \Delta^*$. The last entry of $E[\bpsi(\bY, A, \bfX, \bM; \underline\btheta)]=\bzero$ shows $E[N\underline\bmu_{\bX}] = E[\bfX^\top \bM] = E[N\bX]$, thereby $\underline\bmu_{\bX} = E[\bX]$. The first two entries of  $E[\bpsi(\bY, A, \bfX, \bM; \underline\btheta)]=\bzero$ imply that
\begin{align*}
    & \bone_n^\top E[\underline\bfV] E[\bY - \underline\beta_0 \bone_n - A \underline\beta_A \bone_n - (\bfX -\bone_n\underline\bmu_{\bX}^\top )\underline\bbeta_{\bX} - A(\bfX -\bone_n\underline\bmu_{\bX}^\top)\underline\bbeta_{A\bX}] = 0, \\
    & \bone_n^\top E[\underline\bfV] E[A(\bY - \underline\beta_0 \bone_n - A \underline\beta_A \bone_n - (\bfX -\bone_n\underline\bmu_{\bX}^\top)\underline\bbeta_{\bX} - A(\bfX -\bone_n\underline\bmu_{\bX}^\top )\underline\bbeta_{A\bX})] = 0.
\end{align*}
Since $E[\bfX] = \bone_n E[\bX]^\top$, we get $ \underline\beta_A = E[(A-\pi)Y]/\pi/(1-\pi) = \Delta^*$.

We next derive the influence function for $\widehat{\Delta}_2$. For $\bfB = E\left[\frac{\partial }{\partial \btheta} \bpsi(\bY, A, \bfX, \bM; \btheta) \big |_{\btheta = \underline\btheta}\right]$, we can partition it into $7\times 7$ block matrices, denoted by $\bfB_{lk}$ according to the seven parameters in $\btheta$. To facilitate the computation, we manipulate the order of parameters in $\btheta$ such that the first three are $\beta_0, \beta_A, \bmu_{\bX}$. Then $\bfB$ is a block-diagonal matrix partitioned by $\beta_0, \beta_A, \bmu_{\bX}$ and $\bbeta_{\bX}, \bbeta_{A\bX}, \sigma^2,\tau^2$ because $\bfB_{lk} = 0$ for $l=1,2,3$ and $k=4,5,6,7$, and $l=4,5,6,7$ and $k=1,2,3$. Since the parameter of interest is in the first block, we only need to compute
\begin{align*}
    -\left(\begin{array}{ccc}
    \bfB_{11}  & \bfB_{12} & \bfB_{13} \\
    \bfB_{21}  & \bfB_{22} & \bfB_{23} \\
    \bfB_{31}  & \bfB_{32} & \bfB_{33}
    \end{array}\right)^{-1} \left(\begin{array}{c}
         \bone_n^\top\bfV(\bY - \bfQ\bbeta)\\ 
         A\bone_n^\top\bfV(\bY - \bfQ\bbeta)\\ 
N \bmu_{\bX} - \bfX^\top \bM         
    \end{array}\right),
\end{align*}
where $\bfB_{11}=-\bone_n^\top E[\underline\bfV]\bone_n$, $\bfB_{21}=\bfB_{12}=\bfB_{22}=-\pi\bone_n^\top E[\underline\bfV]\bone_n$, $\bfB_{31} = \bfB_{32} = 0$, $\bfB_{13} = \bone_n^\top E[\underline\bfV]\bone_n (\bbeta_{\bX}+\pi \bbeta_{A\bX})^\top$, $\bfB_{23} = \pi\bone_n^\top E[\underline\bfV]\bone_n (\bbeta_{\bX}+ \bbeta_{A\bX})^\top$, and $\bfB_{33} = E[N]\bfI_p$. Direct algebra shows that its second entry is 
\begin{equation}\label{if:mixed-ANCOVA2}
 IF(\bY, A, \bfX, \bM;\underline\btheta) = \frac{A-\pi}{\pi(1-\pi)\bone_n^\top E[\underline\bfV] \bone_n}\bone_n^\top \underline\bfV (\bY - \bfQ \underline\bbeta) + \frac{1}{E[N]}\underline\bbeta_{A\bX}^\top (\bfX^\top \bM - NE[\bX]).       
\end{equation}
The consistency of the sandwich variance estimator is implied similarly.
\end{proof}

\subsection{Proof of Theorem 3}
By Assumption 1 (a-b), Assumption 2, and regularity conditions in the Appendix, Theorem 1 of \cite{wang2021model} implies the consistency and asymptotic normality of $\widehat{\Delta}_1$ and $\widehat{\Delta}_2$ under stratified randomization. Furthermore
\begin{align*}
    \widetilde{v}_1 &= v_1 - \frac{1}{\pi(1-\pi)}E[E\{(A-\pi)IF(\bY, A, \bfX,\bM;\underline{\btheta})|S\}^2],
\end{align*}
where $IF(\bY, A, \bfX,\bM;\underline{\btheta})$ is defined in Equation~(\ref{eq:if}) for the mixed-model ANCOVA1 estimator. Similarly, 
\begin{align*}
    \widetilde{v}_2 &= v_2 - \frac{1}{\pi(1-\pi)}E[E\{(A-\pi)IF(\bY, A, \bfX,\bM;\underline{\btheta})|S\}^2],
\end{align*}
where $IF(\bY, A, \bfX,\bM;\underline{\btheta})$ is defined in Equation~(\ref{eq:if}) for the mixed-model ANCOVA2 estimator.

If $S$ is adjusted in the mixed-model ANCOVA1 model, then  $E[\bpsi(\bY, A, \bfX, \bM; \underline\btheta)] = \bzero$ implies that
\begin{displaymath}
 E[\bone_n^\top \underline\bfV (\bY - \bfQ \underline\bbeta)|S] = 0.
\end{displaymath}
Hence
\begin{align*}
    \widetilde{v}_1 &= v_1 - \frac{1}{\pi(1-\pi)}E\left[E\left\{\frac{(A-\pi)^2}{\pi(1-\pi)\bone_n^\top E[\underline\bfV] \bone_n}\bone_n^\top \underline\bfV (\bY - \bfQ \underline\bbeta)\bigg|S\right\}^2\right] \\
    &= v_1 - \frac{(1-2\pi)^2}{\pi^3(1-\pi)^3(\bone_n^\top E[\underline\bfV] \bone_n)^2}E[E\{A\bone_n^\top \underline\bfV (\bY - \bfQ \underline\bbeta)|S\}^2],
\end{align*}
which implies $\widetilde{v}_1 = v_1$ if $\pi = 0.5$.

If $S$ is adjusted in the mixed-model ANCOVA2 model, then  $E[\bpsi(\bY, A, \bfX, \bM; \underline\btheta)] = \bzero$ implies that $ E[\bone_n^\top \underline\bfV (\bY - \bfQ \underline\bbeta)|S] = 0$ and $ E[A\bone_n^\top \underline\bfV (\bY - \bfQ \underline\bbeta)|S] = 0$. Then
\begin{align*}
    \widetilde{v}_2 &= v_2 - \frac{1}{\pi(1-\pi)}E\left[E\left\{\frac{(A-\pi)^2}{\pi(1-\pi)\bone_n^\top E[\underline\bfV] \bone_n}\bone_n^\top \underline\bfV (\bY - \bfQ \underline\bbeta)\bigg|S\right\}^2\right] \\
    &= v_2 - \frac{(1-2\pi)^2}{\pi^3(1-\pi)^3(\bone_n^\top E[\underline\bfV] \bone_n)^2}E[E\{A\bone_n^\top \underline\bfV (\bY - \bfQ \underline\bbeta)|S\}^2],\\
    &=v_2.
\end{align*}

\section{Asymptotic variances of considered estimators}\label{sec:variance}
\subsection{The mixed-model ANCOVA1 estimator}
We inherit the notation from the proof of Theorem 1.  We have shown that
\begin{align*}
    \bfV = \bfD_{\bM}\left( \frac{1}{\sigma^2} \bfI_N - \frac{\tau^2}{\sigma^2(\sigma^2 + N \tau^2)} \bone_N\bone_N^\top \right)\bfD_{\bM}^\top,
\end{align*}
which implies $\bfV \bone_n = \frac{1}{\sigma^2 + N \tau^2} \bM$ and $\bone_n^\top\bfV \bone_n = \frac{N}{\sigma^2 + N \tau^2} $. When $N_i = \widetilde{n}$, the influence function of the mixed-model ANCOVA estimator becomes 
\begin{align*}
    IF(\bY, A, \bfX, \bM;\underline\btheta) = \frac{A-\pi}{\pi(1-\pi) \widetilde{n}}\bM^\top (\bY - \bfQ \underline\bbeta).
\end{align*}
Letting $\bc_{\bX} = Var(\overline{\bX}^o)^{-1} Cov(\overline{\bX}^o, \overline{Y}^o(0))$ and $\bc_{A\bX} = Var(\overline{\bX}^o)^{-1} Cov(\overline{\bX}^o, \overline{Y}^o(1)-\overline{Y}^o(0))$, then we have
\begin{align*}
    v_1 &=  E\left[\left\{\frac{A-\pi}{\pi(1-\pi) \widetilde{n}}\bM^\top (\bY - \bfQ \underline\bbeta)\right\}^2\right] \\
    &=  \frac{1}{\pi} Var\{\overline{Y}^o(1)\} + \frac{1}{1-\pi} Var\{\overline{Y}^o(0)\} \\
    &\quad - \frac{1}{\pi(1-\pi)} \{\bc_{\bX} + (1-\pi)\bc_{A\bX}\}^\top Var(\overline{\bX}^o)\{\bc_{\bX} + (1-\pi)\bc_{A\bX}\} \\
    &\quad + \frac{1}{\pi(1-\pi)} \{\bc_{\bX} - \underline{\bbeta}_{\bX} + (1-\pi)\bc_{A\bX}\}^\top Var(\overline{\bX}^o)\{\bc_{\bX} - \underline{\bbeta}_{\bX} + (1-\pi)\bc_{A\bX}\}.
\end{align*}
Here
\begin{align*}
    \underline{\bbeta}_{\bX} &= E\left[(\bfX - E[\bfX])^\top \underline{\bfV}(\bfX - E[\bfX])\right]^{-1} E\left[(\bfX - E[\bfX])^\top \underline{\bfV}(\bY - E[\bY])\right] \\
    % &= \left\{\frac{\widetilde{n}}{\underline\sigma^2}Var(\bX) - \frac{\underline\tau^2}{\underline\sigma^2+\widetilde{n}\underline\tau^2 }Var(\bfX^o{}^\top\bone_{\widetilde{n}})\right\}^{-1}\left\{\frac{\widetilde{n}}{\underline\sigma^2}Cov(\bX, Y) - \frac{\underline\tau^2}{\underline\sigma^2+\widetilde{n}\underline\tau^2 }Cov(\bfX^o{}^\top\bone_{\widetilde{n}}, \bY^o{}^\top\bone_{\widetilde{n}})\right\} \\
    % &= \left\{\frac{\widetilde{n}}{\underline\sigma^2}Var(\bX) - \frac{\widetilde{n}^2\underline\tau^2}{\underline\sigma^2+\widetilde{n}\underline\tau^2 }Var(\overline\bX^o)\right\}^{-1}\left\{\frac{\widetilde{n}}{\underline\sigma^2}Cov(\bX, Y) - \frac{\widetilde{n}^2\underline\tau^2}{\underline\sigma^2+\widetilde{n}\underline\tau^2 }Cov(\overline{\bX}^o, \overline{Y}^o)\right\} \\
    &=  \left\{Var(\bX) - \frac{\widetilde{n}\underline\tau^2}{\underline\sigma^2+\widetilde{n}\underline\tau^2 }Var(\overline\bX^o)\right\}^{-1}\left\{Cov(\bX, Y) - \frac{\widetilde{n}\underline\tau^2}{\underline\sigma^2+\widetilde{n}\underline\tau^2 }Cov(\overline{\bX}^o, \overline{Y}^o)\right\}.
\end{align*}
To prove the above expression for $ \underline{\bbeta}_{\bX}$, we observe that
\begin{align*}
    & E\left[(\bfX - E[\bfX])^\top \underline{\bfV}(\bfX - E[\bfX])\right] \\
    &= E[\bfX^\top \underline{\bfV} \bfX] - E[\bfX]^\top E[\underline{\bfV}] E[\bfX] \\
    &= \frac{1}{\underline\sigma^2}E[\bfX^\top \bfD_{\bM} \bfD_{\bM}^\top \bfX] -   \frac{\underline\tau^2}{\underline\sigma^2(\underline\sigma^2 +  \widetilde{n} \underline\tau^2)} E[\bfX^\top\bfD_{\bM}\bone_{ \widetilde{n}}\bone_{ \widetilde{n}}^\top \bfD_{\bM}^\top\bfX] - E[\bX] \bone_n^\top E[\underline{\bfV}] \bone_n  E[\bX]^\top.
\end{align*}
Since $\bfD_{\bM} \bfD_{\bM}^\top = diag\{\bM\}$, then $\bfX^\top \bfD_{\bM} \bfD_{\bM}^\top \bfX = \sum_{j=1}^n M_{.j} \bX_{.j}\bX_{.j}^\top$. Therefore, $E[\bfX^\top \bfD_{\bM} \bfD_{\bM}^\top \bfX] = \sum_{j=1}^n E[M_{.j} \bX_{.j}\bX_{.j}^\top] =  E[\sum_{j=1}^n M_{.j}] E[\bX_{.j}\bX_{.j}^\top] = \widetilde{n} E[\bX\bX^\top]$. Since $\bfD_{\bM}\bone_N = \bM$, we also have $\bfX^\top\bfD_{\bM}\bone_N = \bfX^\top \bM = \widetilde{n} \overline{\bX}^o$. As a result, 
\begin{align*}
     & E\left[(\bfX - E[\bfX])^\top \underline{\bfV}(\bfX - E[\bfX])\right] \\
     &= \frac{\widetilde{n}}{\underline\sigma^2} E[\bX\bX^\top] -\frac{\widetilde{n}^2\underline\tau^2}{\underline\sigma^2(\underline\sigma^2 +  \widetilde{n} \underline\tau^2)} E[\overline{\bX}^o\overline{\bX}^o{}^\top] - \frac{\widetilde{n}}{\sigma^2 +  \widetilde{n} \underline\tau^2} E[\bX]E[\bX]^\top \\
     &= \frac{\widetilde{n}}{\underline\sigma^2} \{Var(\bX) +  E[\bX]E[\bX]^\top\}  - \frac{\widetilde{n}^2\underline\tau^2}{\underline\sigma^2(\underline\sigma^2 +  \widetilde{n} \underline\tau^2)}\{ Var(\overline{\bX}^o) +E[\overline{\bX}^o]E[\overline{\bX}^o]^\top\} - \frac{\widetilde{n}}{\underline\sigma^2 +  \widetilde{n} \underline\tau^2}E[\bX]E[\bX]^\top\\
     &= \frac{\widetilde{n}}{\underline\sigma^2}Var(\bX) - \frac{\widetilde{n}^2\underline\tau^2}{\underline\sigma^2(\underline\sigma^2 +  \widetilde{n} \underline\tau^2)}Var(\overline{\bX}^o),
\end{align*}
where the last equation results from $E[\overline{\bX}^o] = \widetilde{n}^{-1} E[\sum_{j=1}^n M_{.j} \bX_{.j}] = \widetilde{n}^{-1} E[\sum_{j=1}^n M_{.j}] E[\bX] = E[\bX]$. Following a similar proof, we have
\begin{align*}
    E\left[(\bfX - E[\bfX])^\top \underline{\bfV}(\bY - E[\bY])\right] =  \frac{\widetilde{n}}{\underline\sigma^2}Cov(\bX, Y) - \frac{\widetilde{n}\underline\tau^2}{\underline\sigma^2(\underline\sigma^2+\widetilde{n}\underline\tau^2) }Cov(\overline{\bX}^o, \overline{Y}^o),
\end{align*}
which implies the desired formula for $\underline{\bbeta}_{\bX}$.
Of note, $\bX \in \mathbb{R}^p$ represents a random variable that has the same distribution as $\bX_{ij}$ for any $i$ and $j$. This is different from $\bfX \in \mathbb{R}^{n\times p}$, which is the random variable for covariate matrix $\bX_i = (\bX_{i1}, \dots, \bX_{in})^\top$.

\subsection{The individual-level ANCOVA1 estimator}
For the individual-level ANCOVA1 estimator that does not model the random effect, the proof of Theorem 1 implies that its influence function is 
\begin{align*}
    \frac{A-\pi}{\pi(1-\pi) E[N]} \bM^\top (\bY-\bfQ \underline{\widetilde{\bbeta}}),
\end{align*}
where $\underline{\widetilde{\bbeta}} = E[\bfQ^\top diag\{M\}\bfQ]^{-1} E[\bfQ^\top diag\{M\} \bY]$, which implies $\underline{\widetilde{\bbeta}}_{\bX} = Var(\bX)^{-1}Cov(\bX,Y)$.
Following a similar proof as in Section B.1, its asymptotic variance is
\begin{align*}
    v_1^{\textup{(ols)}}&=\frac{1}{E[N]^2}\left[\frac{1}{\pi} Var(Y^+(1)) + \frac{1}{1-\pi} Var(Y^+(0))\right.\\
    &- \boldsymbol{q}^\top Var(\bX^+) \boldsymbol{q} + \left.\{\boldsymbol{q} -\frac{1}{\pi(1-\pi)}\underline{\widetilde{\bbeta}}_{\bX}\}^\top Var(\bX^+) \{\boldsymbol{q} -\frac{1}{\pi(1-\pi)}\underline{\widetilde{\bbeta}}_{\bX}\} \right],
\end{align*}
where $Y^+(a) = \bM^\top \bY(a) - NE[Y(a)]$ for $a =0,1$, $\bX^+ =\bM^\top \bfX - NE[\bX]$ and $\boldsymbol{q} = Var(\bX^+)^{-1} Cov\{\bX^+, \frac{1}{\pi}Y^+(1)+\frac{1}{1-\pi}Y^+(0)\} $.
Therefore, when $N_i = \widetilde{n}$,  $v_1^{\textup{(ols)}}$ has the same formula as $v_1$ if $\pi = 0.5$ except that
$\underline{\widetilde{\bbeta}}_{\bX} = Var(\bX)^{-1}Cov(\bX,Y)$.

\subsection{The mixed-model ANCOVA2 estimator}
When $N_i = \widetilde{n}$, the influence function of the mixed-model ANCOVA2 estimator becomes 
\begin{align*}
    IF(\bY, A, \bfX, \bM;\underline\btheta) = \frac{A-\pi}{\pi(1-\pi) \widetilde{n}}\bM^\top (\bY - \bfQ \underline\bbeta) + \frac{1}{\widetilde{n}}\underline\bbeta_{A\bX}^\top (\bfX^\top \bM - \widetilde{n}E[\bX]).   
\end{align*}
Then $ v_2 = E[IF(\bY, A, \bfX, \bM;\underline\btheta)^2] = (a) + (b) + (c)$, where
\begin{align*}
    (a) &=  E\left[\left\{\frac{A-\pi}{\pi(1-\pi) \widetilde{n}}\bM^\top (\bY - \bfQ \underline\bbeta)\right\}^2\right],\\
    (b) &= 2 E\left[\frac{A-\pi}{\pi(1-\pi) \widetilde{n}^2}\bM^\top (\bY - \bfQ \underline\bbeta)\bbeta_{A\bX}^\top (\bfX^\top \bM - \widetilde{n}E[\bX])\right], \\
    (c) &=  \underline\bbeta_{A\bX}^\top Var(\overline{\bX}^o)  \underline\bbeta_{A\bX}.
\end{align*}
Letting $\bc_{\bX} = Var(\overline{\bX}^o)^{-1} Cov(\overline{\bX}^o, \overline{Y}^o(0))$ and $\bc_{A\bX} = Var(\overline{\bX}^o)^{-1} Cov(\overline{\bX}^o, \overline{Y}^o(1)-\overline{Y}^o(0))$, then we have
\begin{align*}
    (a) &= \frac{1}{\pi} Var\{\overline{Y}^o(1) - \overline{\bX}^o (\underline{\bbeta}_{\bX} + \underline{\bbeta}_{A\bX})\} + \frac{1}{1-\pi} Var\{\overline{Y}^o(0) - \overline{\bX}^o \underline{\bbeta}_{\bX}\} \\
    &= \frac{1}{\pi} Var\{\overline{Y}^o(1)\} -  \frac{1}{\pi}(\bc_{\bX} + \bc_{A\bX})^\top Var(\overline{\bX}^o)  (\bc_{\bX} + \bc_{A\bX})\\
    &\quad + \frac{1}{\pi}(\bc_{\bX} + \bc_{A\bX} - \underline{\bbeta}_{\bX} - \underline{\bbeta}_{A\bX})^\top Var(\overline{\bX}^o)  (\bc_{\bX} + \bc_{A\bX} - \underline{\bbeta}_{\bX} - \underline{\bbeta}_{A\bX}) \\
    &\quad + \frac{1}{1-\pi} Var\{\overline{Y}^o(0)\} - \frac{1}{1-\pi}\bc_{\bX}^\top Var(\overline{\bX}^o) \bc_{\bX} + \frac{1}{1-\pi}(\bc_{\bX}  - \underline{\bbeta}_{\bX})^\top Var(\overline{\bX}^o)  (\bc_{\bX} - \underline{\bbeta}_{\bX}), \\
    (b) &= 2Cov(\overline{Y}^o(1) - \underline{\bbeta}_{A\bX}^\top \overline{\bX}^o, \overline{\bX}^o) \underline{\bbeta}_{A\bX} - 2Cov(\overline{Y}^o(0), \overline{\bX}^o) \underline{\bbeta}_{A\bX} \\
    &= 2 \bc_{A\bX}^\top  Var(\overline{\bX}^o)\underline{\bbeta}_{A\bX} - 2(c).
\end{align*}
Putting together, we have
\begin{align*}
    v_2 &= \frac{1}{\pi} Var\{\overline{Y}^o(1)\} + \frac{1}{1-\pi} Var\{\overline{Y}^o(0)\} \\
    &\quad - \frac{1}{\pi(1-\pi)} \{\bc_{\bX} + (1-\pi)\bc_{A\bX}\}^\top Var(\overline{\bX}^o)\{\bc_{\bX} + (1-\pi)\bc_{A\bX}\} \\
    &\quad + \frac{1}{\pi(1-\pi)} \{(\bc_{\bX} - \underline{\bbeta}_{\bX}) + (1-\pi)(\bc_{A\bX}-\underline{\bbeta}_{A\bX})\}^\top Var(\overline{\bX}^o)\{(\bc_{\bX} - \underline{\bbeta}_{\bX}) + (1-\pi)(\bc_{A\bX}-\underline{\bbeta}_{A\bX})\}.
\end{align*}
Here 
\begin{align*}
    \underline{\bbeta}_{\bX}&=\left\{Var(\bX) - \frac{\widetilde{n}\underline\tau^2}{\underline\sigma^2+\widetilde{n}\underline\tau^2 }Var(\overline\bX^o)\right\}^{-1}\left\{Cov(\bX, Y(0)) - \frac{\widetilde{n}\underline\tau^2}{\underline\sigma^2+\widetilde{n}\underline\tau^2 }Cov(\overline{\bX}^o, \overline{Y}^o(0))\right\} \\
 \underline{\bbeta}_{A\bX}&=\left\{Var(\bX) - \frac{\widetilde{n}\underline\tau^2}{\underline\sigma^2+\widetilde{n}\underline\tau^2 }Var(\overline\bX^o)\right\}^{-1}\left\{Cov(\bX, Y(1)-Y(0)) - \frac{\widetilde{n}\underline\tau^2}{\underline\sigma^2+\widetilde{n}\underline\tau^2 }Cov(\overline{\bX}^o, \overline{Y}^o(1)-\overline{Y}^o(0))\right\}.    
\end{align*}
\subsection{The individual-level ANCOVA2 estimator}
For the individual-level ANCOVA2 estimator that does not model the random effect, the proof of Theorem 2 implies that its influence function is 
\begin{align*}
    \frac{A-\pi}{\pi(1-\pi) E[N]} \bM^\top (\bY-\bfQ \underline{\widetilde{\bbeta}}) + \frac{1}{E[N]}\underline\bbeta_{A\bX}^\top (\bfX^\top \bM - NE[\bX]),
\end{align*}
where $\underline{\widetilde{\bbeta}} = E[\bfQ^\top diag\{M\} \bfQ]^{-1} E[\bfQ^\top diag\{M\} \bY]$ with $\bfQ$ defined in the proof of Theorem 2, which implies $\underline{\widetilde{\bbeta}}_{\bX} = Var(\bX)^{-1}Cov(\bX,Y(0))$ and $\underline{\widetilde{\bbeta}}_{A\bX} = Var(\bX)^{-1}Cov(\bX,Y(1)-Y(0))$. 
Following a similar proof to Section B.3, we can derive its asymptotic variance is
\begin{align*}
    v_2^{\textup{(ols)}}&=\frac{1}{E[N]^2}\left[\frac{1}{\pi} Var(Y^+(1)) + \frac{1}{1-\pi} Var(Y^+(0))- \boldsymbol{q}^\top Var(\bX^+) \boldsymbol{q} \right.\\\
    &+ \left.\{\boldsymbol{q} - \frac{1}{\pi}(\underline{\widetilde{\bbeta}}_{\bX}+\underline{\widetilde{\bbeta}}_{A\bX})-\frac{1}{1-\pi}\underline{\widetilde{\bbeta}}_{\bX}\}^\top Var(\bX^+) \{\boldsymbol{q}  - \frac{1}{\pi}(\underline{\widetilde{\bbeta}}_{\bX}+\underline{\widetilde{\bbeta}}_{A\bX})-\frac{1}{1-\pi}\underline{\widetilde{\bbeta}}_{\bX}\} \right],
\end{align*}
where $Y^+(a) = \bM^\top \bY(a) - NE[Y(a)]$ for $a =0,1$, $\bX^+ =\bM^\top \bfX - NE[\bX]$ and $\boldsymbol{q} = Var(\bX^+)^{-1} Cov\{\bX^+, \frac{1}{\pi}Y^+(1)+\frac{1}{1-\pi}Y^+(0)\}$.
Therefore, when $N_i = \widetilde{n}$, $v_2^{\textup{(ols)}}$ has the same formula as $v_2$ except that
$\underline{\widetilde{\bbeta}}_{\bX} = Var(\bX)^{-1}Cov(\bX,Y(0))$ and $\underline{\widetilde{\bbeta}}_{A\bX} = Var(\bX)^{-1}Cov(\bX,Y(1)-Y(0))$.

Furthermore, when $\pi = 0.5$, it is easy to verify that $ \frac{1}{\pi}(\underline{\widetilde{\bbeta}}_{\bX}+\underline{\widetilde{\bbeta}}_{A\bX})+\frac{1}{1-\pi}\underline{\widetilde{\bbeta}}_{\bX} = 4 Var(\bX)^{-1} Cov(\bX, Y)$, which is also the formula for $\underline{\widetilde{\bbeta}}_{\bX}$ based on individual-level ANCOVA1. By comparing the expression of $v_1^{\textup{(ols)}}$ and $v_2^{\textup{(ols)}}$, we obtain $v_1^{\textup{(ols)}} = v_2^{\textup{(ols)}}$ if $\pi = 0.5$.

\subsection{The cluster-level ANCOVA1 
 estimator and ANCOVA2 estimator}
By \cite{tsiatis2007}, we get the influence function for cluster-level ANCOVA  estimator based on cluster-averages as
\begin{align*}
    \frac{A-\pi}{\pi(1-\pi)} (\overline{Y}^o - \underline\alpha_0 -  A \underline\alpha_A- \underline\balpha_{\bX}^\top\overline{\bX}^o),
\end{align*}
and its variance is 
\begin{align*}
    v_1^{\textup{(cl)}}     &=  \frac{1}{\pi} Var\{\overline{Y}^o(1)\} + \frac{1}{1-\pi} Var\{\overline{Y}^o(0)\} \\
    &\quad - \frac{1}{\pi(1-\pi)} \{\bc_{\bX} + (1-\pi)\bc_{A\bX}\}^\top Var(\overline{\bX}^o)\{\bc_{\bX} + (1-\pi)\bc_{A\bX}\} \\
    &\quad + \frac{1}{\pi(1-\pi)} \{\bc_{\bX} - \underline{\balpha}_{\bX} + (1-\pi)\bc_{A\bX}\}^\top Var(\overline{\bX}^o)\{\bc_{\bX} - \underline{\balpha}_{\bX} + (1-\pi)\bc_{A\bX}\},
\end{align*}
where $\underline{\balpha}_{\bX} = Var(\overline{X}^o)^{-1} Cov(\overline{X}^o, \overline{Y}^o)$.

The influence function for cluster-level ANCOVA2 estimator based on cluster averages is
\begin{align*}
    \frac{A-\pi}{\pi(1-\pi)} (\overline{Y}^o - \underline\alpha_0 -  A \underline\alpha_A- \underline\balpha_{\bX}^\top\overline{\bX}^o - \underline\balpha_{A\bX}^\top A\overline{\bX}^o ) + \underline\balpha_{A\bX}^\top(\overline{\bX}^o  - E[\bX]),
\end{align*}
and its variance is
\begin{align*}
    v_2^{\textup{(cl)}} &= \frac{1}{\pi} Var\{\overline{Y}^o(1)\} + \frac{1}{1-\pi} Var\{\overline{Y}^o(0)\} \\
    &\quad - \frac{1}{\pi(1-\pi)} \{\bc_{\bX} + (1-\pi)\bc_{A\bX}\}^\top Var(\overline{\bX}^o)\{\bc_{\bX} + (1-\pi)\bc_{A\bX}\}.
\end{align*}

\section{Comparison of variances} \label{sec: comparison}
\subsection{Comparing the mixed models with proof for Proposition 1}
Compared to the unadjusted estimator, the variance reduction of covariate adjustment by the mixed-model ANCOVA1 estimator (assuming constant cluster sizes) is
\begin{align*}
    & \frac{1}{\pi(1-\pi)} \{\bc_{\bX} + (1-\pi)\bc_{A\bX}\}^\top Var(\overline{\bX}^o)\{\bc_{\bX} + (1-\pi)\bc_{A\bX}\} \\
    &\quad - \frac{1}{\pi(1-\pi)} \{\bc_{\bX} - \underline{\bbeta}_{\bX} + (1-\pi)\bc_{A\bX}\}^\top Var(\overline{\bX}^o)\{\bc_{\bX} - \underline{\bbeta}_{\bX} + (1-\pi)\bc_{A\bX}\}
\end{align*}
which is unnecessarily non-negative. For example, consider $Y_{ij} = X_{ij} - n^{-1} \sum_{j=1}^n X_{ij} + \varepsilon_i$. Then $\bc_{\bX}= \bc_{A\bX} = \bzero$. However, $\underline{\bbeta}_{\bX}$ is nonzero since $cov(\bX,Y) = \frac{n-1}{n} Var(\bX)$. Plugging into the above formula, we have a negative quantity, indicating a variance increase. The same example also applies to the mixed-model ANCOVA2 estimator. 

Comparing mixed-model ANCOVA1 and mixed-model ANCOVA2, we first observe that, when $\pi = 0.5$, for mixed-model ANCOVA2 (assuming constant cluster sizes), 
\begin{align*}
    &\underline{\bbeta}_{\bX} + (1-\pi) \underline{\bbeta}_{A\bX}\\
    &=\left\{Var(\bX) - \frac{\widetilde{n}\underline\tau^2}{\underline\sigma^2+\widetilde{n}\underline\tau^2 }Var(\overline\bX^o)\right\}^{-1}\\
    &\quad \left\{Cov(\bX, 0.5Y(1)+0.5Y(0)) - \frac{\widetilde{n}\underline\tau^2}{\underline\sigma^2+\widetilde{n}\underline\tau^2 }Cov(\overline{\bX}^o, 0.5\overline{Y}^o(1)+0.5\overline{Y}^o(0))\right\} \\
    &= \left\{Var(\bX) - \frac{\widetilde{n}\underline\tau^2}{\underline\sigma^2+\widetilde{n}\underline\tau^2 }Var(\overline\bX^o)\right\}^{-1}\left\{Cov(\bX, Y) - \frac{\widetilde{n}\underline\tau^2}{\underline\sigma^2+\widetilde{n}\underline\tau^2 }Cov(\overline{\bX}^o, \overline{Y}^o)\right\}, 
\end{align*}
which is the $\underline{\bbeta}_{\bX}$ in mixed-model ANCOVA1 if the two models yield the same $\underline\sigma^2$ and $\underline{\tau}^2$. Then, they yield the same variance by comparing $v_1$ and $v_2$. 
Unfortunately, mixed-model ANCOVA1 and ANCOVA2 generally lead to different $\underline\sigma^2$ and $\underline{\tau}^2$ due the treatment-covariate interaction terms. To proceed, we first consider condition (a) that the mixed-model ANCOVA2 is correctly specified. By the model of $Y$, we have $Cov(\bX,Y) = Var(\bX) \{\bbeta_{\bX} + \pi \bbeta_{A\bX}\}$ and $Cov(\overline{\bX}^o,\overline{Y}^o) = Var(\overline{\bX}^o) \{\bbeta_{\bX} + \pi \bbeta_{A\bX}\}$, where $\bbeta_{\bX}$ and $\bbeta_{A\bX}$ are the true parameters in mixed-model ANCOVA2. Therefore, the quantity $ \underline{\bbeta}_{\bX} + 0.5 \underline{\bbeta}_{A\bX} = \bbeta_{\bX} + 0.5\bbeta_{A\bX}$  for $\underline{\bbeta}_{\bX}$, $\underline{\bbeta}_{A\bX}$ induced by the working mixed-model ANCOVA2. Similarly, if we use mixed-model ANCOVA1 for estimation, we will have $\underline{\bbeta}_{\bX} = \bbeta_{\bX} + 0.5\bbeta_{A\bX}$, and therefore $v_1=v_2$. We next consider condition (b) where $\bX$ only has cluster-level covariates, which implies $Var(\bX) = Var(\overline{\bX}^o)$ and $Cov(\bX,Y) = Cov(\overline{\bX}^o,\overline{Y}^o)$. As a result $\underline{\bbeta}_{\bX} + 0.5 \underline{\bbeta}_{A\bX} = Var(\bX)^{-1}Cov(\bX,Y)$ in the above derivation. This is also equal to  $\underline{\bbeta}_{\bX}$ from mixed-model ANCOVA1, which implies $v_1=v_2$. Finally, condition (c) that $(\bX_{ij},Y_{ij})$ are independent across $j$, we $Var(\bX) = \widetilde{n}Var(\overline{\bX}^o)$ and $Cov(\bX,Y) = \widetilde{n}Cov(\overline{\bX}^o,\overline{Y}^o)$. This result leads to the same formula for $\underline{\bbeta}_{\bX} + 0.5 \underline{\bbeta}_{A\bX}$ given condition (b), and hence yields $v_1=v_2$.

When $\pi \ne 0.5$, their variance comparison is indeterminate. For example, consider $Y_{ij} = A_i(X_{ij} - n^{-1} \sum_{j=1}^n X_{ij}) + \varepsilon_i$. Then $\bc_{\bX}= \bc_{A\bX} = Cov(\bX, Y(0))=\bzero$. Their variance difference is $q \{\pi^2 - (1-\pi)^2\}$ for a constant $q$. In this example, the mixed-model ANCOVA estimator is more precise if $\pi < 0.5$, while the other is more precise if $\pi > 0.5$.
\subsection{Proof of Proposition 2}
Let $v^{\textup{(ols)}}_2$ denote the variance of the individual-level ANCOVA2 estimator.  We further denote
\begin{align*}
    T &= \pi Y(0) + (1-\pi) Y(1)\\
    \overline{T} &= \pi \overline{Y}^o(0) + (1-\pi)\overline{Y}^o(1) \\
    d &= \bc_{\bX} + (1-\pi) \bc_{A\bX} = Var(\overline{\bX}^o)^{-1} Cov(\overline{\bX}^o, \overline{T}) \\
    \zeta &= \underline{\bbeta}_{\bX} + (1-\pi)\underline{\bbeta}_{A\bX} =\left\{Var(\bX) - \frac{\widetilde{n}\underline\tau^2}{\underline\sigma^2+\widetilde{n}\underline\tau^2 }Var(\overline\bX^o)\right\}^{-1}\left\{Cov(\bX, T) - \frac{\widetilde{n}\underline\tau^2}{\underline\sigma^2+\widetilde{n}\underline\tau^2 }Cov(\overline{\bX}^o, \overline{T})\right\} \\
    \widetilde{\zeta} &= \widetilde{\underline{\bbeta}}_{\bX} + (1-\pi)\widetilde{\underline{\bbeta}}_{A\bX} = Var(\bX)^{-1} Cov(\bX, T).
\end{align*}
Then the definition of $\zeta$ implies $Var(\overline\bX^o)d = \frac{\underline\sigma^2+\widetilde{n}\underline\tau^2}{\widetilde{n}\underline\tau^2}Var(\bX)(\widetilde{\zeta} - \zeta) + Var(\overline\bX^o)\zeta$ (assuming $\underline\tau^2 > 0$). If $\underline\tau^2 = 0$, then it is trivial that mixed-model ANCOVA2 and individual-level ANCOVA2 are equivalent asymptotically.

Using the variance formula derived in Sections B.4.3 and B.4.4 (assuming constant cluster sizes), we obtain that
\begin{align*}
    &\pi(1-\pi)(v_2 - v^{\textup{(ols)}}_2)\\
    &=  (d-\zeta)^\top Var(\overline{\bX}^o) (d-\zeta) -  (d-\widetilde\zeta)^\top Var(\overline{\bX}^o) (d-\widetilde\zeta) \\
    &= \zeta^\top Var(\overline{\bX}^o)\zeta - \widetilde\zeta^\top Var(\overline{\bX}^o)\widetilde\zeta + 2(\widetilde\zeta - \zeta)^\top Var(\overline{\bX}^o)d \\
    &= \zeta^\top Var(\overline{\bX}^o)\zeta - \widetilde\zeta^\top Var(\overline{\bX}^o)\widetilde\zeta + 2(\widetilde\zeta - \zeta)^\top \left\{\frac{\underline\sigma^2+\widetilde{n}\underline\tau^2}{\widetilde{n}\underline\tau^2}Var(\bX)(\widetilde{\zeta} - \zeta) + Var(\overline\bX^o)\zeta\right\} \\
    &= - (\widetilde{\zeta} - \zeta)^\top Var(\overline{\bX}^o) (\widetilde{\zeta} - \zeta) +\frac{2\underline\sigma^2+2\widetilde{n}\underline\tau^2}{\widetilde{n}\underline\tau^2}(\widetilde\zeta - \zeta)^\top  Var(\bX)(\widetilde{\zeta} - \zeta) \\
    &=(\widetilde\zeta - \zeta)^\top  \left\{\frac{2\underline\sigma^2+2\widetilde{n}\underline\tau^2}{\widetilde{n}\underline\tau^2}Var(\bX) - Var(\overline{\bX}^o)\right\}(\widetilde{\zeta} - \zeta) \\
    &\ge 0,
\end{align*}
where the last inequality comes from the fact that $Var(\bX) - \frac{\widetilde{n}\underline\tau^2}{\underline\sigma^2+\widetilde{n}\underline\tau^2 }Var(\overline\bX^o)$ is positive-definite (since it is proportional to $E\left[(\bfX - E[\bfX])^\top \underline{\bfV}(\bfX - E[\bfX])\right]$, which is invertible by the regularity condition that $\underline{\bbeta}_{\bX}$ is unique).

For mixed-model ANCOVA1 and individual-level ANCOVA1, we have the same results under $\pi = 0.5$ following a similar proof.

\subsection{Proof of Proposition 3}
When $\pi = 0.5$, we have
\begin{align*}
    v_1^{\textup{(cl)}}     &=  \frac{1}{\pi} Var\{\overline{Y}^o(1)\} + \frac{1}{1-\pi} Var\{\overline{Y}^o(0)\} \\
    &\quad - \frac{1}{\pi(1-\pi)} \{\bc_{\bX} + (1-\pi)\bc_{A\bX}\}^\top Var(\overline{\bX}^o)\{\bc_{\bX} + (1-\pi)\bc_{A\bX}\}
\end{align*}
and hence $v_1^{\textup{(cl)}} \le v_1^{\textup{(ols)}}$ and $v_1^{\textup{(cl)}} \le v_1$ from the variance formulas of $v_1^{\textup{(ols)}}$ and $v_1$ assuming constant cluster sizes. 
Following a similar proof, we have $v_2^{\textup{(cl)}} \le v_2^{\textup{(ols)}}$ and $v_2^{\textup{(cl)}} \le v_2$.

\subsection{An example showing the mixed-model ANCOVA1 estimator can be more efficient than the cluster-level ANCOVA1 estimator}
Suppose that $E[\bY_i|A_i,\bfX_i] = \beta_0 \bone_n + A_i\beta_A \bone_n + \bbeta_{\bX}^\top \bfX_i$ and $Var(\bY_i - E[\bY_i|A_i,\bfX_i]) = \sigma^2 \bfI_n + \tau^2 \bone_n \bone_n^\top$ for some $\btheta^* = (\beta_0, \beta_A, \bbeta_{\bX}, \sigma^2, \tau^2) \in \boldsymbol{\Theta}$. Then it is straightforward to show that $\underline{\btheta} = \btheta^*$. Then the asymptotic variance of $\widehat{\Delta}_1$ is $v_1 = 4 \{E\left[\frac{N}{\sigma^2 + N \tau^2}\right]\}^{-1}$. By Jensen's inequality, we have 
$$v_1 \ge 4\left\{ \frac{E[N]}{\sigma^2 + E[N] \tau^2}\right\}^{-1},$$ 
which indicates that the variation of cluster sizes will result in precision loss compared to constant cluster sizes.

For comparison between $v_1$ and $v_1^{\textup{(cl)}}$, we have $\underline{\balpha}_{\overline\bX^o} = {\bbeta}_{\bX}$ and
\begin{align*}
    v_1^{\textup{(cl)}} &= 4Var(\overline{Y}^o-\Delta^* A - \underline{\balpha}_{\overline\bX^o}^t\overline\bX^o) \\
    &= 4 Var\left(\frac{1}{N}\bM^\top\{\bY - E[\bY|A,\bfX]\}\right) \\
    &= 4 E\left[Var\left(\frac{1}{N}\bM^\top\{\bY - E[\bY|A,\bfX]\}\bigg | N\right)\right] + 4Var\left(E\left[\frac{1}{N}\bM^\top\{\bY - E[\bY|A,\bfX]\}\bigg | N\right]\right) \\
    &= 4E\left[\frac{\sigma^2 + N \tau^2}{N}\right].
\end{align*}
Hence by the H\"older's inequality, we have
\begin{align*}
    \frac{v_1^{\textup{(cl)}}}{v_1} = E\left[\frac{\sigma^2 + N \tau^2}{N}\right]E\left[\frac{N}{\sigma^2 + N \tau^2}\right] \ge 1,
\end{align*}
which indicates that $v_1^{\textup{(cl)}} \ge v_1$. In the special case that $N$ is fixed, then $v_1^{\textup{(cl)}} = v_1$.

\section{The cluster-level ANCOVA2 estimator  based on scaled cluster totals}\label{ssec: cluster-totals}
Consider a regression model
\begin{equation}
   \frac{Y_i^+}{\widehat{\mu}_{N}} \sim \beta_0 + \beta_A A_i + (\beta_N + \beta_{AN} A_i) (\frac{N_i}{\widehat{\mu}_{N}} - 1) + (\bbeta_{\bX} + \bbeta_{A\bX} A_i)^\top (\frac{\bX_i^+}{\widehat{\mu}_{N}} - \frac{N_i\widehat{\bmu}_{\bX}}{\widehat{\mu}_{N}}),
\end{equation}
where $Y_i^+ = \sum_{j \in \mathcal{O}_i} Y_{ij}$, $\bX_i^+ = \sum_{j \in \mathcal{O}_i} \bX_{ij}$, $\widehat{\mu}_{N} = \frac{1}{m} \sum_{i=1}^m N_i$ and $\widehat{\bmu}_{\bX} = \frac{1}{m\widehat{\mu}_{N}} \sum_{i=1}^m\bX_i^+$.  This model is referred to as adjusted OLS based on cluster totals in \cite{su2021model}.

The related estimating equation is 
\begin{displaymath}
    \bpsi(\bY, A,\bfX, \bM; \btheta) = \left(\begin{array}{c}
    \boldsymbol{Z}(Y^+/\mu_N - \boldsymbol{Z}^\top \bbeta)  \\
    N - \mu_N\\
   N \bmu_{\bX} - \bX^+
   \end{array}\right),
\end{displaymath}
where $\bbeta = (\beta_0, \beta_A, \beta_N, \beta_{AN}, \bbeta_{\bX}^\top, \bbeta_{A\bX})$ and $\boldsymbol{Z} = (1, A, (N/\mu_N-1), A(N/\mu_N-1), (\bX^+{}^\top - N\bmu_{\bX})/\mu_N, A(\bX^+{}^\top - N\bmu_{\bX})/\mu_N)^\top$. The influence function for $\widehat{\beta}_A$ is
\begin{align*}
    \frac{A-\pi}{\pi(1-\pi)} (Y^+/\underline\mu_N - \boldsymbol{Z}^\top \underline\bbeta)  - \frac{N-\underline\mu_N}{\underline\mu_N} (\Delta - \underline\beta_{AN}) + \frac{1}{\underline\mu_N}  \underline{\bbeta}_{A\bX}^\top (\bX^+ - N E[\bX]).
\end{align*}
Since we have $\Delta = \underline\beta_{AN} = \underline\beta_{A}$, $E[Y(0)] = \underline{\beta}_N = \underline{\beta}_0$, $\underline\mu_N = E[N]$, the above influence function can be simplified to
\begin{align*}
    \frac{A-\pi}{\pi(1-\pi)E[N]} \{Y^+ - N\underline\beta_0 - AN\underline\beta_A - (\underline\bbeta_{\bX} + \underline\bbeta_{A\bX} A)^\top (\bX^+ - N E[\bX])\} + \frac{1}{E[N]}  \underline{\bbeta}_{A\bX}^\top (\bX^+ - N E[\bX]),
\end{align*}
with $\underline\bbeta_{\bX} = Var(\bX^+)^{-1} Cov(\bX^+, Y^+(0))$ and $\underline\bbeta_{A\bX} = Var(\bX^+)^{-1} Cov(\bX^+, Y^+(1) - Y^+(0))$. 
Of note, this influence function is the same as it for the individual-level ANCOVA2 estimator except that $\underline{\widetilde\bbeta}_{\bX} = Var(\bX)^{-1} Cov(\bX, Y(0))$ $\underline{\widetilde\bbeta}_{A\bX} = Var(\bX)^{-1} Cov(\bX, Y(1)-Y(0))$ there.

For any vector $\bbeta_{\bX}, \bbeta_{A\bX}$, the variance of the above influence function is 
\begin{align*}
    &\frac{1}{E[N]^2}\left[\frac{1}{\pi} Var(Y^+(1) - NE[Y(1)]) + \frac{1}{1-\pi} Var(Y^+(0) - NE[Y(0)])\right.\\
    &- \boldsymbol{q}^\top Var(\bX^+) \boldsymbol{q} \\
    &+ \left.\{\boldsymbol{q} - \frac{1}{\pi}(\bbeta_{\bX}+\bbeta_{A\bX})-\frac{1}{1-\pi}\bbeta_{\bX}\}^\top Var(\bX^+) \{\boldsymbol{q}  - \frac{1}{\pi}(\bbeta_{\bX}+\bbeta_{A\bX})-\frac{1}{1-\pi}\bbeta_{\bX}\} \right],
\end{align*}
where $\boldsymbol{q} = Var(\bX^+)^{-1} Cov\{\bX^+, \frac{1}{\pi}Y^+(1)+\frac{1}{1-\pi}Y^+(0)\} $. Therefore, the optimal $\bbeta_{\bX}$ and $\bbeta_{A\bX}$ should be $Var(\bX^+)^{-1} Cov(\bX^+, Y^+(0))$ and $Var(\bX^+)^{-1} Cov(\bX^+, Y^+(1) - Y^+(0))$, which are $\underline\bbeta_{\bX}$ and $\underline\bbeta_{A\bX}$ produced by the cluster-level ANCOVA2 with scaled cluster totals. This result implies the superiority of the precision of $\widehat{\Delta}^{\textup{(sct)}}_2$ than $\widehat{\Delta}^{\textup{(ols)}}_2$

When $N$ is a constant, then the influence function of cluster-level ANCOVA2 based on scaled cluster totals is exactly the same as that based on cluster averages; therefore they yield the same asymptotic variance. However, if $N$ varies, their variance comparison is indeterminate in general. 
Consider the simplest case where $\bX$ is an empty set and $Y(1)=Y(0)=Y$ with $E[Y(0)]=0$. Then the asymptotic variance for $\widehat{\Delta}^{\textup{(sct)}}_2$ is $\{\pi(1-\pi)\}^{-1} E[N]^{-2} Var(Y^+)$ and the asymptotic variance for $\widehat{\Delta}^{\textup{(cl)}}_2$ is $\{\pi(1-\pi)\}^{-1} Var(Y^+/N)$. 
To see why the variance comparison result is indeterminate, consider (i) $Y^+ = NY$, i.e., all outcomes are equal (perfectly correlated), which implies $Var(Y^+)/E[N]^2 = E[N^2]/E[N]^2 Var(Y) \ge  Var(Y)=Var(Y^+/N)$, and (ii) $Y_{ij}$ are independent conditional on $N$, which implies $Var(Y^+)/E[N]^2 = \frac{1}{E[N]} Var(Y) \le E[1/N] Var(Y) = E[\sum_{j=1}^N E[Y_{ij}^2|N]/N^2]= E[Var(Y^+/N|N)]=  Var(Y^+/N)$, where we use the assumption that $N$ is independent of $Y_{ij}$ and $E[Y_{ij}]=0$ as we assumed. Therefore, we can see that $\widehat{\Delta}^{\textup{(sct)}}_2$ can be either more precise or less precise than $\widehat{\Delta}^{\textup{(cl)}}_2$.

Comparing cluster-level ANCOVA2 based on scaled cluster totals and mixed-model ANCOVA2, the latter can be more efficient than the former, as we demonstrate in an example below. Let the mixed-model ANCOVA2 model be correctly specified with $\pi = 0.5$, then the asymptotic variance of mixed-model ANCOVA2 estimator is $4\{E[\frac{N}{\sigma^2 + N\tau^2}]\}^{-1}$, while the asymptotic variance of cluster-level ANCOVA2 based on  scaled cluster totals is $4 E[N]^{-2} E[N\sigma^2 + N^2 \tau^2]$. Through Holder's inequality, we have $E[\frac{N}{\sigma^2 + N\tau^2}] E[N\sigma^2 + N^2 \tau^2] \ge E[\sqrt{\frac{N}{\sigma^2 + N\tau^2}}\sqrt{N\sigma^2 + N^2 \tau^2}]^2 = E[N]^2$, which implies that $4\{E[\frac{N}{\sigma^2 + N\tau^2}]\}^{-1} \le 4 E[N]^{-2} E[N\sigma^2 + N^2 \tau^2]$. Hence, the mixed-model ANCOVA2 estimator has a smaller variance than in this special case. In more general cases, their variance comparison is indeterminate and can depend on features of the data-generating process.

\section{Empirical comparison of REML and ML}\label{sec:REML}

As discussed in Section 6 of the main manuscript, here we provide additional numerical results (in the following Table 1 and Table 2) to compare the maximum restricted maximum likelihood (REML) and maximum likelihood (ML) estimation for the mixed-model ANCOVA parameters; the former is known to reduce the bias of the variance component estimators, while the latter is a more standard approach which we consider to prove our main results. The purpose of this additional comparison is provide some preliminary evidence that the our theoretical results may also hold for REML-based mixed-model ANCOVA analysis of CRTs. A formal proof of the robustness of the REML estimator under arbitrary model misspecification is subject to additional research. 

\begin{figure}
	\centering
	\includegraphics[width=0.01\textwidth]{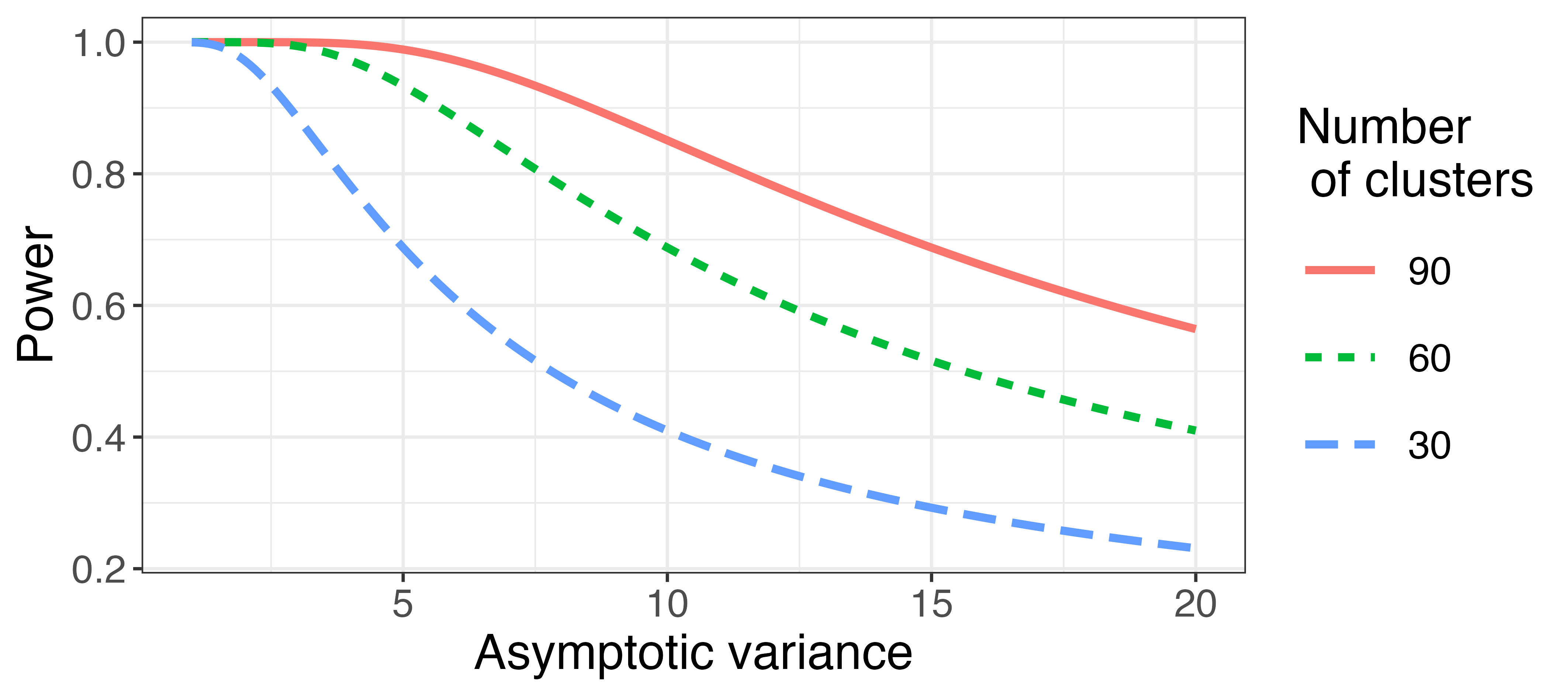}
\end{figure}

\begin{table}[ht]
\centering
\caption{Simulation results for Scenarios 1--3 comparing ML and REML estimation in mixed models. The performance metrics are bias, empirical standard error (Emp SE), averaged model-based standard error (ASE), coverage probability of nominal 0.95 confidence intervals based on normal approximation and model-based standard error (CP), and relative efficiency to the mixed-model unadjusted ML estimator (RE).}
\renewcommand{\arraystretch}{0.8}
\begin{tabular}{ccrrrrrrr}
  \hline
& & Mixed models & Method & Bias & Emp SE & ASE & CP & RE\\ 
  \hline
   \multirow{8}{*}{Scenario 1} & \multirow{4}{*}{$m=20$} & \multirow{2}{*}{unadjusted} & ML & 0.00& 0.95 & 1.00 & 0.94 &1.00\\ 
   & & & REML &  0.00& 0.96 & 0.98 & 0.95 &1.00\\ 
    \cline{3-9}
    &  & \multirow{2}{*}{ANCOVA1} & ML & -0.01 & 0.97 & 1.01 & 0.93  & 0.96\\ 
   & & & REML &  0.00& 0.97 & 0.98 & 0.94 &0.96\\    
 \cline{2-9}
& \multirow{4}{*}{$m=200$} & \multirow{2}{*}{unadjusted} & ML & 0.01 & 0.29 & 0.30 & 0.95 &1.00\\ 
   & & & REML &  0.01 & 0.29 & 0.30 & 0.95 &1.00\\ 
       \cline{3-9}
    &  & \multirow{2}{*}{ANCOVA1} & ML & 0.01 & 0.30 & 0.30 & 0.95 &0.96\\ 
   & & & REML &  0.01 & 0.30 & 0.30 & 0.95 &0.96\\  
 \hline
   \multirow{8}{*}{Scenario 2} & \multirow{4}{*}{$m=20$} & \multirow{2}{*}{unadjusted} & ML &  -0.02 & 1.29 & 1.29 & 0.93  &1.00\\ 
   & & & REML & -0.02 & 1.29 & 1.28 & 0.94&1.00 \\ 
    \cline{3-9}
    &  & \multirow{2}{*}{ANCOVA1} & ML & 0.00& 1.09 & 1.09 & 0.92   & 1.40\\ 
   & & & REML &  0.00& 1.09 & 1.08 & 0.94 &1.40\\    
 \cline{2-9}
& \multirow{4}{*}{$m=200$} & \multirow{2}{*}{unadjusted} & ML &-0.01 & 0.41 & 0.41 & 0.95 &1.00\\ 
   & & & REML & -0.01 & 0.41 & 0.41 & 0.95 &1.00\\ 
       \cline{3-9}
    &  & \multirow{2}{*}{ANCOVA1} & ML &  0.00& 0.34 & 0.34 & 0.95 &1.40\\ 
   & & & REML & 0.00& 0.34 & 0.34 & 0.95 &1.40\\  
 \hline
  \multirow{8}{*}{Scenario 3} & \multirow{4}{*}{$m=20$} & \multirow{2}{*}{unadjusted} & ML &  -0.01 & 1.00 & 1.02 & 0.94 &1.00\\
   & & & REML & -0.01 & 1.00 & 1.01 & 0.95 &1.00\\ 
    \cline{3-9}
    &  & \multirow{2}{*}{ANCOVA1} & ML & -0.01 & 0.95 & 0.99 & 0.94 &1.11\\ 
   & & & REML & -0.01 & 0.95 & 0.95 & 0.94 &1.11\\ 
 \cline{2-9}
& \multirow{4}{*}{$m=200$} & \multirow{2}{*}{unadjusted} & ML &0.01 & 0.31 & 0.31 & 0.95 &1.00\\ 
   & & & REML & 0.01 & 0.31 & 0.31 & 0.95 &1.00\\ 
       \cline{3-9}
    &  & \multirow{2}{*}{ANCOVA1} & ML &  0.01 & 0.29 & 0.29 & 0.95 &1.12 \\ 
   & & & REML &  0.01 & 0.29 & 0.29 & 0.95 &1.12\\ 
 \hline
\end{tabular}
\end{table}

\begin{table}[ht]
\centering
\caption{Summary of data analyses results comparing ML and REML estimation in mixed models: point estimate of the average treatment effect (Est), model-based estimator for standard error (SE), 95\% confidence interval (CI), and proportion variance reduction compared to the unadjusted ML estimator (PVR). Positive (negative) PVR indicates that covariate adjustment leads to variance reduction (inflation).}
\begin{tabular}{crrrrrr}
  \hline
Study name & Estimators & method & Est & SE & 95\% CI & PVR \\ 
  \hline
\multirow{4}{*}{TSSSH}& \multirow{2}{*}{mixed-model unadjusted} & ML & -1.29 & 2.08 & (-5.36, 2.78) &- \\ 
 & & REML & -1.29 & 2.08 & (-5.36, 2.78) & 0\%\\
\cline{2-7}
& \multirow{2}{*}{mixed-model ANCOVA} & ML &  -2.22 & 2.09 & (-6.32, 1.87) & -1\% \\ 
& & REML & -2.22 & 1.98 & (-6.11, 1.67) & 9\% \\ 
   \hline
\multirow{4}{*}{IECDZ}& \multirow{2}{*}{mixed-model unadjusted} & ML & 0.08 & 0.14 & (-0.19, 0.35) & - \\
 & & REML & 0.08 & 0.14 & (-0.19, 0.35) & 0\% \\ 
\cline{2-7}
& \multirow{2}{*}{mixed-model ANCOVA} & ML & 0.08 & 0.15 & (-0.22, 0.37) & -21\% \\ 
& & REML & 0.08 & 0.14 & (-0.20, 0.36) & -9\% \\ 
   \hline
\multirow{4}{*}{WFHS}& \multirow{2}{*}{mixed-model unadjusted} & ML & 0.16 & 0.07 & (0.01, 0.30) & - \\
 & & REML & 0.16 & 0.07 & (0.01, 0.30) & -1\% \\ 
\cline{2-7}
& \multirow{2}{*}{mixed-model ANCOVA} & ML &  0.21 & 0.05 & (0.12, 0.31) & 56\% \\ 
& & REML & 0.21 & 0.05 & (0.11, 0.31) & 55\% \\ 
   \hline
\end{tabular}
\end{table}

% \subsection{Regularity conditions when the cluster size is constant}
% Given assumptions 1-2 and $N_i = n$, we show that the following conditions imply the regularity conditions given in the Appendix:
% (a) the distribution $\mathcal{P}^{\bW}$ has bounded fourth moments, (b) $Var(\bX)$ is non-singular, 
% (c) for any $\bbeta_{\bX} \in \mathbb{R}^p$, $Y_{\bullet,j} - \bbeta_{\bX} \bX_{\bullet,j}$ are not identical for $j = 1,\dots, n$ with probability 1, and (d) $Var\{Y(a) - E[Y(a)|X]\} > 0$ for $a = 0,1$.

% We first derive $\underline{\btheta}$. Direct calculation shows that
% \begin{align*}
%     \underline{\sigma}^2 &= \frac{n}{n-1} \{Var(Y-  A\Delta^* - \underline{\bbeta}_{\bX}^\top\bX) - Var(\overline{Y} - A\Delta^* - \underline{\bbeta}_{\bX}^\top\overline\bX )\} \\
%     \underline{\tau}^2 &= \frac{n}{n-1}Var(\overline{Y} - A\Delta^* - \underline{\bbeta}_{\bX}^\top\overline\bX ) - \frac{1}{n-1}Var(Y-  A\Delta^* - \underline{\bbeta}_{\bX}^\top\bX), \\
%     \underline{\beta}_0 &= E[Y - A\Delta^* - \underline{\bbeta}_{\bX}^\top\bX],
% \end{align*}
% and $\underline{\bbeta}_{\bX}$ is given in the proof of Theorem 2.

% We next show the regularity condition (2). For $\underline{\sigma}^2$, we have 
% $$Var(Y-  A\Delta^* - \underline{\bbeta}_{\bX}^\top\bX) \ge Var(\overline{Y} - A\Delta^* - \underline{\bbeta}_{\bX}^\top\overline\bX ),$$
%  where the equality holds if and only if $Y_{\bullet,j}-  A\Delta^* - \underline{\bbeta}_{\bX}^\top\bX_{\bullet,j}$ is the same across $j$ with probability one. Since we assume that

\bibliographystyle{apalike}
\bibliography{references}